%% file: jthaler_TASI_Master.tex
\documentclass{article}
\usepackage{jheppub}

\pdfoutput=1

\usepackage{epstopdf}

\usepackage{mathtools}
\usepackage{dsfont}
\usepackage[all]{xy}

\usepackage{amsmath}

\numberwithin{equation}{section}
\numberwithin{figure}{section}
\numberwithin{table}{section}

\usepackage{alphalph}
\makeatletter
\newalphalph{\AlphMult}[mult]{\@alph}{26}
\makeatother

\makeatletter
\def\@fpheader{\\}    %This is to kill "Prepared for Submission for JHEP"
\makeatother

\usepackage{color}
\definecolor{darkgreen}{rgb}{0,0.5,0}

\newcommand{\Sec}[1]{Sec.~\ref{#1}}
\newcommand{\Secs}[2]{Secs.~\ref{#1} and \ref{#2}}

\newcommand{\Tab}[1]{Table~\ref{#1}}

\newcommand{\Fig}[1]{Fig.~\ref{#1}}

\newcommand{\Eq}[1]{Eq.~(\ref{#1})}
\newcommand{\Eqs}[2]{Eqs.~(\ref{#1}) and (\ref{#2})}
\newcommand{\Eqss}[3]{Eqs.~(\ref{#1}), (\ref{#2}) and  (\ref{#3})}
\newcommand{\Ref}[1]{Ref.~\cite{#1}}   %??
\newcommand{\Refs}[1]{Refs.~\cite{#1}}  %??

\newcommand{\nn}{\nonumber}

\newcommand{\bone}{{\boldsymbol 1}}
\newcommand{\btwo}{{\boldsymbol 2}}
\newcommand{\bthree}{{\boldsymbol 3}}

\newcommand{\D}{D}
\newcommand{\U}{\textrm{U}}
\newcommand{\SU}{\textrm{SU}}

\newcommand{\epsilonbar}{\overline{\epsilon}}
\newcommand{\Dbar}{\overline{D}}
\newcommand{\Q}{Q}
\newcommand{\Qbar}{\overline{Q}}
\newcommand{\thetabar}{\overline{\theta}}
\newcommand{\sigmabar}{\overline{\sigma}}
\newcommand{\alphadot}{{\dot{\alpha}}}
\newcommand{\betadot}{{\dot{\beta}}}
\newcommand{\bV}{{\boldsymbol V}}
\newcommand{\bF}{{\boldsymbol f}}
\newcommand{\bVcomp}{{\boldsymbol V}_{\! \rm comp}}
\newcommand{\bR}{{\boldsymbol R}}
\newcommand{\bX}{{\boldsymbol X}}
\newcommand{\bY}{{\boldsymbol Y}}
\newcommand{\bK}{{\boldsymbol K}}
\newcommand{\bQ}{{\boldsymbol Q}}
\newcommand{\bU}{{\boldsymbol U}}
\newcommand{\bD}{{\boldsymbol D}}
\newcommand{\bL}{{\boldsymbol L}}
\newcommand{\bN}{{\boldsymbol N}}
\newcommand{\bE}{{\boldsymbol E}}
\newcommand{\bH}{{\boldsymbol H}}
\newcommand{\bPhi}{{\boldsymbol \Phi}}
\newcommand{\bOmega}{{\boldsymbol \Omega}}

\newcommand{\bPhicomp}{{\boldsymbol \Phi}_{\rm comp}}

\newcommand{\bS}{{\boldsymbol S}}
\newcommand{\bW}{{\boldsymbol W}}
\newcommand{\bWalpha}{{{\boldsymbol W}_{\! \alpha}}}
\renewcommand{\i}{{\bar{\imath}}}
\renewcommand{\j}{{\bar{\jmath}}}
\renewcommand{\k}{{\bar{k}}}
\renewcommand{\l}{{\bar{l}}}
\newcommand{\vev}[1]{\left\langle #1 \right\rangle}

\newcommand{\hc}{\text{h.c.}}

\newcommand{\be}{\begin{equation}}
\newcommand{\ee}{\end{equation}}
\newcommand{\bea}{\begin{eqnarray}}
\newcommand{\eea}{\end{eqnarray}}

\newcommand{\jbox}[1]{*=<6.5ex>[Fo]{#1}}
\newcommand{\kbox}[1]{*=<6.5ex>[]{#1}}

\begin{document}

\title{TASI 2012: Super-Tricks for Superspace}
\author{Daniele Bertolini,}
\author{Jesse Thaler,}
\author{and Zoe Thomas}
\affiliation{Center for Theoretical Physics, Massachusetts Institute of Technology, Cambridge, MA 02139, USA}
\emailAdd{danbert@mit.edu}
\emailAdd{jthaler@mit.edu}
\emailAdd{ztt@mit.edu}

\abstract{These lectures from the TASI 2012 summer school outline the basics of supersymmetry (SUSY) in 3+1 dimensions.  Starting from a ground-up development of superspace, we develop all of the tools necessary to construct SUSY lagrangians.  While aimed at an introductory level, these lectures incorporate a number of ``super-tricks'' for SUSY aficionados, including SUSY-covariant derivatives, equations of motion in superspace, background field methods, and non-linear realizations of goldstinos.}

\preprint{MIT-CTP {4444}}
\maketitle

%\bodymatter

%%% Needed to make aa,bb,cc footnotes
%\renewcommand*{\thefootnote}{\AlphMult{\value{footnote}}}

\section{Introduction}

Supersymmetry (SUSY) is a well-motivated extension of the Standard Model (SM), with rich implications for collider physics and cosmology.  This, however, will not be the topic of these lectures.  At TASI 2012, Pierce and Papucci went into considerable depth on topics of direct phenomenological relevance, including the Minimal SUSY SM (MSSM), SUSY at colliders, and aspects of SUSY model building.  Instead, the goal of these lectures is to introduce the basics of SUSY in 3+1 dimensions using the language of superspace \cite{Salam:1974yz,Ferrara:1974ac}, an essential tool for any serious student of SUSY.  In addition, these lectures will present a number of super-tricks which we have found useful in our own SUSY research.

In general, the role of symmetries is to relate the properties of different physical states.  This is so familiar that we often forget how important it is.  For example, an electron moving in the $y$-direction is distinct from an electron moving in the $z$-direction, and only because of rotational symmetry can we describe both modes in terms of a common electron field. Relativistic quantum field theory exhibits the full Poincar\'{e} symmetry, which relates particles traveling with different momenta and joins particles and antiparticles into a common multiplet.  By packaging all of the electron creation and annihilation operators into a single quantum field $\Psi(x)$ that lives in space-time, we make the Poincar\'{e} symmetry manifest.

The amazing feature of SUSY is that it relates the properties of bosons and fermions.  For example, a spin-1/2 fermionic quark field in the SM could have a spin-0 bosonic ``squark" superpartner with the same gauge quantum numbers but different spin-statistics.  While it is possible to describe quarks and squarks as separate quantum fields with couplings related by SUSY, it is more convenient to  describe them as components of a common SUSY multiplet.  Superspace introduces an auxiliary fermionic coordinate $\theta$ (and its complex conjugate $\thetabar$) such that bosonic and fermionic degrees of freedom can be packaged into a single superfield $\bPhi(x,\theta,\thetabar)$, extending space-time to make SUSY manifest. 

Whether or not SUSY is actually realized in nature, there are a variety of reasons one might want to relate the properties of bosons and fermions.  For physics beyond the SM, SUSY can be used to regulate quantum corrections to the spin-0 Higgs potential.  The mass parameter for a spin-1/2 fermion is always protected by a chiral symmetry, whereas the mass parameter for a generic spin-0 boson is quadratically sensitive to high-scale physics.  When spin-1/2 and spin-0 fields are part of the same SUSY multiplet, the spin-0 boson inherits the spin-1/2 chiral mass protection.  For understanding generic features of quantum field theories, SUSY theories have powerful constraints like holomorphy that make it possible to robustly predict the properties of theories even at strong coupling.  Even for weakly-coupled theories, SUSY relates complicated calculations involving fermions to simpler calculations involving scalars.

Beyond the intrinsic motivations for SUSY, these lectures will emphasize how superspace simplifies the construction and manipulation of SUSY lagrangians.  As we will see, the structure of $\mathcal{N} = 1$ SUSY can be largely understood from the transformation properties of the superspace coordinates $\theta$ and $\thetabar$.  We will briefly touch on supergravity (SUGRA) as well.

\subsection{About These Lectures}
\label{subsec:aboutlectures}

These lecture notes mimic the original four TASI 2012 lectures, but with some embellishments and corrections compared to what you can find on the online videos.  The topics covered are:
\begin{itemize}
\item \textbf{Introducing Superspace}.
\item \textbf{Fermions and Sfermions}.
\item \textbf{SUSY Gauge Theories}.  
\item \textbf{SUSY Breaking and Goldstinos}.
\end{itemize}
For the aficionados, each lecture will feature one super-trick, namely, a technique that can be explained at the introductory level but often features prominently in advanced SUSY research.  These tricks are:
\begin{itemize}
\item \#1: SUSY-covariant derivatives.  (\Sec{subsec:IntroD})
\item \#2: Equations of motion in superspace.  (\Sec{subsec:eominsuperspace})
\item \#3: Background field methods.  (\Sec{subsec:BackgroundField})
\item \#4: Non-linear goldstino multiplets.  (\Sec{subsec:nonlineargoldstino})
\end{itemize}
Of course, superspace itself is the real super-trick, and we hope these lecture notes will help both beginning and advanced students gain familiarity and dexterity with superspace methods.

While writing these lecture notes, we relied on a number of helpful SUSY references, including Wess \& Bagger \cite{Wess:1992cp}, Martin's Supersymmetry Primer \cite{Martin:1997ns}, Terning's Modern Supersymmetry \cite{Terning:2006bq}, Luty's 2004 TASI lecture notes \cite{Luty:2005sn}, and the notes of Dumitrescu and Komargodski \cite{Dumitrescu:2011zz}.  We have also found \Refs{Weinberg:2000cr,Freedman:2012zz,Binetruy:2006ad,Signer:2009dx,Drees:1996ca,Baer:2006rs,Argyres,Dine:2007zp} to be helpful SUSY resources.  
Unfortunately, each of these references uses a different set of  conventions (and the three of us have our own differing preferences on factors of $-1$, $i$, $2$, and $\sqrt{2}$), but we have tried to make these notes self-consistent.  

Compared to standard treatments of SUSY, these lectures are more ``ground-up''.  Instead of starting with the SUSY algebra, we start with the task of simply trying to put bosons and fermions into multiplets.  We take the point of view that SUSY should be thought of in the context of effective lagrangians, so we make an effort to point out important dimension-5 and dimension-6 interactions that are generically present.  When talking about SUSY breaking, instead of talking about specific SUSY-breaking models, we focus on the special role of the goldstino. 

As a disclaimer, here are some topics that we will \emph{not} tell you about in these lectures.  We will not tell you how SUSY can solve the gauge hierarchy problem.\footnote{In particular, we will not tell you that SUSY ensures that the Higgs mass parameter is only logarithmically sensitive to the scale of SUSY-breaking.}
We will not tell you how SUSY naturally furnishes a dark matter candidate.\footnote{In particular, we will not tell you that in the MSSM a $\mathbf{Z}_2$ symmetry called $R$-parity is invoked to protect against proton decay, which in turn renders the lightest SUSY particle stable, and thus a dark matter candidate.}  We will not discuss the phenomenology of the MSSM.\footnote{Among other things, we will not tell you about gauge coupling unification, the need for two Higgs doublets for anomaly cancellation, or sparticle phenomenology at the LHC.}  We will, however, tell you that in order for SUSY to be realized in nature, it must be spontaneously broken in a hidden sector (see \Sec{subsec:paradigm}).   And we will pontificate in the conclusion about the status of SUSY after the 8 TeV LHC run.

\subsection{About Us}

To give you some context for these lecture notes, we want to briefly say something about our SUSY background.  

Jesse wrote his first paper on SUSY in 2005 \cite{ArkaniHamed:2005px}.  Since then, he has found it enormously helpful to have a bag of tricks on hand to simplify SUSY calculations and develop SUSY intuition.  The specific tricks included in these lectures were motivated by some of his recent papers.
\begin{itemize}
\item \textit{Goldstini} \cite{Cheung:2010mc,Cheung:2011jq} --- Goldstini arise when SUSY is broken independently by more than one sector.  Deriving the spectrum and couplings of goldstini is made simpler using \textbf{non-linear goldstino multiplets}.
\item \textit{Flavor Mediation} \cite{Craig:2012yd,Craig:2012di} --- The (approximate) global flavor symmetries of the SM can be lifted to (spontaneously broken) gauged flavor symmetries.  These flavor gauge groups can mediate SUSY breaking, and the resulting soft spectrum is most easily calculated using \textbf{background field methods}.
\item \textit{Anomaly Mediation from AdS$_4$ SUSY} \cite{D'Eramo:2012qd} --- Anomaly-mediated SUSY breaking arises from uplifting AdS$_4$ SUSY to flat space broken SUSY.  This physics is easiest to understand in the \textbf{conformal compensator formalism of SUGRA} (which is unfortunately a super-trick that was not covered in the original TASI 2012 lectures, but is explained fairly concisely in \Ref{Cheung:2011jp}).  Various cross-checks are easiest to perform using \textbf{SUSY-covariant derivatives}.
\end{itemize}
Finally, the super-trick of \textbf{equations of motion in superspace} is a classic technique for constructing effective SUSY lagrangians.  

Daniele and Zoe were participants in the TASI 2012 summer school, and graciously agreed to help Jesse write and edit these notes.  At TASI, Daniele gave a presentation about his work on visible sector SUSY breaking \cite{Bertolini:2011tw}, and Zoe gave a presentation about her work on AdS$_4$ SUSY and anomaly mediation \cite{D'Eramo:2012qd}.

Without further ado, let us begin our study of superspace.

\input{jthaler_TASI_Lecture1.tex}

\input{jthaler_TASI_Lecture2.tex}

\input{jthaler_TASI_Lecture3.tex}

\input{jthaler_TASI_Lecture4.tex}

\section{Summary}

We hope you have found these TASI lectures notes useful, regardless of your previous background in SUSY.  For the complete neophyte, we have outlined the basic structure and motivation for SUSY:
\begin{itemize}
\item SUSY is a symmetry that relates the properties of bosons and fermions.  Specially, SUSY establishes relationships between masses and couplings in the lagrangian. 
\item Chiral multiplets package a spin-0 complex scalar with a spin-1/2 Weyl fermion.
\item Vector multiplets package a spin-1 gauge boson with a spin-1/2 gaugino fermion.
\item In order to be realized in nature, SUSY must be spontaneously broken in a hidden sector, leading to mass splittings between particles and sparticles in the visible sector.
\end{itemize} 
For the serious SUSY student, we have presented the following tools for constructing SUSY lagrangians:
\begin{itemize}
\item One can conveniently package SUSY components into a superfield $\bS(x^\mu, \theta^\alpha, \thetabar^{\alphadot})$ that lives in superspace.
\item Superspace consists of the ordinary space-time coordinate $x^\mu$ augmented by a Grassmann coordinate $\theta^\alpha$ (and its complex conjugate $\thetabar^{\alphadot}$) with the following transformation properties
\begin{align}
\theta^\alpha &\to \theta^\alpha + \epsilon^\alpha,\\
\thetabar^{\alphadot} &\to \thetabar^{\alphadot} + \overline{\epsilon}^{\alphadot},\\
x^\mu &\to x^\mu + i\epsilon\sigma^\mu\thetabar+i\bar{\epsilon}\sigmabar^\mu\theta.
\end{align}
\item A generic SUSY lagrangian of chiral and vector multiplets can be written as
\be
\begin{split}
\mathcal{L}=&\int d^4\theta \, \bK(\bPhi^{\dagger\i} e^{\bV},\bPhi^j)\\
&+\int d^2\theta \, \bW(\bPhi^i)+\hc\\
&+\int d^2\theta \, \bF_{ab}(\bPhi^i)\bW^{\alpha a}\bW_{\! \alpha}^b +\hc\\
&+D_\alpha,\partial_\mu\textrm{ terms}.
\end{split}
\ee
While the leading effects can be captured by the K\"ahler potential $\bK$, superpotential $\bW$, and gauge kinetic function $\bF_{ab}$, woe unto the student who forgets the possible presence of $\partial_\mu, \D_\alpha, \Dbar_{\alphadot}$ terms in an effective SUSY lagrangian.
\item SUSY breaking (in flat space) inevitably leads to a massless fermion called the goldstino.  Crucially, SUSY-breaking mass splittings are always accompanied by corresponding goldstino-particle-sparticle couplings.
\end{itemize}
And for the SUSY experts, we hope that you use are able to use these super-tricks in your own research:
\begin{itemize}
\item SUSY-covariant derivatives, for constructing new superfields out of old superfields and simplifying the construction of SUSY lagrangians.
\item Equations of motion in superspace, for finding the vacuum structure of the theory and integrating out heavy (SUSY-preserving) thresholds at tree-level.
\item Background field methods, for analytically continuing static background fields to full superfields, thereby capturing important one-loop (and sometimes two-loop) effects.
\item Non-linear multiplets for goldstinos, for abstracting the important features of the hidden sector and relating SUSY-breaking mass-splittings to their corresponding goldstino couplings. 
\end{itemize}
To fully understand the implications of SUSY and SUSY-breaking, one needs to learn about SUGRA (unfortunately not covered in depth in these lectures).  We recommend the conformal compensator formalism of SUGRA \cite{Kugo:1982mr,Kugo:1982cu,Gates:1983nr}, which makes it possible to capture the leading effects of SUGRA in \emph{global} superspace \cite{Cheung:2011jp}.  At minimum, you should be aware that if SUSY is realized in nature, then it must be SUSY in AdS$_4$ space, spontaneously broken to yield the (nearly) flat space SUSY-breaking vacuum we see today.

By design, these lectures have only briefly mentioned the phenomenological consequences of SUSY, since these were covered in depth in other TASI lectures.  That said, it is no secret that recent analyses of 7 TeV and 8 TeV LHC data have placed stringent bounds on TeV-scale SUSY, so it is worth talking about the status of SUSY in 2012.  
\begin{itemize}
\item First, symmetries are a powerful tool for understanding the behavior of quantum field theories, and finding genuinely new types of symmetries like SUSY is rare.  Even if SUSY is not realized in nature, SUSY allows us to better understand the generic features of quantum field theory, as evidenced in these lectures by the power of SUSY background field methods.
\item Second, by learning about superspace, one learns how to make symmetries manifest by introducing new ``fake'' coordinates.  An excellent recent example of this is explained in  Sundrum's notes on the AdS/CFT correspondence \cite{Sundrum:2011ic}, where an extra spatial dimension ``emerges'' from trying to make conformal symmetries manifest.  It is always valuable to stretch one's notion of space-time, and superspace in particular offers a new way to think about the relationship between bosons and fermions.  
\item Third, SUSY is by now the lingua franca for physics beyond the SM, and even non-SUSY extensions of the SM often share phenomenology features with SUSY.  As with the language of QCD (confinement, asymptotic freedom, chiral symmetry breaking, and so on), the language of SUSY allows one to quickly explain the features of many quantum field theories.  
\item Finally, it is not at all clear (as of 2012) whether SUSY is really as tightly constrained as one might naively expect.  The canonical flavor-blind SUSY theories with low fine-tuning of the Higgs potential are  strongly disfavored, but more exotic (and some might say more generic) SUSY theories with novel flavor structures or some degree of fine-tuning are only starting to be tested.  Therefore, some of the pessimism surrounding SUSY is surely misplaced, and we prefer to remain optimistic that hints of a SUSY-like theory will emerge in the 14 TeV LHC data (or elsewhere).
\end{itemize}
With that, we would like to thank all of the participants at TASI 2012 for a fun week!

\section*{Acknowledgments}

We thank Martin Schmaltz for encouraging us to write these lecture notes.  We thank Yoni Kahn, Grant Larsen, Matthew McCullough, and Yiming Xu for helpful comments on the manuscript.  This work is supported by the U.S. Department of Energy (DOE) under cooperative research agreement DE-FG02- 05ER-41360.  J.T. is supported by the DOE Early Career research program DE-FG02-11ER-41741. D.B. is partly supported by Istituto Nazionale di Fisica Nucleare (INFN) through a �Bruno Rossi� Fellowship.

\bibliographystyle{JHEP}
\bibliography{jthaler_TASI_Bib}

\end{document}

%% file: jthaler_TASI_Lecture1.tex
\section{Introducing Superspace}
\label{sec:lecture1}

In this lecture, we will introduce the motivation and structure of superspace.  After a discussion of Weyl and Grassmann notation, we  show how translations in superspace can be used to determine the SUSY algebra.  En route to constructing a generic SUSY lagrangian, we will introduce our first super-trick involving SUSY-covariant derivatives.

\subsection{Why Superspace?}

\label{subsec:WhySuperspace}

SUSY relates the properties of bosons and fermions, but in ordinary relativistic quantum field theory, bosons and fermions are represented by very different objects.  For example, a spin-0 boson is described by a complex-valued scalar field $\phi(x)$, while a spin-1/2 fermion is described by a Grassmann-valued Weyl field $\psi_\alpha(x)$ (with a Lorentz spinor index, no less).  In order to make SUSY manifest, we want to somehow package bosons and fermions into a single object.

To do so, we introduce the technique of superspace, which augments the ordinary  space-time coordinates with an additional Grassmann spinor $\theta^\alpha$ (and its complex conjugate $\thetabar^{\alphadot}$):
\be
x^\mu \rightarrow \{ x^\mu, \theta^\alpha, \thetabar^{\alphadot}\}.
\ee
A field that depends on $\{ x^\mu, \theta^\alpha, \thetabar^{\alphadot}\}$ is called a superfield,\footnote{A true superfield must also have well-behaved properties under a SUSY transformation. This is discussed in more detail in \Sec{subsec:whatissuper}.} which automatically packages boson and fermion fields into multiplets.  While it is possible to describe SUSY theories using ordinary space-time alone, superspace makes it simpler to identify SUSY-invariants and write SUSY lagrangians. 

On one level, the Grassmann spinor $\theta^\alpha$ simply serves as a placeholder.  As an analogy, consider the case of a vector with components $v_x$, $v_y$, and $v_z$.  This vector can be expressed either as a list of separate components or as a single vector object with the help of placeholders $\hat{x}$, $\hat{y}$, and $\hat{z}$:
\be
\{v_x, v_y, v_z\} \quad \text{vs.} \quad \vec{v} = v_x \hat{x} + v_y \hat{y} + v_z \hat{z}.
\ee
The second notation allows objects with different formal properties to be summed together into a common object $\vec{v}$.  Similarly, given a SUSY multiplet that contains a spin-0 boson $\phi$ and a spin-1/2 fermion $\psi_\alpha$, we can write the SUSY multiplet either as a list of components or in a superfield:
\be
\{\phi, \psi_\alpha, \ldots\} \quad \text{vs.} \quad \boldsymbol{\Phi} = \phi + \theta^\alpha \psi_\alpha + \ldots
\ee
Because both $\theta^\alpha$ and $\psi_\alpha$ are Grassmann spinors, the two terms in $\boldsymbol{\Phi}$ have the same statistics and can therefore be summed together.  

This analogy between vectors and superfields goes even a bit deeper.  In the case of vectors, one can describe rotations either in terms of active rotations (rotation of components) or passive rotations (rotation of basis vectors).  If one knows how $\hat{x}$, $\hat{y}$, and $\hat{z}$ transform under rotations, then one can easily determine how $v_x$, $v_y$, and $v_z$ transform.  In the case of SUSY, one can describe SUSY transformations in terms of manipulations of $\theta^\alpha$, and we will ``derive'' the SUSY algebra by considering the possible transformation properties of $\theta^\alpha$.

In terms of how $\theta^\alpha$ transforms under SUSY, however, $\theta^\alpha$ behaves less like a $\hat{x}$-like unit vector and more like a genuine coordinate.  For example, one can consider arbitrary functions of $\theta^\alpha$ just like one can have arbitrary functions of $x_\mu$ (though there are important differences since $\theta^\alpha$ is a Grassmann spinor).  In addition, SUSY transformations act like translations on $\theta^\alpha$
\be
\theta^\alpha \to \theta^\alpha + \epsilon^\alpha,
\ee
just as translations on space-time coordinates act as $x^\mu \to x^\mu + \delta^\mu$.  We will see explicitly how this works in \Sec{subsec:SUSYtranslations}.

\subsection{Invitation to Two-Component Notation}

In these lectures we will use exclusively two-component spinor notation, also known as Weyl spinors.  While it is possible to do superspace manipulations using four-component notation (as in \Refs{Weinberg:2000cr,Freedman:2012zz}), Weyl spinors are far more convenient, since they are true irreducible representations of the Lorentz group.  For the uninitiated, we recommend the excellent notes of Dreiner, Haber, and Martin \cite{Dreiner:2008tw}.  We will give a quick motivation for why two-component notation is the ``natural'' notation to use, with the hope that this will give you sufficient motivation to learn this topic on your own.  Throughout these notes we use the $(+,-,-,-)$ metric convention.

Consider the following way of packaging the four-vector $x_\mu = \{t,x,y,z\}$ into a two-by-two matrix
\be
X = \sigma^\mu x_\mu = \left(\begin{array}{cc} t + z & x - i y \\ x+ i y &  t -z \end{array}\right),
\ee
where
\be
\label{eq:sigmamatrices}
\sigma^\mu_{\alpha \alphadot}=\{\mathds{1},\sigma_1,\sigma_2,\sigma_3\}
\ee 
is defined in terms of the Pauli matrices, and we will explain the reason for the dotted index in a moment.  The matrix $X$ is hermitian ($X^\dagger = X$), and its determinant is the familiar Lorentz-invariant norm\footnote{This unambiguously shows that $(+,-,-,-)$ is the ``correct'' metric signature.}
\be
\det X = t^2 - x^2 - y^2 - z^2 = x_\mu x^\mu.
\ee

The Lorentz group is the set of transformations that leave $x_\mu x^\mu$ invariant, or equivalently, the set of transformations that leave the determinant of $X$ fixed while keeping $X = X^\dagger$.  The most general Lorentz transformation acting on $X$ is matrix multiplication by an arbitrary complex matrix $P$ with determinant $1$:
\be
\label{eq:Lorentztransformation}
X \to P X P^\dagger.
\ee
Clearly, this transformation leaves $X$ hermitian.  Note that
\be
\det{P X P^\dagger} = |\!\det{P}|^2 \det{X},
\ee
so $\det P$ could in principle have an arbitrary phase factor, but this phase can be pulled out of the matrix $P$ since it has no effect on the components of $X$. This (special linear) transformation in \Eq{eq:Lorentztransformation} is the group $\mathrm{SL}(2,\mathbb{C})$ which is the (double) covering group of the Lorentz group $\mathrm{SO}(3,1)$.

While \Eq{eq:Lorentztransformation} might seem cumbersome for describing the transformation of Lorentz four-vectors, it makes the transformations of spinors manifest.  Spinors are simply objects that transform under $P$ as
\be
\label{eq:alphaindex}
%\psi'{}^\alpha = \psi^\beta P_\beta{}^\alpha ,
\psi'_\alpha = P_{\alpha}{}^{\beta} \psi_\beta,
\ee
where $\alpha = \{1,2\}$ is called an undotted index (i.e.\ corresponding to the $(1/2,0)$ representation of the Lorentz group).  (Anti-)spinors with a dotted index transform in the complex conjugate representation under $P^*$ as
\be
\label{eq:alphadotindex}
\overline{\psi}'{}_{\alphadot} = P^*_{\alphadot}{}^{\betadot} \overline{\psi}_\betadot = \overline{\psi}_{\betadot} P^{\dagger \betadot}{}_{\alphadot},
\ee
(i.e.\ the $(0,1/2)$ representation).  We can build arbitrary representations of the Lorentz group by considering objects with different numbers of dotted and undotted indices.  Ordinary four-vectors with a $\mu$ Lorentz index can be converted into objects with two spinor indices with the help of the $\sigma^\mu$ matrices from \Eq{eq:sigmamatrices}:
\be
x_{\alpha \dot{\beta}} \equiv x_\mu \sigma^\mu_{\alpha \dot{\beta}}.
\ee
We see immediately that \Eqs{eq:alphaindex}{eq:alphadotindex} then imply \Eq{eq:Lorentztransformation}, which is a nice consistency check.

The antisymmetric matrix
\be
\label{eq:epsilonmatrix}
\epsilon_{\alpha\beta} =\epsilon_{\alphadot\betadot}=
\begin{pmatrix*}[r]
0 & -1\\
1 & 0
\end{pmatrix*}
\ee 
is invariant under Lorentz transformations, so we can use the $\epsilon_{\alpha \beta}$ (and its inverse $\epsilon^{\alpha \beta}$) to lower (and raise) indices
\be
\chi_\alpha = \epsilon_{\alpha \beta} \chi^\beta, \qquad \chi^\alpha = \epsilon^{\alpha \beta} \chi_\beta,
\ee
and analogously for dotted indices.\footnote{For raising and lowering indices on the $\epsilon$ tensor itself, note that $\epsilon^{\alpha \beta} \epsilon_{\beta \gamma} = \delta^\alpha{}_\gamma$, but $\epsilon^{\alpha \beta} \epsilon_{\gamma \beta} = - \delta_\gamma{}^\alpha$.} When raising and lowering indices on the $\sigma^\mu$ matrix, it is convenient to define 
\be
\sigmabar^{\mu\alphadot\alpha}\equiv\epsilon^{\alphadot\betadot}\epsilon^{\alpha\beta}\sigma^\mu_{\beta\betadot}=\left\{\mathds{1},-\sigma_1,-\sigma_2,-\sigma_3\right\}.
\ee
Because there is a sign flip when raising and lowering indices, we have to define a convention for suppressing indices, and we use descending undotted indices and ascending dotted indices as shown below:
\be
\begin{tabular}{c c}
$\alpha$ &\\
&$\alpha$
\end{tabular}
\qquad
\begin{tabular}{c c}
&$\alphadot$\\
$\alphadot$ &\\
\end{tabular}
\ee

The power of two-component notation is that one can write Lorentz-invariant objects simply by matching indices (in particular avoiding the contortions necessary in Dirac notation to write down a Majorana mass).  The easiest Lorentz invariants we can build are
\be
\psi^\alpha\chi_\alpha\equiv\psi\chi, \qquad \overline{\psi}_{\alphadot}\overline{\chi}^\alphadot\equiv\overline{\psi}\overline{\chi}.
\ee
Since fermions anti-commute, we have
\be
\psi\chi=\psi^\alpha\chi_\alpha=-\chi_\alpha\psi^\alpha=+\chi^\alpha\psi_\alpha=\chi\psi.
\ee
However there is an annoying minus sign when manipulating the following Lorentz-covariant combination:
\be
\overline{\psi}\sigmabar^\mu\chi=\overline{\psi}_{\alphadot}\sigmabar^{\mu\alphadot\beta}\chi_\beta=-\chi^\alpha\sigma^\mu_{\alpha\betadot}\overline{\psi}^{\betadot}=-\chi\sigma^\mu\overline{\psi}.
\ee
It is often convenient to use the notation
\be
\label{eq:sigmamunu}
\sigma^{\mu\nu} \equiv \frac{1}{4}(\sigma^\mu\sigmabar^\nu-\sigma^\nu\sigmabar^\mu), \qquad \sigmabar^{\mu\nu}\equiv\frac{1}{4}(\sigmabar^\mu\sigma^\nu-\sigmabar^\nu\sigma^\mu).
\ee

When taking the hermitian conjugate of expressions involving multiple spinors, it is conventional for all the spinors to reverse order.  Combined with the hermiticity of the sigma matrices, this implies
\begin{align}
(\psi \chi)^\dagger & = \overline{\chi} \overline{\psi}, \\
(\psi \sigma^\mu \overline{\chi})^\dagger & = \chi \sigma^\mu \overline{\psi}, \\
(\psi \sigma^\mu \sigmabar^\nu \chi)^\dagger & = \overline{\chi} \sigmabar^\nu \sigma^\mu \overline{\psi},
\end{align}
and so forth.

One can often simplify complicated expressions using the following Schouten and completeness identities:
\begin{align}
A_\alpha B_\beta - A_\beta B_\alpha & = \epsilon_{\alpha \beta} A^\gamma B_\gamma, \label{eq:Schouten}\\
\sigma_{\mu \alpha \alphadot} \sigmabar^{\mu \beta \betadot} & = 2 \delta_\alpha{}^\beta \delta_{\alphadot}{}^\betadot.  \label{eq:completeness} 
\end{align}
Together, these imply many useful Fierz identities, including
\begin{align}
(\chi \psi) (\eta \lambda) & = - (\chi \eta) (\lambda \psi) - (\chi \lambda) (\psi \eta), \label{eq:Fierz1} \\
(\overline{\chi} \sigmabar^\mu \psi) (\eta \sigma_\mu \overline{\lambda} ) & = - 2 (\chi^\dagger \lambda^\dagger) (\psi \eta), \label{eq:Fierz2} \\
(\theta \chi) (\theta \psi) & = -\frac{1}{2} (\theta \theta) (\chi \psi). \label{eq:Fierz3}
\end{align}

The sigma matrices satisfy identities reminiscent of the familiar relation $\{\gamma^\mu,\gamma^\nu\} = 2 \eta^{\mu \nu}$ of four-component notation, namely
\begin{align}
\sigma^\mu_{\alpha \alphadot} \sigmabar^{\nu \alphadot \beta} + \sigma^{\nu}_{\alpha \alphadot} \sigmabar^{\mu \alphadot \beta} & = 2 \eta^{\mu \nu} \delta_\alpha{}^\beta, \\
\sigmabar^{\mu \alphadot \alpha} \sigma^\nu_{\alpha \betadot} + \sigmabar^{\nu \alphadot \alpha} \sigma^\mu_{\alpha \betadot} & = 2 \eta^{\mu \nu} \delta^{\alphadot}{}_{\betadot}.
\end{align}

Finally, given a Dirac (four-component) spinor, one can decompose it into two Weyl spinors via
\be
\Psi_D=
\begin{pmatrix}
\psi_\alpha\\
\overline{\psi^c}^{\alphadot}
\end{pmatrix},
\ee
and gamma matrices can be similarly decomposed
\begin{align}
\gamma^\mu=
\begin{pmatrix}
0 & \sigma^\mu\\
\sigmabar^\mu &0
\end{pmatrix},&\qquad
\gamma_5=
\begin{pmatrix*}[r]
-1 & ~0~\\
0 & ~1~
\end{pmatrix*}.
\end{align}
Note that ``$c$" in $\psi^c$ is just a label ($\psi$ and $\psi^c$ are two separate quantum fields, completely unrelated if one does not invoke equations of motion), while the bar stands for complex conjugation.  Both $\psi$ and $\psi^c$ are (left-handed) Weyl spinors.

\subsection{Grassmann Coordinates}
\label{subsec:Grassmann}

Now that we have settled on two-component notation, we can talk about the Grassmann coordinate $\theta^\alpha$ (and its complex conjugate $\thetabar^{\alphadot}$).  Here are the basics of Grassmann manipulation.

A generic Grassmann variable $\eta$ (without a Lorentz index) is an anti-commuting object
\be
\{\eta, \eta\} = 0 \quad \Rightarrow \quad \eta^2 = 0.
\ee
That means an arbitrary function of $\eta$ can be expressed in terms of its (finite) Taylor expansion as
\be
f(\eta) = a + b \eta.
\ee
There are a variety of methods to extract the components of $f(\eta)$.  We will often use the notation $|_0$ to indicate setting all Grassmanns to zero
\be
f(\eta)|_0 = a.
\ee
We define derivatives and integrals as acting the same way on Grassmanns:
\be
\label{eq:etaderivative}
\frac{\partial}{\partial \eta} (a + b \eta) = b, \qquad \int d\eta \, (a + b \eta) = b.
\ee
Note that the integral definition implies translational invariance
\be
\int d\eta \, f(\eta + \epsilon) = \int d\eta f(\eta),
\ee
which has been extensively used, for example, in \Ref{Cheung:2011jq}.
Also, note that
\be
\int d\eta \, \frac{\partial}{\partial \eta} (a + b \eta) = 0,
\ee
which will allow us to use integration by parts in superspace.

In order to package bosons and fermions into a common multiplet, we need Grassmann variables with spinor indices: 
\be
\theta_\alpha=\{\theta_1,\theta_2\}, \qquad \thetabar_{\alphadot}=\{\thetabar_1,\thetabar_2\}.
\ee  
These objects anti-commute with each other such that
\be
\theta_1\theta_1=\theta_2\theta_2=0, \qquad \theta_1 \theta_2 = - \theta_2 \theta_1.
\ee
We can use the $\epsilon$ matrix in \Eq{eq:epsilonmatrix} to form Lorentz-invariant combinations of the $\theta$s:
\begin{align}
\theta^2&\equiv\theta^\alpha\theta_\alpha=\epsilon^{\alpha \beta} \theta_\beta \theta_\alpha = -2 \theta_1 \theta_2,\\
\thetabar^2&\equiv\thetabar_{\alphadot}\thetabar^{\alphadot}=\epsilon^{\alphadot\betadot}\thetabar_\alphadot \thetabar_\betadot=2\thetabar_1\thetabar_2,\\
\theta^4&\equiv\thetabar^2\theta^2.
\end{align}
Note that the superscripts in the above relations denote powers, not components.  There is a potential factor of 2 confusion that arises from the relation
\be
\label{eq:ThetaSquare}
\theta_\alpha\theta_\beta=\frac{1}{2}\epsilon_{\alpha\beta}\theta^2.
\ee

We can define derivatives and integrals just as in \Eq{eq:etaderivative}
\be
\label{eq:GrassDerInt}
\frac{\partial}{\partial\theta^\alpha}\theta^\beta=\delta_\alpha^\beta, \qquad  \frac{\partial}{\partial\thetabar_{\alphadot}}\thetabar_{\betadot}=\delta^{\alphadot}_{\betadot}, \qquad \int d\theta_\alpha(a+b\theta^\beta)=b\delta_\alpha^\beta.
\ee
We will often use the combinations
\be
\label{eq:GrassDiff}
d^2\theta\equiv -\frac{1}{4}d\theta^\alpha d\theta^\beta\epsilon_{\alpha\beta}, \qquad
d^2\thetabar \equiv -\frac{1}{4}d\thetabar_\alphadot d\thetabar_\betadot\epsilon^{\alphadot\betadot}, \qquad
d^4\theta\equiv d^2\theta d^2\thetabar,
\ee
such that
\be
\label{eq:integrals}
\int d^2\theta~\theta^2=1,\qquad \int d^2\thetabar~\thetabar^2=1,\qquad \int d^4\theta~\theta^4=1.
\ee
With all the notation out of the way, we are ready now to talk about SUSY multiplets.

\subsection{Generic SUSY Multiplets}
\label{subsec:SUSYmultiplets}

A generic scalar supermultiplet is
\be
\bS(x^\mu, \theta^\alpha, \thetabar^{\alphadot}),
\ee
which depends on the Grassmann spinor placeholders/coordinates $\theta^\alpha$. Throughout this paper we will use boldface letters to indicate superfields. This object is an overall Lorentz scalar, but it contains spin-0, spin-1/2, and spin-1 components.  Because of the Grassmann nature of our placeholders,
the Taylor expansion is exact:
\be
\begin{array}{lcclclcl}
\label{eq:GenericMultiplet}
\bS&=&&a&+&\theta\xi&+&\theta^2b\\
&&+&\thetabar\overline{\chi}&+&\thetabar\sigmabar^\mu\theta v_\mu&+&\theta^2\thetabar\overline{\zeta}\\
&&+&\thetabar^2c&+&\thetabar^2\theta\eta&+&\theta^4d.
\end{array}
\ee
We can express this more visually as
\be
\bS=\quad
\begin{array}{c|ccc}
&\theta^{0}&\theta^{1} &\theta^{2}\\
\hline
\rule[-3mm]{0mm}{.8cm}
\thetabar^{0}&a&\xi_\alpha &b\\
\thetabar^{1}&\overline{\chi}^\alphadot &v_\mu&\overline{\zeta} ^\alphadot\\
\thetabar^{2}&c&\eta_\alpha &d\\
\end{array}
\ee
where $a,b,c,d$ are complex scalars, $\xi,\chi,\zeta,\eta$ are Weyl spinors, $v_\mu$ is a complex vector,
and $\theta^{n}$ and $\thetabar^{n}$  stand for $n$ powers of $\theta$ or $\thetabar$.
Note that $\bS$ contains exactly eight (complex) fermionic  and eight (complex) bosonic degrees of freedom.
One might think that the superfield could include additional objects like $\theta_\alpha\theta_\beta z^{\alpha\beta}$ 
but one can easily show that such a term is equivalent to  $\theta^2 b$ by using \Eq{eq:ThetaSquare}.
We use the language  ``lower'' (``higher'') to refer to components that involve fewer (more) factors of $\theta$ or $\thetabar$, such that $a$ ($d$) is the lowest (highest) component of $\bS$.

The above discussion is for a Lorentz-scalar superfield, but superfields can carry additional Lorentz structure.  Indeed, there is no problem with 
\be
\boldsymbol{S}_\nu (x^\mu, \theta^\alpha, \thetabar^{\alphadot}),
\ee
as long as the components of the superfields have extra Lorentz indices as well.  When the superfield has a spinor index, this can lead to terms where the indices on $\theta$ are uncontracted due to reducible Lorentz structure. For example, the $\xi_{\alpha \beta}$ component of a superfield $\bS_\beta$ could contain $\omega \epsilon_{\alpha \beta}$, leading to an expansion like
\be
\boldsymbol{S}_\beta = \lambda_\beta - \omega \theta_\beta  + \ldots 
\ee
When discussing gauge theories in \Sec{sec:lecture3}, we will make regular use of superfields with spinor indices.

In these notes, we will restrict ourselves to one set of Grassmann coordinates $\{\theta^\alpha, \thetabar^{\alphadot}\}$, which is known as $\mathcal{N}=1$ SUSY.  Adding more sets of $\theta$ yields higher $\mathcal{N}$ SUSY, which will not be covered here.

\subsection{Translations in Superspace}
\label{subsec:SUSYtranslations}

In the previous subsection, we successfully packaged bosons and fermions into a common multiplet.  However, as we will see in \Sec{subsec:whatissuper}, there is a crucial difference between a package of bosons/fermions and a true superfield, which is a package of bosons/fermions that have the correct transformation properties under SUSY.  This is the same distinction between a list (i.e.~a collection of components) and a vector (i.e.~a collection of components that transform into each other in a certain way under rotations).  The real meat of SUSY is in how the components of a supermultiplet are related to each other, and we will now ``derive'' SUSY by considering translations in superspace.

In ordinary space-time, translational invariance
\be
x^\mu \rightarrow x^\mu + \delta^\mu
\ee
is an important subset of the full Poincar\'{e} space-time symmetry.  Using the Taylor expansion for infinitesimal $\delta^\mu$, this translation acts on fields as:
\begin{align}
\label{eq:xTaylorexpand}
\phi(x^\mu) &\to \phi( x^\mu + \delta^\mu) \\&= \phi(x^\mu) +  \delta^\mu \partial_\mu \phi(x^\mu),
\end{align}
where we are using  the notation $\partial_\mu \equiv \frac{\partial}{\partial x^\mu}$. 

It is natural to define translational invariance in superspace via
\be
\theta^\alpha \to \theta^\alpha + \epsilon^\alpha,
\ee
where $\epsilon^\alpha$ is an infinitesimal Grassmann parameter.  This (passive) transformation of the coordinate can be interpreted instead as an (active) transformation of the components.  For example, starting with
\be
\label{eq:SFshift}
\boldsymbol{\Phi}(\theta^\alpha) = \phi + \theta^\alpha \psi_\alpha + \ldots,
\ee
translations yield
\begin{align}
\label{eq:thetashift}
\boldsymbol{\Phi}(\theta^\alpha + \epsilon ^\alpha) &= \phi + (\theta^\alpha + \epsilon ^\alpha) \psi_\alpha + \ldots\\
&= (\phi  + \epsilon ^\alpha \psi_\alpha) + \theta^\alpha \psi_\alpha + \ldots
\end{align}
so the components transform as $\phi \to \phi + \epsilon ^\alpha \psi_\alpha$ (with $\psi_\alpha$ left fixed).  As desired, we have related bosons to fermions!  However, we know that boson and fermion kinetic terms have differing numbers of derivatives, so in order to successfully build a SUSY lagrangian, we must somehow combine $\theta^\alpha$ translations with space-time derivatives.

The key to SUSY is that the shift of the fermionic coordinate $\theta^\alpha$ is accompanied by a translation of the ordinary bosonic coordinate $x_\mu$ as well
\be
\begin{split}
\label{eq:SUSYtransformations}
\theta^\alpha &\to \theta^\alpha + \epsilon^\alpha,\\
\thetabar^{\alphadot} &\to \thetabar^{\alphadot} + \overline{\epsilon}^{\alphadot},\\
x^\mu &\to x^\mu + \Delta^\mu,
\end{split}
\ee
where
\be
\Delta^\mu \equiv i\epsilon\sigma^\mu\thetabar+i\overline{\epsilon}\sigmabar^\mu\theta .
\ee
We could have guessed the form of $\Delta^\mu$, since this is the unique real four-vector object one can build that is linear in $\epsilon$ and has the right dimension.\footnote{Note that from \Eq{eq:SFshift}, $\theta$ (and thus $\epsilon$) must have mass dimension $[\theta]=-1/2$.} 

Let us now act this SUSY coordinate shift on our generic supermultiplet from \Eq{eq:GenericMultiplet}:
\begin{align}
 \label{eq:SuperfieldSUSY}
\bS(x^\mu, \theta^\alpha, \thetabar^{\alphadot}) & \to \bS(x^\mu + \Delta x^\mu, \theta^\alpha  + \epsilon^\alpha, \thetabar^{\alphadot} + \overline{\epsilon}^{\alphadot}) \\
&= \bS(x^\mu, \theta^\alpha, \thetabar^{\alphadot}) \nn \\
& \quad + \left(\Delta^\mu \partial_\mu+\epsilon^\alpha\partial_\alpha+\overline{\epsilon}_\alphadot\overline{\partial}^\alphadot\right)\bS(x^\mu, \theta^\alpha, \thetabar^{\alphadot}) \nn,
\end{align}
where we have used the Taylor expansion up to the first order, both for ordinary and Grassmann coordinates.
Here, we are using the notation $\partial_\alpha\equiv\frac{\partial}{\partial\theta^\alpha}$ and $\overline{\partial}^\alphadot\equiv\frac{\partial}{\partial\thetabar_\alphadot}$.

We see immediately that translations in superspace act in two different ways.  First, the shift $\theta^\alpha \to \theta^\alpha  + \epsilon^\alpha$ relates higher components of the superfield to lower components as in \Eq{eq:thetashift}.  Second, because $\Delta^\mu$ contains factors of $\theta^\alpha$, the translation $x^\mu \to x^\mu + \Delta^\mu$ relates \emph{derivatives} of lower components to higher components.  This is crucial for relating the kinetic terms for bosons and fermions.

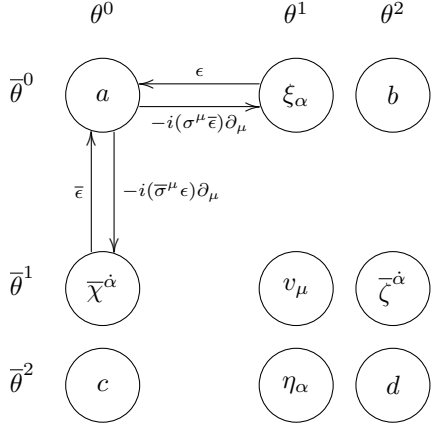
\begin{figure}
\begin{center}
\makebox{\xymatrix@R=2ex@C=2ex{& \theta^{0} & &  \theta^{1} &  \theta^{2}\\
\thetabar^{0} & \jbox{a} \ar@<-1ex>[rr]_{-i(\sigma^\mu\overline{\epsilon})\partial_\mu} \ar@<1ex>[dd]^{-i(\sigmabar^\mu \epsilon)\partial_\mu} & \kbox{} & \jbox{\xi_\alpha} \ar@<-1ex>[ll]_{\epsilon}   &  \jbox{b}\\ 
 & \kbox{} \\
\thetabar^{1} &  \jbox{\overline{\chi}^\alphadot} \ar@<1ex>[uu]^{\overline{\epsilon}} &&  \jbox{v_\mu} &  \jbox{\overline{\zeta}^\alphadot}\\
\thetabar^{2} &  \jbox{c} &&  \jbox{\eta_\alpha} &  \jbox{d}
}}
\end{center}
\caption{Visual representation of superspace translations acting on components.}
\label{fig:SUSYtransformation}
\end{figure}

Visually, the SUSY translations act on components as in \Fig{fig:SUSYtransformation}.  Getting all the factors of two and minus signs correct in the component transformations is an exercise best done in the woodshed. The answers are
\be
\label{eq:alltransformations}
\begin{array}{rcl}
\delta a&=&\xi\epsilon+\overline{\chi}\epsilonbar,\\
\delta\xi&=&2b\epsilon-(v_\mu+i\partial_\mu a)\sigma^\mu\epsilonbar,\\
\delta b&=&\overline{\zeta}\epsilonbar+\frac{i}{2}\partial_\mu\xi\sigma^\mu\epsilonbar,\\
\delta\overline{\chi}&=&2c\epsilonbar +(v_\mu-i\partial_\mu a)\sigmabar^\mu\epsilon,\\
\delta v_\mu&=&\overline{\zeta}\sigmabar_\mu\epsilon-\eta\sigma_\mu\epsilonbar+\frac{i}{2}\partial_\nu\left(\xi\sigma_\mu\sigmabar^\nu\epsilon-\overline{\chi}\sigmabar_\mu\sigma^\nu\epsilonbar\right),\\
\delta\overline{\zeta}&=&2d\epsilonbar-i\partial_\mu b\sigmabar^\mu\epsilon+\frac{i}{2}\partial_\mu v_\nu\sigmabar^\nu\sigma^\mu\epsilonbar,\\
\delta c&=&\epsilon\eta+\frac{i}{2}\partial_\mu\overline{\chi}\sigmabar^\mu\epsilon,\\
\delta\eta&=&2d\epsilon-i\partial_\mu c\sigma^\mu\epsilonbar-\frac{i}{2}\partial_\mu v_\nu\sigma^\nu\sigmabar^\mu\epsilon,\\
\delta d&=&\frac{i}{2}\partial_\mu\left(\overline{\zeta}\sigmabar^\mu\epsilon+\eta\sigma^\mu\epsilonbar\right), 
\end{array}
\ee
where we have extensively used the Fierz identities of \Eqs{eq:Fierz2}{eq:Fierz3} to simplify expressions like $(\thetabar \sigmabar^\mu \theta) (\psi \sigma_\mu \overline{\chi})$ and $(\overline{\psi} \sigmabar^\mu \theta)(\thetabar \sigmabar^\nu \theta)$.

We see now very clearly that SUSY ``rotates'' bosons into fermions, and vice versa.  Note that the highest ($\theta^4$) component $d$ transforms as a total derivative under SUSY, which will be particularly useful for building lagrangians in \Sec{subsec:SUSYactions}.

\subsection{The SUSY Algebra}
\label{subsec:SUSYalgebra}

Thus far, we have talked about SUSY transformations without ever mentioning the underlying SUSY algebra.  Indeed, one advantage of superspace is that \Eq{eq:SUSYtransformations} contains the full structure of SUSY.  However, it is instructive to turn the superspace translation picture into an operator picture to show that the SUSY algebra closes.

Recall that ordinary space-time translations are generated by the energy-momentum operator
\be
\label{eq:ordinarytranslations}
e^{ia_\mu P^\mu}f(x^\mu)=f(x^\mu+a^\mu),
\ee 
where
\be
P_\mu \equiv -i\partial_\mu.
\ee
Translations are part of the full Poincar\'e group that includes Lorentz transformations generated by $M_{\mu \nu}$, with algebra
\begin{align}
[M_{\mu \nu}, M_{\rho \tau} ] & = i \left( \eta_{\nu \rho} M_{\mu \tau} +  \eta_{\mu \tau} M_{\nu \rho} - \eta_{\mu \rho} M_{\nu \tau} - \eta_{\nu \tau} M_{\mu \rho} \right), \label{eq:MMcoMM}\\
[P_\mu, M_{\nu \rho} ] & = i \left( \eta_{\mu \nu} P_\rho - \eta_{\mu \rho} P_\nu \right), \label{eq:PcoMM} \\
[P_\mu, P_\nu ] & = 0. 
\end{align}
Note that the explicit expression of $M_{\mu\nu}$ depends on the spin of the field it acts on. For a scalar field, for example,
\be
M_{\mu\nu}=i\left(x_\mu\partial_\nu-x_\nu\partial_\mu\right).
\ee 

We want to introduce new SUSY generators that implement \Eq{eq:SuperfieldSUSY}, namely operators $Q$ and $\Qbar$ such that 
\be
\label{eq:operatorshift}
e^{-i\epsilon Q-i\overline{\epsilon}\Qbar} \bS(x^\mu, \theta^\alpha, \thetabar^{\alphadot}) = \bS(x^\mu+\Delta^\mu, \theta^\alpha+\epsilon^\alpha,\thetabar^\alphadot+\overline{\epsilon}^\alphadot).
\ee
In analogy with \Eq{eq:ordinarytranslations}, we see immediately that
\begin{align}
&Q_\alpha=i\frac{\partial}{\partial\theta^\alpha}-(\sigma^\mu\thetabar)_\alpha\partial_\mu,\\
&\Qbar^\alphadot=i\frac{\partial}{\partial\thetabar_\alphadot}-(\sigmabar^\mu\theta)^\alphadot\partial_\mu.
\end{align}
It is straightforward to show that these generators satisfy the SUSY algebra
\be
\label{eq:SUSYalgebra}
\begin{split}
\{Q_\alpha,\Qbar_\betadot\} &=-2\sigma^\mu_{\alpha\betadot}P_\mu,\\
\{Q_\alpha,Q_\beta\} &=\{\Qbar_\alphadot,\Qbar_\betadot\}=0,\\
\left[Q_\alpha,P_\mu\right] &=\left[\Qbar_\alphadot,P_\mu\right]=0,\\
\left[ M_{\mu \nu}, Q_\alpha \right] &= i \sigma_{\mu \nu \alpha}{}^\beta Q_\beta, \\
\left[ M_{\mu \nu}, \Qbar^{\alphadot} \right] &= i \sigmabar_{\mu \nu}{}^{\alphadot}{}_{\betadot} \Qbar^{\betadot},
\end{split}
\ee 
thus extending the Poincar\'e algebra.  In this way, two SUSY translations are equivalent to one ordinary space-time translation, and we sometimes refer to SUSY as being the ``square root'' of translations.  The non-trivial commutator between SUSY and Lorentz generators just indicates that the SUSY generator is a Lorentz spinor.   The SUSY algebra indeed closes, and accounting for the possibility of higher $\mathcal{N}$, one can show that SUSY is the unique extension of the Poincar\'{e} algebra \cite{Haag:1974qh}.

It is worth mentioning that SUSY can be present in more general space-time geometries.  As we will discuss in \Sec{subsec:AdSSUSY}, one can also define a SUSY algebra relevant for anti-de Sitter space (which counterintuitively is highly relevant for describing our approximately flat universe).

\subsection{What is a Superfield?}
\label{subsec:whatissuper}

It is important to make the distinction between a true superfield and a random collection of bosons and fermions.  Functions of $\{x^\mu,\theta^\alpha,\thetabar^\alphadot\}$ are superfields, in the sense that they have well-behaved properties under a SUSY transformation:  
\be
\left\{x^\mu, \theta^\alpha, \thetabar^{\alphadot} \right\} \to \left\{x^\mu + \Delta^\mu, \theta^\alpha  + \epsilon^\alpha, \thetabar^{\alphadot} + \overline{\epsilon}^{\alphadot} \right\}.
\ee
Equivalently, a (Lorentz-scalar) superfield is any object whose components transform as \Eq{eq:alltransformations}.   However, if we take a superfield $\bS$ and randomly manipulate its components (for example by zeroing out various components), the resulting object is generically \emph{not} a superfield, since its (manipulated) components do not transform as \Eq{eq:alltransformations}.

Since superfields are the building blocks of SUSY lagrangians, it is important to know which objects are superfields and which are not.  \Eq{eq:SuperfieldSUSY} immediately implies that any function of any number of superfields is also a superfield
\be
f(\boldsymbol{S_1}(x^\mu,\theta^\alpha,\thetabar^\alphadot),\boldsymbol{S_2}(x^\mu,\theta^\alpha,\thetabar^\alphadot),\ldots),
\ee
since each argument of the function $f$ has well-defined transformation properties under the shifts in \Eq{eq:SUSYtransformations}.  In particular, the sum of two superfields is also a superfield, as is the product.  

Because $[P_\mu,Q_\alpha]=0$, ordinary space-time derivatives of superfields are also superfields
\be
\label{eq:derivatives}
\partial_\mu \bS,\quad \Box \bS, \quad \ldots
\ee
To see this, consider the SUSY transformation from \Eq{eq:operatorshift} acting on $\partial_\mu \bS$:
\be
\label{eq:SUSYtranslationcommute}
e^{-i\epsilon Q-i\overline{\epsilon}\Qbar} \partial_\mu \bS = \partial_\mu e^{-i\epsilon Q-i\overline{\epsilon}\Qbar} \bS.
\ee
The overall space-time derivative acts on each of the components, so \Eq{eq:alltransformations} is preserved.

However, there are plenty of manipulations of superfields that do not result in another superfield.  For example, if we multiply a superfield by a function of $x^\mu$ alone
\be
f(x_\mu) \bS,
\ee
the resulting object is \emph{not} a superfield, because $f(x_\mu)$ does not have any corresponding fermionic components to rotate into.  Similarly, we cannot multiply by a generic function of $\theta^\alpha$ or $\thetabar^\alphadot$ alone.  Relevant for the next subsection is the fact that
\be
\label{eq:dthetanotasuper}
\frac{\partial}{\partial\theta_\alpha} \bS
\ee
is not a superfield.  While $\frac{\partial}{\partial\theta_\alpha} \bS$ is a collection of bosonic and fermionic components, these components do not transform as \Eq{eq:alltransformations}, which can be readily understood because $\{\frac{\partial}{\partial\theta_\alpha}, \Qbar^\alphadot\} \not= 0$ so the equivalent manipulation in \Eq{eq:SUSYtranslationcommute} is not possible.

\subsection{Super-trick \#1: The SUSY-Covariant Derivative}
\label{subsec:IntroD}

Our first super-trick is to introduce the SUSY-covariant derivative $D_\alpha$.  This operator is incredibly useful for SUSY manipulations, because it acts on a superfield to yield another superfield.  

One way to motivate $D_\alpha$ is that we want an object whose  lowest component is the second-lowest component of $\bS$.  The natural choice would be $\frac{\partial}{\partial\theta_\alpha} \bS$, but we argued in \Eq{eq:dthetanotasuper} that this was not a superfield. We can, however, construct such an object with the help of the SUSY-covariant derivatives 
\begin{align}
D_\alpha & =\frac{\partial}{\partial\theta_\alpha}-i(\sigma^\mu\thetabar)_\alpha\partial_\mu,\\
\Dbar^\alphadot &=\frac{\partial}{\partial\thetabar^\alphadot}-i(\sigmabar^\mu\theta)^\alphadot\partial_\mu.
\end{align}
It is a straightforward exercise to show that these commute with $Q_\alpha$ and $\Qbar_\betadot$.  Note also that
\be
\label{eq:twoDsmakeaP}
\{ D_\alpha, \Dbar_{\betadot} \} = 2 i \sigma_{\alpha \betadot}^\mu \partial_\mu. 
\ee
Like any sensible derivative, these obey a Leibniz (product) rule
\begin{align}
\D_\alpha (\bX \bY) & = (\D_\alpha \bX) \bY + \bX (\D_\alpha \bY ).  \label{eq:leibniz}
\end{align}
There is one subtlety, however, because the $D$s pick up a minus sign if you move them across an odd number of spinor indices:
\begin{align}
\D_\alpha (\bX_\beta \bY) & = (\D_\alpha \bX_\beta ) \bY - \bX_\beta (\D_\alpha \bY).
\end{align}
Note that $D^3 \bS=\Dbar^3\bS=0$, because $\{D_\alpha, \D_\beta\} = 0$.\footnote{This fact is not true in $\text{AdS}_4$ space or SUGRA, which is part of the reason why SUGRA is so complicated.}

These SUSY-covariant derivatives are useful in a variety of circumstances, justifying our designation of them as a super-trick.  Let us count the ways.

\begin{enumerate}
\item \textbf{Making new superfields.}  As advertised, a SUSY-covariant derivative acting on a superfield yields another superfield.  Therefore a SUSY lagrangian can depend on
\be
\bS, \quad D_\alpha \bS, \quad D^2 \bS, \quad \Dbar_\alphadot \bS, \quad \Dbar^2 \bS,\quad\text{etc.}
\ee
Space-time derivatives can also be written using $D$s,
since
\be
\partial^\mu \bS =-\frac{i}{4}\sigmabar^{\mu\betadot\alpha} \{ D_\alpha, \Dbar_{\betadot} \} \bS.
\ee

\item \textbf{Extracting components of a superfield}.   Using the notation that ``$|_0$" stands for the lowest component of a supermultiplet, then
\be
\bS |_0=a, \quad D_\alpha \bS |_0=\xi, \quad -\frac{1}{4}D^2 \bS|_0=b, \quad\text{etc}\dots
\ee
This gives us an alternative way to write the integrals in \Eqss{eq:GrassDerInt}{eq:GrassDiff}{eq:integrals} as
\be
\int d^2\theta \, \bS = -\frac{1}{4}D^2 \bS|_0, \quad \int d^2\thetabar \, \bS = -\frac{1}{4}\Dbar^2 \bS|_0.
\ee
We will make regular use of the equivalence
\be
\label{eq:Dthetaequivalence}
\int d^4\theta \, \bS = \int d^2\theta\left(-\frac{1}{4}\Dbar^2 \bS\right) = \frac{1}{16}D^2\Dbar^2 \bS|_0.
\ee
\item \textbf{Constructing SUSY-invariant actions}.  As a special case of the last point, we can use $D$s to identify the highest component of a superfield:\footnote{The choice of $\D^2 \Dbar^2 \bS$ rather than $\Dbar^2 \D^2 \bS$ here (related to our convention for $d^4 \theta$) is rather arbitrary. Thankfully, the only differences between the two are total derivative terms, which we will ruthlessly ignore henceforth.}
\be
\frac{1}{16} \D^2\Dbar^2 \bS|_0 = d.
\ee
As we showed in \Eq{eq:alltransformations} and  as we will discuss further in \Sec{subsec:SUSYactions}, the highest component of a superfield transforms as a total derivative under SUSY.  Therefore, for any superfield $\bS$, we can construct a SUSY-invariant action via
\be
\mathcal{S} = \int d^4x \, \frac{1}{16} \D^2\Dbar^2 \bS|_0 + \hc=\int d^4x\, d^4\theta\, \bS+\hc
\ee

\item \textbf{Restricting superfields}. The generic superfield in \Eq{eq:GenericMultiplet} is unwieldy for most purposes since it corresponds to a reducible representation of the SUSY algebra.  We usually work with irreducible representations, two of which are defined using $D$s.  The key multiplets are 
\begin{align}
\text{Chiral superfield}:&   \quad \Dbar \bPhi = 0, \\
\text{Vector superfield}:&   \quad \bV = \bV^\dagger, \\
\text{Linear superfield}: & \quad D^2 \bL = 0 \text{ and } \bL = \bL^\dagger.  \label{eq:linearmultiplet}
\end{align}
Chiral multiplets are the workhorses of $\mathcal{N} = 1$ SUSY and will be the topic of \Sec{sec:lecture2}.  Vector multiplets are needed to describe SUSY gauge theories, as discussed in \Sec{sec:lecture3}.  Linear multiplets show up as conserved currents (see \Ref{Dumitrescu:2011zz}, for example), but will not appear again (with one exception) in these lectures.

\item \textbf{Building chiral multiplets}.  Given a generic superfield $\bS$, one can construct a chiral superfield by acting twice with $\Dbar$:
\be
\bPhi = \Dbar^2 \bS.
\ee
We immediately see that $\Dbar \bPhi = 0$ (the defining characteristic of a chiral superfield) because $\Dbar^3 = 0$.

\item \textbf{Constructing the chiral projector}.  Finally, a trick using $D$s for true SUSY aficionados.  Consider the following operator acting on a generic superfield:
\be
- \frac{\Dbar^2 \D^2}{16\Box} \bS.
\ee
Because $\Dbar^3 = 0$, this is guaranteed to be a chiral multiplet (although the object is generically non-local).  More interestingly, if we replace $\bS$ with a chiral multiplet $\bPhi$ that already satisfies $\Dbar{\bPhi} = 0$, then
\be
\label{eq:chiralprojector}
- \frac{\Dbar^2 D^2}{16 \square} \bPhi = \bPhi.
\ee
To see this, we make repeated use of \Eq{eq:twoDsmakeaP}.  The operator $- \frac{\Dbar^2 \D^2}{16 \square}$ is known as a chiral projector, and can be useful when manipulating chiral multiplets.
\end{enumerate}

\noindent We will get a chance to use each $D$ super-trick in these notes.

\subsection{SUSY-Invariant Actions}
\label{subsec:SUSYactions}

Thus far, we have defined superfields as objects living in $\{x^\mu,\theta^\alpha,\thetabar^\alphadot\}$ superspace that transform nicely under \Eq{eq:SUSYtransformations}.  We have seen that products and sums of superfields are superfields, as are space-time and SUSY-covariant derivatives acting on superfields.  We are now ready to construct SUSY-invariant actions in terms of superfields.

From \Eq{eq:alltransformations}, no component of a superfield is invariant under SUSY.  For the purposes of defining a SUSY-invariant theory, though, all we need is a lagrangian that transforms as a total derivative under SUSY.  Indeed,  the highest component of a generic multiplet $\bS$ shifts as a total derivative,
\be
(-i\epsilon Q-i\overline{\epsilon}\Qbar)\left(\frac{1}{16}D^2 \Dbar^2 \bS \right)\bigg|_0 = \text{total derivative}.
\ee
This is obvious from \Fig{fig:SUSYtransformation}, which shows that the highest ($\theta^4$) component only gets contributions from the derivatives of lower components, and it is shown explicitly in \Eq{eq:alltransformations}.

Because the lagrangian must be hermitian, we can build an arbitrary SUSY lagrangian out of the $\theta^4$ component of a vector multiplet $\bV$ satisfying $\bV^\dagger = \bV$, 
\be
\mathcal{L} = \frac{1}{16} D^2 \Dbar^2 \bVcomp|_0 = \int d^4 \theta \, \bVcomp,
\ee
where we have used the equivalence in \Eq{eq:Dthetaequivalence}.  The notation $\bVcomp$ reminds us that the vector multiplet is generically composite, meaning that it consists of products and sums of other superfields.  

Certain superfield manipulations will have no effect whatsoever on the action.  For example, imagine shifting $\bVcomp$ by a chiral multiplet $\bOmega$,
\be
\label{eq:shiftByChiral}
\bVcomp \rightarrow \bVcomp + \bOmega + \bOmega^\dagger.
\ee
This transformation leaves $\bVcomp$ as a vector multiplet, but because $\Dbar^{\alphadot} \bOmega = 0$, this shift changes the action by a total derivative.  Thus, we can remove any purely chiral plus anti-chiral pieces of $\bVcomp$.\footnote{This is only true in global SUSY.  The equivalent manipulation in SUGRA (forming part of a K\"ahler transformation) is more complicated.}

While it is true that we can build an arbitrary SUSY lagrangian using vector multiplets alone, such constructions are not in general  local in superspace.  For example, consider a particular non-local choice for $\bVcomp$ 
\be
\bVcomp = \frac{1}{4\Box} D^2 \bPhicomp + \text{\hc}
\ee
where $\bPhicomp$ is a (composite) chiral multiplet.   Using our super-trick chiral projector from \Eq{eq:chiralprojector}, the $\theta^4$ component of the first term is
\be
\label{eq:Fchiral}
\frac{1}{16} D^2 \Dbar^2 \frac{1}{4\Box} D^2 \bPhicomp |_0  = - \frac{1}{4}D^2 \bPhicomp |_0 = \int d^2 \theta \, \bPhicomp,
\ee
which is just the $\theta^2$ component of the chiral multiplet $\bPhicomp$! Indeed, both the $\theta^4$ component of a vector multiplet and the $\theta^2$ component of a chiral multiplet transform as total derivatives under SUSY, so both can be used to build SUSY-invariant actions.  (SUSY aficionados  should be mildly impressed that were able to deduce this without ever determining the components of a chiral multiplet $\bPhicomp$.)
Note that, as already mentioned, the operator $- \frac{\Dbar^2 \D^2}{16 \square}$ acts on $\bPhicomp$ as a projector, crucially because $\bPhicomp$ is chiral. This allows us to rewrite the $\theta^4$ component of this non-local $\bVcomp$ as a local term.

This saturates all the possibilities for a local action, and the most generic SUSY-invariant action one can build can be written using the lagrangian
\begin{align}
\label{eq:genericlagrangian}
\mathcal{L} & = \int d^4 \theta \, \bVcomp + \left( \int d^2 \theta \, \bPhicomp + \int d^2 \thetabar \, \bPhicomp^\dagger \right),
\end{align}
where $\bVcomp = \bVcomp^\dagger$ is a (composite) vector superfield and $\bPhicomp$ is a (composite) chiral superfield satisfying $\Dbar \bPhicomp = 0$.

%% file: jthaler_TASI_Lecture2.tex
\section{Fermions and Sfermions}
\label{sec:lecture2}

In this lecture, we construct SUSY lagrangians for chiral multiplets.  After introducing the basic properties of chiral multiplets and the free lagrangian for chiral superfields, we use effective field theory power counting to write down the most relevant interactions.  We explain the super-trick of equations of motion in superspace, as well as how global symmetries work in SUSY.

\subsection{Chiral Multiplets}

In order to write down concrete SUSY lagrangians, it is instructive to restrict our attention to chiral multiplets which satisfy the constraint
\be
\Dbar^{\alphadot} \bPhi  = 0.
\ee
The complex conjugate equation
\be
D^{\alpha} \bPhi^\dagger  = 0
\ee
defines an anti-chiral superfield $\bPhi^\dagger$.  The physical states contained in a chiral multiplet are a spin-1/2 Weyl fermion and a spin-0 complex scalar (often called a sfermion).  Chiral multiplets are needed to describe the quark, lepton, and Higgs sectors of the SUSY SM (and their corresponding squark, slepton, and higgsino superpartners).

One way to automatically construct a chiral multiplet is to introduce a modified space-time coordinate
\be
\label{eq:y}
y^\mu  \equiv x^\mu + i \thetabar \sigmabar^\mu \theta.
\ee
A generic chiral superfield can be expressed as an arbitrary function of $y^\mu$ and $\theta_\alpha$ (but not of $\thetabar^\alphadot$):
\be
\bPhi (y^\mu, \theta_\alpha).
\ee
We can check that this automatically satisfies the constraint $\Dbar^{\alphadot} \bPhi = 0$, via 
\be
\begin{split}
\Dbar^{\alphadot} \bPhi(y^\mu,\theta) & = \left( \frac{\partial}{\partial \thetabar_{\alphadot} } - i (\sigmabar^\mu \theta)^{\alphadot} \partial_\mu \right) \bPhi (y^\mu, \theta) \\
& = \left( i (\sigmabar^\mu \theta)^{\alphadot} \frac{\partial}{\partial y^\mu} - i (\sigmabar^\mu \theta)^{\alphadot} \frac{\partial}{\partial y^\mu} \right) \bPhi(y^\mu,\theta) \\
& = 0,
\end{split}
\ee
recalling that $\bPhi$ has no $\thetabar$ dependence except through $y^\mu$.

A chiral superfield has a compact expression in $y$-dependent components
\begin{align}
\label{eq:chiralcomponents}
\bPhi(y^\mu,\theta) & = \phi(y) + \sqrt{2} \theta \psi(y) + \theta^2 F(y).
\end{align}
We will often represent such chiral superfields schematically as
\begin{align}
\bPhi & = \{ \phi, \psi, F \}.
\end{align}
This is the smallest representation of SUSY containing a fermion ($\psi$) and a sfermion ($\phi$).  The field $F$ is an auxiliary field which will turn out to be non-propagating; its existence is necessary to ensure that there are an equal number of bosonic and fermionic components off-shell.  We can extract these components from the superfield by applying SUSY-covariant derivatives:
\begin{align}
\bPhi|_0 &= \phi,\\
\D_\alpha \bPhi|_0 &= \sqrt{2} \psi_\alpha, \\
-\frac{1}{4} \D^2 \bPhi|_0 &= F.
\end{align}
By making the replacement $y^\mu \rightarrow x^\mu + i \thetabar \sigmabar^\mu \theta$ in \Eq{eq:chiralcomponents} and Taylor expanding as in \Eq{eq:xTaylorexpand}, we can express these as components with ordinary $x^\mu$ dependence as 
\be
\bPhi =\quad
\begin{array}{c|ccc}
& \theta^{0} & \theta^{1} & \theta^{2} \\
\hline
\rule[-3mm]{0mm}{.8cm}
\thetabar^{0} & \phi & \sqrt{2} \psi_\alpha & F \\
\thetabar^{1} & 0 & i \partial_\mu \phi & - \frac{1}{\sqrt{2}} i \left(\sigmabar^\mu \partial_\mu \psi\right)^\alphadot \\
\thetabar^{2} & 0 & 0 & - \frac{1}{4} \Box \phi \\
\end{array}
\label{eq:gridcomponentschiral}
\ee 
However, it is usually more convenient to stick with the $y^\mu$ coordinates.

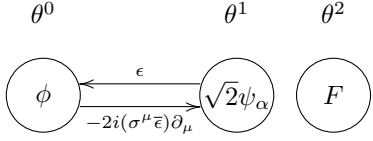
\begin{figure}[t]
\begin{center}
\makebox{
\xymatrix@R=2ex@C=2ex{
\theta^{0} && \theta^{1} & \theta^{2}\\
\jbox{\phi}  \ar@<-1ex>[rr]_{-2i(\sigma^\mu\overline{\epsilon})\partial_\mu}
& \kbox{} & \jbox{\sqrt{2}\psi_\alpha}  \ar@<-1ex>[ll]_{\epsilon}  & \jbox{F} 
}}
\end{center}
\caption{Visual representation of superspace translations on a chiral multiplet.}
\label{fig:Chiralvisualization}
\end{figure}

The SUSY transformations of a chiral multiplet follow from the transformations of $y^\mu$ and $\theta^\alpha$ as derived from \Eq{eq:SUSYtransformations}:
\begin{align}
\theta^\alpha & \rightarrow \theta^\alpha + \epsilon^\alpha,\\
y^\mu & \rightarrow y^\mu + 2 i \overline{\epsilon} \sigmabar^\mu \theta.
\end{align}
Visually, this transformation acts on the components as in \Fig{fig:Chiralvisualization}, giving rise to 
\be
\begin{array}{rcl}
\delta\phi &=& \sqrt{2} \epsilon \psi,\\
\delta\psi &=&  -i \sqrt{2} (\sigma^\mu \bar{\epsilon}) \partial_\mu \phi + \sqrt{2} \epsilon F,\\
\delta F &=&  - i \sqrt{2} \bar{\epsilon} \bar{\sigma}^\mu \partial_\mu \psi.
\end{array}
\ee
As advertised in \Sec{subsec:SUSYactions}, the highest component (i.e.~$F$ component) of a chiral superfield transforms as a total derivative, and can therefore be used to construct a SUSY-invariant action.

%In the SUSY Standard Model, the $\psi$ are matter fermions (quarks/leptons), and $\phi$ are sfermions (squarks/sleptons).

Using the super-trick from \Sec{subsec:IntroD}, we can construct new superfields from chiral superfields by applying SUSY-covariant derivatives.  Of course, $\Dbar^{\alphadot} \bPhi$ vanishes by definition, but $\D_\alpha \bPhi$ and $\D^2 \bPhi$ do not and can appear in SUSY lagrangians.  We will be able to avoid a lot of messy algebra in the future by taking the anti-chiral superfield $\bPhi^\dagger$ and constructing a chiral superfield via
\be
- \frac{1}{4} \Dbar^2 \bPhi^\dagger = F^\dagger (y) - i \sqrt{2} \theta \sigma^\mu \partial_\mu \overline{\psi}(y) - \theta^2 \Box \phi^\dagger(y). \label{eq:xi}
\ee
which in components is $- \frac{1}{4} \Dbar^2 \bPhi^\dagger = \{ F^\dagger, - i \sigma^\mu \partial_\mu \overline{\psi}, - \Box \phi^\dagger \}$.  Crucially, the components are a function of $y$ (not $y^*$).  We see that this is indeed a chiral superfield by recalling that $\Dbar^3 \bPhi = 0$.  Also, it is helpful to know that products of chiral multiplets are also chiral
\be
\label{eq:chiralproductischiral}
\Dbar^{\alphadot} (\bPhi_1 \bPhi_2 \cdots \bPhi_n)  = 0,
\ee
which is easy to prove because $\bPhi_1 \bPhi_2 \cdots \bPhi_n$ is a function only of $y^\mu$ and $\theta^\alpha$, or from the Leibniz rule of \Eq{eq:leibniz}.

\subsection{A Free SUSY Lagrangian}

We are now (finally) ready to construct our first SUSY lagrangian: a free theory of a single chiral superfield.  First, let us conduct some dimensional analysis.  The lowest component of a chiral multiplet $\phi$ has mass dimension $1$, while the fermionic component has mass dimension $3/2$.  In order for $\bPhi$ to have a well-defined mass dimension, we must have the following mass dimension assignments:
\be
\begin{split}
[\bPhi] = [\phi] &= 1,\\
[\psi] &= 3/2,\\
[F] &= 2, \\
[\theta]  &= -1/2,\\
[d\theta^\alpha] = [\D_\alpha]  &= 1/2.
\end{split}
\ee
In particular, $[d^4 \theta] = 2$ and $[d^2 \theta] = 1$, so in the generic lagrangian
\be
\label{eq:genericL}
\mathcal{L} = \int d^4 \theta \, \bV_{\rm comp} + \left( \int d^2 \theta \, \bPhi_{\rm comp} + \int d^2 \thetabar \, \bPhi_{\rm comp}^\dagger \right),
\ee
we must have $[\bV_{\rm comp}] = 2$ and $[\bPhi_{\rm comp}] = 3$.

A free theory is quadratic in fields, so given a single chiral superfield $\bPhi$, the only possible choice for a free lagrangian is
\begin{align}
\mathcal{L}_{\rm free} & = \int d^4 \theta \, \bPhi^\dagger \bPhi + \left( \int d^2 \theta \,  \frac{1}{2} m \bPhi^2 + \textrm{h.c.} \right), \label{eq:quadraticaction}
\end{align}
where $m$ is a mass parameter.  The only other possible quadratic term $\int d^4 \theta \, \bPhi^2$ does not contribute to the action because it is a total derivative (see \Eq{eq:shiftByChiral}).\footnote{By adding space-time derivatives, one can obtain other quadratic terms, such as $\int d^4 \theta\, \bPhi^\dagger \Box \bPhi/\Lambda^2$, where $\Lambda$ is some mass scale.  These will always lead to terms in the component Lagrangian featuring more than two derivatives.  In fact, these terms turn the auxiliary field $F$ into a propagating field (i.e.~a field with a kinetic term), suggesting that important physics was integrated out at the scale $\Lambda$.  We will come back to this in \Sec{subsec:eominsuperspace}.}   The first term in \Eq{eq:quadraticaction} is called the kinetic term and the second term is called the mass term, such that $\mathcal{L}_{\rm free} = \mathcal{L}_{\rm kinetic} + \mathcal{L}_{\rm mass}$.

To check that $\mathcal{L}_{\rm free}$ corresponds with our expectations, it is instructive to find the action in components.  The kinetic term is easiest to analyze by noting that
\begin{align}
\mathcal{L}_{\rm kinetic} = \int d^4 \theta \, \bPhi^\dagger \bPhi & = \int d^2 \theta \left( - \frac{1}{4} \Dbar^2 \bPhi^\dagger \right) \bPhi.
\end{align}
We already found the components of the superfield in parentheses in \Eq{eq:xi}.  Finding the component lagrangian is then just a matter of finding the $\theta^2$ component of the product of two chiral multiplets.  Arranging them in a grid
\be
\label{eq:chiralgrid}
\begin{tabular}{r | c c c }
 & $\theta^{0}$ & $\theta^{1}$ & $\theta^{2}$ \\
 \hline
$\bPhi$& $\phi$ & $\sqrt{2} \psi$ & $F$ \\
$-\frac{1}{4} \Dbar^2 \bPhi^\dagger$ & $F^\dagger$ & $- i \sqrt{2} \sigma^\mu \partial_\mu \overline{\psi}$ & $-\Box \phi^\dagger$
\end{tabular}
\ee
we need only multiply the $\theta^{n}$ component of the first line with the $\theta^{2-n}$ component of the second line, for $n = 0,1,2$.\footnote{The Fierz identity of \Eq{eq:Fierz3} tells us that $(\theta \chi) (\theta \psi) = - \frac{1}{2} \theta^2 \chi \psi$, so when multiplying the two $\theta^1$ components together, we lose the factors of $\sqrt{2}$ and pick up a sign.}
\begin{align}
\mathcal{L}_{\rm kinetic} & = - \phi \Box \phi^\dagger + i \overline{\psi} \sigmabar^\mu \partial_\mu \psi + F^\dagger F \label{eq:Lkinetic} \\
\mathcal{L}_{\rm mass} & = \frac{1}{2} m \left( \phi F - \psi \psi + F \phi \right) + \rm h.c. 
\end{align}
The first line contains the usual kinetic terms for $\phi$ and $\psi$, but $F$ is an auxiliary field since it does not have a kinetic term.  The second line is not yet illuminating, but we can do the exact path integral over $F$ by finding its equation of motion
\be
F^\dagger = - m \phi.
\ee
Inserting this back into $\mathcal{L}_{\rm free}$ yields our final answer
\be
\mathcal{L}_{\rm free} = - \phi (\Box + m^2) \phi^\dagger + i \bar{\psi} \sigmabar^\mu \partial_\mu \psi - \frac{1}{2} m (\psi \psi + \overline{\psi} \overline{\psi} ). \label{eq:LfreeMajorana}
\ee

Thus, the free theory of a chiral multiplet is the theory of a free complex spin-0 scalar and a free Weyl fermion.  Moreover, the mass terms for the scalar and fermion are related by SUSY!  The $m \rightarrow 0$ limit has enhanced chiral symmetry acting on $\bPhi$; this protects the masses of both the fermion and the scalar at the same time.  This effect helps protects radiative corrections to the Higgs boson mass in the SUSY SM.

The lagrangian in \Eq{eq:LfreeMajorana} contains a Majorana mass term for the single Weyl fermion.  To obtain a Dirac mass term, we would need two chiral multiplets:
\begin{align}
\mathcal{L}_{\rm Dirac} & = \int d^4 \theta \, \left( \bPhi^\dagger \bPhi + \bPhi^{c \dagger} \bPhi^c \right) + \left( \int d^2 \theta \, m \bPhi \bPhi^c + \textrm{h.c.} \right).
\end{align}
As before, the superscript $c$ is just a label, and $\bPhi$ and $\bPhi^c$ are independent fields off-shell.  The action in components is straightforward  to find given our earlier results.  We simply have two copies of the kinetic action of \Eq{eq:Lkinetic} (one for $\bPhi$ and one for $\bPhi^c$), while the mass terms are
\begin{align}
\mathcal{L}_{\rm mass} & = - m^2 \phi^\dagger \phi - m^2 \phi^{c \dagger} \phi^c  - m ( \chi \chi^c + \chi^\dagger \chi^{\dagger c} )
\end{align}
after integrating out auxiliary fields.  This theory has one massive Dirac fermion (composed of two Weyl fermions) and two complex scalars, with all fields having the same mass.

\subsection{A Generic Lagrangian}

Having successfully constructed a free SUSY lagrangian, we can construct more generic lagrangians using chiral multiplets.  However, it is immediately clear that the number of possibilities is rather large.  In the generic lagrangian in \Eq{eq:genericlagrangian}, $\bVcomp$ and $\bPhicomp$ can involve all sorts of functions and combinations of superfields, including ordinary derivatives and SUSY-covariant derivatives:
\be
\bPhi,  \quad
\D_\alpha \bPhi, \quad \D^2 \bPhi,   \quad \partial_\mu \bPhi, \quad \Box \bPhi,  \quad \partial_\mu \D_\alpha \bPhi, \quad \text{etc.}
\ee
\be
\bPhi^\dagger, \quad \Dbar_{\alphadot} \bPhi^\dagger,  \quad \Dbar^2 \bPhi^\dagger, \quad \partial_\mu \bPhi^\dagger, \quad \Box \bPhi^\dagger, \quad \partial_\mu \Dbar_{\alphadot} \bPhi^\dagger, \quad  \text{etc.}
\ee
Clearly we need some kind of organizing principle to sort out the multitude of possible terms.  

Just as in non-SUSY theories, we can use effective field theory power counting to organize terms in a SUSY lagrangian by mass dimension.  We will discuss relevant and marginal interactions in \Sec{subsec:dim4} and higher dimensional interactions in \Sec{subsec:dim56}.

Before doing so, it is worth emphasizing that some terms will not contribute to the SUSY action.   For example, we saw in \Eq{eq:shiftByChiral} that a purely chiral piece of $\bVcomp$ does not contribute to the action.  Similarly, we can ignore terms that only change the lagrangian by a total derivative, such as
\be
\int d^4 \theta \, \Box ( \cdots ) \to \text{total derivative}.
\ee
Less obvious is that we can ignore terms that correspond to a total SUSY-covariant derivative
\be
\int d^4 \theta \, D_\alpha ( \cdots ) \to \text{total derivative}, \label{eq:superintbyparts}
\ee
which one can derive by using the fact that $D^3 = 0$.  This implies that we can always do integration by parts with SUSY-covariant derivatives, which we will use to show the equivalence of different SUSY operators.

\subsection{Relevant and Marginal Interactions}
\label{subsec:dim4}

Relevant and marginal interactions correspond to couplings with positive or zero mass dimension.  From \Eq{eq:genericL}, we argued that $[\bVcomp] = 2$ and $[\bPhi_{\rm comp}] = 3$.  Since a linear term in $\bVcomp$ would be pure chiral, \Eq{eq:shiftByChiral} tells us that it would not contribute to the action.  Thus, the only choice for a renormalizable $\bVcomp$ is quadratic in chiral superfields
\be
\label{eq:chiralkinetic}
\bVcomp = \bPhi^{\dagger}_i \bPhi^i,
\ee
where we now have a number of chiral superfields $\bPhi^i$ labeled by $i$, and we sum over repeated indices.  Note that we can always rotate and rescale the $\bPhi^i$ to bring $\bVcomp$ into this form.  For $\bPhicomp$, there are three possible terms
\be
\label{eq:WZsuperpotential}
\bPhicomp = f_i \bPhi^i + \frac{1}{2} m_{i j} \bPhi^i \bPhi^j + \frac{1}{6} \lambda_{i j k} \bPhi^i \bPhi^j \bPhi^k \equiv \bW(\bPhi^i),
\ee
where the object $\bW$ is called the superpotential.  By \Eq{eq:chiralproductischiral}, $\bW$ itself is a (composite) chiral multiplet, and we say that $\bW$ is a holomorphic function, since it only depends on chiral and not anti-chiral superfields.  The mass dimensions of these couplings are $[f_i] = 2$, $[m_{ij}] = 1$, and $[\lambda_{i j k}] = 0$.

The most general renormalizable theory of chiral multiplets is often called the Wess-Zumino model:
\begin{align}
\mathcal{L}_{\rm WZ} & = \int d^4 \theta \, \bPhi_i^{\dagger} \bPhi^i + \left(\int d^2 \theta \, \bW(\bPhi^i) + \textrm{h.c.} \right).
\end{align}
By Taylor expanding $\bW$ to extract the $\theta^2$ component, we can easily write down the component form of the lagrangian 
\be
\begin{split}
\mathcal{L}_{\rm WZ} & = - \phi^i \Box \phi_i^\dagger + i \bar{\psi}_i \sigmabar^\mu \partial_\mu \psi^i + F_i^\dagger F^i  \\
& \quad \, + W_i F^i - \frac{1}{2} W_{i j} \psi^i \psi^j + \rm h.c.
\end{split}
\ee
where
\be
W_i(\phi^i) \equiv \frac{\partial W}{\partial \phi^i}, \qquad W_{i j}(\phi^i) \equiv \frac{\partial^2 W}{\partial \phi^i \phi^j},
\ee
and $W_i$ and $W_{ij}$ should now be thought of as scalar functions of the lowest (scalar) component of the various $\bPhi^i$ fields. The equations of motion for the auxiliary fields are 
\begin{align}
\label{eq:eomaux}
F^i = - W^{\dagger i}, \qquad F^\dagger_i & = - W_i,
\end{align}
so the final component lagrangian is
\begin{align}
\label{eq:finalWZ}
\mathcal{L}_{\rm WZ} & = - \phi^i \Box \phi^\dagger_i + i \overline{\psi}_i \sigmabar^\mu \partial_\mu \psi_i  \nonumber \\
& \quad \, - W^{\dagger i} W_i - \frac{1}{2} (W_{i j} \psi^i \psi^j + W^{\dagger i j} \overline{\psi}_i \overline{\psi}_j ).
\end{align}
The $W^{\dagger i} W_i$ term corresponds to the scalar potential of the theory, which in general has linear, quadratic, cubic, and quartic interactions.  The $W_{i j} \psi^i \psi^j$ term contains both mass terms for the fermions as well as Yukawa interactions between a scalar and two fermions.

We can now interpret the various terms in \Eq{eq:WZsuperpotential} one by one.
\begin{itemize}
\item $f_i$ is a source term for $F_i$.  It generates a constant in the scalar potential but no fermion mass terms.  This type of term will be relevant for SUSY-breaking in \Sec{subsec:polonyi}.  
\item $m_{ij}$ is a mass term for bosons and fermions.  The fermion mass matrix is $m_{ij}$ directly, but the boson mass-squared matrix is $M^2_{ij} = m_{ik} m^{k}{}_{j}$.
\item $\lambda_{i j k}$ yields a quartic term for bosons and Yukawa interactions between one boson and two fermions.  These two types of couplings are related in a non-trivial way by SUSY, with the quartic interaction related to the square of the Yukawa interaction.
\end{itemize}
Of course, additional terms in the scalar potential are generated when all of the $f_i$, $m_{ij}$, and  $\lambda_{i j k}$ are present.  Apart from gauge fields (which we will discuss in the next lecture), the Wess-Zumino model accounts for all interactions present in the MSSM.

\subsection{Dimension Five and Six}
\label{subsec:dim56}

Continuing with our power counting, the first irrelevant operators we encounter are at dimension 5.  Recalling that $\D^2$ has mass dimension 1, the terms which show up at dimension 5 are
\be
\label{eq:extrasuper}
\int d^2 \theta \, \frac{1}{\Lambda} \bPhi_1 \bPhi_2 \bPhi_3 \bPhi_4,
\ee
\be
\label{eq:extrakahler}
\int d^4 \theta \, \frac{1}{\Lambda} \bPhi_1^\dagger \bPhi_2 \bPhi_3,
\ee
\be
\int d^4 \theta \, \frac{1}{\Lambda} \bPhi_1 \D^2 \bPhi_2=4\int d^2 \theta \, \frac{1}{\Lambda} \bPhi_1 \Box \bPhi_2.
\ee
The last equality follows from the chirality of $\bPhi_1$ and $\bPhi_2$. This exhausts all the possible terms at dimension 5.\footnote{You can convince yourself that $\int d^4 \theta \, \bPhi_1^\dagger \D^2 \bPhi_2  = 0$ by doing integration by parts.}   At low energies, $\Box$ is small, so the first two terms are often the most important deformations.

Terms like \Eq{eq:extrasuper} are already captured by the superpotential (in the sense that \Eq{eq:finalWZ} still holds).  In order to describe the effect of terms like \Eq{eq:extrakahler}, it is helpful to introduce the K\"ahler potential $\bK$:
\begin{align}
\label{eq:KWsuperspace}
\mathcal{L} & = \int d^4 \theta \, \bK(\bPhi^i, \bPhi^{\dagger \j}) + \int d^2 \theta \, \bW(\bPhi^i) + \rm h.c. \\
& \quad + \D_\alpha, \partial_\mu \textrm{ terms.} \nonumber
\end{align}
To emphasize, the K\"ahler potential and superpotential by themselves \emph{do not} give the most general SUSY effective actions, but often contain the most important effects at low energies.  The K\"ahler potential and superpotential are most useful when scalar fields can get large vacuum expectation values but space-time derivatives are small.  

To derive the Lagrangian of \Eq{eq:KWsuperspace} in components, the following trick is helpful.  If you write a term in the K\"ahler potential as
\be
\begin{split}
& \int d^4 \theta (\bPhi_1 \bPhi_2 \cdots) (\bPhi_n \bPhi_{n+1} \dots)^\dagger \\
& = \int d^2 \theta (\bPhi_1 \bPhi_2 \cdots) \left(-\frac{1}{4} \Dbar^2 (\bPhi_n \bPhi_{n+1} \cdots)^\dagger \right),
\end{split}
\ee
then we recognize $(\bPhi_1 \bPhi_2 \cdots)$ and $(\bPhi_n \bPhi_{n+1} \cdots)$ as being chiral multiplets, whose components are easy to find.  Using the expression for $-\frac{1}{4} \Dbar^2 \bPhi^\dagger$ from \Eq{eq:xi}, it is straightforward to find the required $\theta^2$ component.  For example, consider the dimension 6 term
\begin{align}
\int d^4 \theta \, \frac{1}{\Lambda^2} (\bPhi_1 \bPhi_2) (\bPhi_3 \bPhi_4)^\dagger.
\end{align}
Making a grid analogous to \Eq{eq:chiralgrid}, we have
\be
\begin{tabular}{r |c c c}
& $\theta^{0}$ & $\theta^{1}$ & $\theta^{2}$ \\
\hline
$\bPhi_1 \bPhi_2$ & $\phi_1 \phi_2$ & $\phi_1 \psi_2 + \phi_2 \psi_1$ & $\phi_1 F_2 + \phi_2 F_1$\\
&&&$- \psi_1 \psi_2$\\
\hline
$-\frac{1}{4} \Dbar^2 (\bPhi_3 \bPhi_4)^\dagger $ & $\phi_3^\dagger F_4^\dagger + \phi_4^\dagger F_3^\dagger$ & $\sigma \cdot \partial(\phi_3^\dagger \overline{\psi}_4$ & $\Box(\phi_3^\dagger \phi_4^\dagger)$\\ 
&$- \overline{\psi}_3 \overline{\psi}_4$&$+\phi_4^\dagger \overline{\psi}_3)$&
\end{tabular}
\ee
Among other things, this contains the four-fermion operator $\frac{1}{\Lambda^2} \psi_1 \psi_2 \overline{\psi}_3 \overline{\psi}_4$, a type of interaction that is not present in the superpotential alone.

There are many possible terms at dimension 6.  Terms like
\be
\int d^2 \theta \frac{1}{\Lambda^2} \bPhi_1 \bPhi_2 \bPhi_3 \bPhi_4 \bPhi_5
\ee
are already captured by the superpotential, and terms like
\be
\int d^4 \theta \frac{1}{\Lambda^2} \bPhi_1 \bPhi_2 \bPhi_3 \bPhi_4^\dagger, \qquad \int d^4 \theta \frac{1}{\Lambda^2} \bPhi_1 \bPhi_2 \bPhi_3^\dagger \bPhi_4^\dagger,
\ee
and their hermitian conjugates are captured by the K\"ahler potential.  Additional terms which show up at dimension 6 include
\be
\int d^2 \theta \frac{1}{\Lambda^2} \bPhi_1 (\Box \bPhi_2) \bPhi_3,
\ee
\be
\int d^4 \theta \frac{1}{\Lambda^2} \D^2 \bPhi_1 \Dbar^2 \bPhi_2^\dagger=-16\int d^4 \theta \frac{1}{\Lambda^2} \bPhi_1 \Box \bPhi_2^\dagger,
\ee
where the last equality follows from integration by parts and the anti-chirality of $\bPhi_2^\dagger$.

There are cases where higher dimensional terms involving $\partial_\mu$ or $D_\alpha$ can be very important.  For example, when we talk about SUSY-breaking, the vacuum expectation value (vev) $\vev{F} \neq 0$.  Thus, terms like $\vev{\frac{\D^2 \bPhi}{\Lambda^2}} \neq 0$, and we have to remember the possibility of order $\frac{F}{\Lambda^2}$ effects not captured by the K\"ahler potential $\bK$ or superpotential $\bW$ alone.

\subsection{The K\"ahler Potential in Components}
\label{subsec:Kahler}

Though one of the goals of these notes is to emphasize  the presence of higher-derivative SUSY interactions, the  K\"ahler potential and superpotential are used so ubiquitously in the literature that one may find it useful to see the expansion in components of a Lagrangian featuring only $\bK$ and $\bW$.  The component expansion of \Eq{eq:KWsuperspace} yields:
\begin{align}
\mathcal{L} & = K_{i \j} \left(\partial^\mu \phi^i \partial_\mu \phi^{\dagger \j} + \frac{i}{2} \chi^i \sigma^\mu \partial_\mu \chi^{\dagger \j} + \frac{i}{2} \chi^{\dagger \j} \sigmabar^\mu \partial_\mu \chi^i + F^i F^{\dagger \j} \right) \nonumber \\
&\quad ~ - \frac{1}{2} K_{i \j k} \left(\chi^i \chi^k F^{* \j} + i \chi^i \sigma^\mu \chi^{\dagger \j} \partial_\mu \phi^k \right) \nonumber \\
&\quad ~ - \frac{1}{2} K_{i \j \k} \left(\chi^{\dagger \j} \chi^{\dagger \k} F^i - i \chi^i \sigma^\mu \chi^{\dagger \j} \partial_\mu \phi^{\dagger \k} \right) \nonumber \\
&\quad ~ + \frac{1}{4} K_{i \j k \l} \chi^i \chi^k \chi^{\dagger \j} \chi^{\dagger \l} \nonumber \\
&\quad ~ + W_i F^i - \frac{1}{2} W_{i j} \chi^i \chi^j + \rm h.c.
\label{eq:Kahlerexpansion}
\end{align}
where the subscripts on $K$ represent derivatives with respect to fields, e.g. $K_{i \j} = \frac{\partial^2 K}{\partial \Phi^i \partial \Phi^{\dagger \j}}$. 

After integrating out the auxiliary fields, this expression can be rewritten in a more compact fashion by thinking of the function $K_{i \j}$ as a metric $g_{i \j}$ in field space \cite{Zumino:1979et}:
\begin{align}
\mathcal{L} & = g_{i \j} \partial^\mu \phi^i \partial_\mu \phi^{\dagger \j} + i g_{i \j} \chi^{\dagger \j} \sigmabar^\mu \D_\mu \chi^i \\
& + \frac{1}{4} R_{i \j k \l} \chi^i \chi^k \chi^{\dagger \j} \chi^{\dagger \l} - g^{i \j} W_i W^\dagger_{\j} \\
& - \frac{1}{2} D_i W_j \chi^i \chi^j + \rm h.c.
\end{align}
where $g^{i \j}$ is the inverse K\"ahler metric ($g^{i \j} g_{i \k} = \delta^{\j}_{\k}$, $g^{i \j} g_{k \j} = \delta^i_k$) and the K\"ahler-covariant derivative $D_i$, Christoffel connection $\Gamma^k_{i j}$, and curvature tensor $R_{i \j k \l}$ are given by:
\begin{align}
D_i V_k & = V_{i k} - \Gamma^k_{i j}V_k, \\
\Gamma^k_{i j} & = g^{k \l} K_{i \l j}, \\
R_{i \j k \l} & = K_{i \j k \l} - g_{m \bar{n}} \Gamma^m_{i k} \Gamma^{\bar{n}}_{\j \l}, \\
\D_\mu \chi^i & = \partial_\mu \chi^i + \Gamma^i_{j k} \chi^k \partial_\mu \phi^k.
\end{align}
In particular, the equations of motion for the auxiliary fields are
\be
F^i=-g^{i\j}W_\j^\dagger+\frac{1}{2}\Gamma^i_{mn}\chi^m\chi^n.
\ee
We will not go into any further detail regarding K\"ahler geometry. For the interested reader, we recommend \Ref{Wess:1992cp} and references therein.  

\subsection{Super-trick \#2:  Equations of Motion in Superspace}
\label{subsec:eominsuperspace}

Our second super-trick involves integrating out heavy mass thresholds in superspace (at tree-level).  In non-SUSY theories, integrating out heavy states yields an effective theory with additional higher-dimension interactions among the light states.  In SUSY theories, we could do the same manipulations with component fields, but it is more convenient to integrate out superfields at tree-level using classical superspace equations of motion.

Recall that in non-SUSY theories, the classical equations of motion are obtained by the Euler-Lagrange procedure of varying the action to find the extrema.  One subtlety in superspace is that the lagrangian involves both $\int d^4 \theta$ and $\int d^2 \theta$ integrations, which require different constraints.   To avoid having to invoke Lagrange multipliers, it is best to express everything in terms of $\int d^2 \theta$ using \Eq{eq:Dthetaequivalence}:
\be
\int d^4 \theta \, \bVcomp = \int d^2 \theta \left(- \frac{1}{4} \Dbar^2 \bVcomp \right).
\ee

For simplicity, consider the Wess-Zumino model:
\begin{align}
\mathcal{L}_{\rm WZ} & = \int d^4 \theta \, \bPhi_i^{\dagger} \bPhi^i + \int d^2 \theta \, \bW(\bPhi^i) + \rm h.c. \\
& = \int d^2 \theta \left(- \frac{1}{4} \Dbar^2 \bPhi^{\dagger}_i \right) \bPhi^i + \int d^2 \theta \, \bW(\bPhi^i) + \int d^2 \thetabar \, \bW^\dagger(\bPhi^\dagger_i).
\end{align}
Written in this form, it is easy to read off the SUSY equation of motion by extremizing the action with respect to $\bPhi^i$, yielding
\begin{align}
 - \frac{1}{4} \Dbar^2 \bPhi^\dagger_i + \bW_i & = 0,
\end{align}
where the first term is familiar from \Eq{eq:xi} and last term should be thought of as a full superfield.  The lowest component of this expression is just the familiar $F_i^\dagger  = - W_i$ from \Eq{eq:eomaux}.  The $\theta$ ($\theta^2$) component yields the fermion (boson) equations of motion.

For a heavy supersymmetric threshold,  $-\frac{1}{4} \Dbar^2 \bPhi^\dagger_i \approx 0$ at low energies:
\be
\vev{F^\dagger_i} = 0, \qquad \vev{\sigma \cdot \partial \overline{\psi}_i} \approx 0,  \qquad \vev{\Box \phi_i} \approx 0.
\ee
Therefore, up to $\D^2$ terms suppressed by the $\bPhi_i$ mass, we can integrate out the field $\bPhi_i$ at tree-level by simply imposing $\bW_i = 0$ on superfields.

Let us try this out with three superfields $\bN$, $\bL$, and $\bH$:
\begin{align}
\bK & = \bN^\dagger \bN+ \bL^\dagger \bL + \bH^\dagger \bH, \\
\bW & = \frac{1}{2} m \bN^2 + \lambda \bN \bL \bH.
\end{align}
Here, $\bN$ is a heavy state we wish to integrate out to get the low energy effective theory for $\bL$ and $\bH$.  Ignoring $-\frac{1}{4} \Dbar^2 \bN^\dagger$, the equation of motion for $\bN$ is simply
\be
\frac{\partial \bW}{\partial \bN} = m \bN + \lambda \bL \bH = 0  \quad \Rightarrow \bN = -\frac{\lambda}{m} \bL \bH.
\ee
Below the scale $m$, the leading theory is 
\begin{align}
\bK_{\rm eff} & = \frac{\lambda^2}{m^2} (\bL \bH)^\dagger (\bL \bH) + \bL^\dagger \bL + \bH^\dagger \bH + \D^2 \textrm{ terms}, \\
\bW_{\rm eff} & = - \frac{\lambda}{m} \bL^2 \bH^2. \label{eq:LHLH}
\end{align}
In the SUSY SM, this effective superpotential can be used to describe (Majorana) neutrino masses.

Here is another example with four superfields:
\begin{align}
\bK & = \bN^\dagger \bN+ \bN^{c\dagger} \bN^c + \bL^\dagger \bL + \bH^\dagger \bH, \\
\bW & = m \bN \bN^c + \lambda \bN \bL \bH.
\end{align}
Here, the superscript $^c$ is just a label to say that $\bN$ and $\bN^c$ are different fields with complex conjugate quantum numbers.  The low energy equations of motion for $\bN$ and $\bN^c$ are
\begin{align}
\bW_N & = m \bN^c + \lambda \bL \bH = 0, \\
\bW_{N^c} & = m \bN = 0.
\end{align}
Below the scale $m$, the effective lagrangian is
\begin{align}
\bK_{\rm eff} & = \frac{\lambda}{m^2} (\bL \bH)^\dagger (\bL \bH)+ \D^2 \textrm{ terms}, \\
\bW_{\rm eff} & = 0.
\end{align}
As we will discuss further in the next subsection, we could have guessed that $\bW_{\rm eff}$ would be zero  because there is a $\U(1)$ symmetry under which the combination $\bL \bH$ has non-zero charge (and the superpotential must be a holomorphic function, so we cannot have $(\bL\bH)(\bL\bH)^\dagger$ in $\bW$).

This trick of superspace equations of motion is often helpful for finding UV completions of low energy effective theories.  Higher-dimension interactions can be generated by integrating out heavy fields, though one must always be mindful of the potential effects of $-\frac{1}{4} \Dbar^2 \bPhi^\dagger_i$ terms which we have neglected in this analysis.  Of course, in order to recover all the relevant physics, we often have to integrate out fields at loop level as well, and we will see how to do that in certain cases with our third super-trick in \Sec{subsec:BackgroundField}.

\subsection{Global Symmetries}
\label{subsec:globalsymmetries}

In the next lecture, we will talk about gauge theories, so it is natural to first think about how global symmetries work in SUSY lagrangians.  One can have a $\U(1)$ transformation acting on the whole superfield:
\be
\bPhi \rightarrow e^{i q \alpha}.
\ee 
This implies every component of $\bPhi$ has charge $q$:  
\be
\begin{tabular}{r | c c c}
& $\theta^{0}$ & $\theta^{1}$ & $\theta^{2}$ \\
\hline
$\bPhi$ & $\phi$ & $\sqrt{2} \psi$ & $F$ \\
$\U(1)$ & $q$ & $q$ & $q$ 
\end{tabular}
\ee
This can be used to constrain the form of the superpotential, for example.  A term $\frac{1}{2} m \bN^2$ in the superpotential has no $\U(1)$ symmetry, but $m \bN \bN^c$ respects a $\U(1)$ symmetry if $\bN$ and $\bN^c$ have equal and opposite charges.  Similarly, one can consider chiral superfields which come in representations of non-Abelian global symmetries.

Another type of global symmetry is a $\U(1)_R$ symmetry, which does not commute with SUSY.  Under a $\U(1)_R$ rotation, the superspace coordinates themselves transform, with $\theta$ ($\thetabar$) having $R$-charge $+1$ ($-1$).  This implies that $\D_\alpha$ ($\Dbar_\alphadot$) has $R$-charge $-1$ ($+1$), that the superpotential $\bW$ has $R$-charge 2, and that different components of $\bPhi$ have different charges:
\be
\begin{tabular}{r | c c c}
& $\theta^{0}$ & $\theta^{1}$ & $\theta^{2}$ \\
\hline
$\bPhi$  & $\phi$ & $\sqrt{2} \psi$ & $F$ \\
$\U(1)_R$ & $r$ & $r-1$ & $r-2$
\end{tabular}
\ee
In the context of SUGRA, $R$-symmetries are always broken, because the gravitino mass  parameter $m_{3/2}$ has $R$-charge $-2$.  But to the extent that $m_{3/2}$ is small (which may or may not be the case), $\U(1)_R$ symmetries can be good symmetries of SUSY lagrangians.

\begin{table}

\centering
\begin{tabular}{@{}|c|ccc|@{}} \hline
& $\SU(3)$ & $\SU(2)$ & $\U(1)$  \\ \hline 
$\bQ$ & $\bthree$ & $\btwo$ & $+\frac{1}{6}$ \\
$\bU^c$ & $\bar{\bthree}$ & $\bone$ & $-\frac{2}{3}$ \\
$\bD^c$ & $\bar{\bthree}$ & $\bone$ & $+\frac{1}{3}$ \\
$\bL$ & $\bone$ & $\btwo$ & $-\frac{1}{2}$ \\
$\bE^c$ & $\bone$ & $\bone$ & $+1$ \\
$\bH_u$ & $\bone$ & $\btwo$ & $+\frac{1}{2}$ \\
$\bH_d$ & $\bone$ & $\btwo$ & $-\frac{1}{2}$ \\ \hline
\end{tabular}
\caption{Quantum numbers of the MSSM.}   %??
\label{tab:MSSMgrid}
\end{table}

As an example of how symmetries can constrain a superpotential, consider a theory with (global) $\SU(3)$, $\SU(2)$, and $\U(1)$ symmetries containing superfields with quantum numbers given in \Tab{tab:MSSMgrid}.  Though these lectures are not supposed to cover topics of phenomenological relevance, this example was carefully chosen (and the fields appropriately named) to have the same symmetry structure as the MSSM, albeit for only one generation.

The leading relevant and marginal interactions are given by terms in the superpotential of up to dimension 3, which could include
\be
\bW \supset \mu \, \bH_u \bH_d  + \lambda^u \bQ \bU^c \bH_u + \lambda^d \bQ \bD^c \bH_d + \lambda^e\bE^c \bL \bH_d.
\ee
The fermions and scalars in the $\bH$ superfields (the Higgs bosons and higgsinos) receive a supersymmetric mass from the $\mu$ mass term.  If $\bH_u$ and $\bH_d$ get vevs in their lowest component, the other three superpotential terms then yield masses for the up-type quarks, down-type quarks, and leptons, respectively.  As the superpotential must be holomorphic, we need at least two Higgs doublets in order to give masses to all fermions.\footnote{In the actual MSSM with gauge interactions, the same argument can be made on the basis of anomaly cancellation.  One can have a single active Higgs doublet in the presence of higher-dimensional interactions \cite{Graham:2009gr,Dobrescu:2010mk,Ibe:2010ig,Davies:2011mp}.}  

As written, this superpotential obeys two additional global $\U(1)$ symmetries: a baryon number symmetry under which $\bQ$ and $\bU^c/\bD^c$ have opposite charges, and a lepton number symmetry under which $\bL$ and $\bE^c$ have opposite charges.  However, we can easily write down terms that do not respect these symmetries:
\be
\label{eq:RPVterms}
\bW  \supset \rho \bL \bH_u + \lambda^{(1)} \bQ \bD \bL + \lambda^{(2)} \bU \bD \bD + \lambda^{(3)} \bL \bL \bE.
\ee
With only one generation, the last two terms vanish (since the $\SU(2)$ and $\SU(3)$ indices are contracted antisymmetrically), but they are relevant in the SUSY SM with three generations.  If one wanted to forbid all of these terms, one could impose baryon and lepton number symmetries explicitly.  

Alternatively, one could use a $\U(1)_R$ symmetry to forbid the terms in \Eq{eq:RPVterms}.  If one gives the Higgs doublets an $R$-charge of $1$, and all other superfields an $R$-charge of $+1/2$, the problematic terms are forbidden since the resulting superpotential would not have an $R$-charge of 2.  We do not even need a full $R$-symmetry to achieve the same effect, which is desirable as we generally expect continuous $R$-symmetries to be broken by SUSY-breaking effects (or by $m_{3/2}$ if nothing else; see \Sec{subsec:AdSSUSY}).  The $R$-symmetry contains a discrete $\mathbf{Z}_2$ subgroup called $R$-parity, under which the Higgs doublets have $R$-parity $+1$ and the other multiplets have $R$-parity $-1$.   This is sufficient to forbid the terms in \Eq{eq:RPVterms}.

Of course, higher-dimension terms may be of interest or pose problems.  The superpotential may contain a $\bL \bH_u \bL \bH_u $ term which gives rise to neutrino masses; we saw in \Eq{eq:LHLH} that this could arise from integrating out a heavy multiplet (i.e.~the right-handed neutrino).  With three generations, the superpotential could also contain a term like $\bQ \bQ \bQ \bL$, which is $R$-parity even but violates both baryon and lepton number.

%% file: jthaler_TASI_Lecture3.tex
\section{SUSY Gauge Theories}
\label{sec:lecture3}

In this lecture, we will derive the lagrangian for SUSY gauge theories, starting from a discussion of gauge redundancy and the vector superfield.  We focus on Abelian gauge theories, with a quick discussion of the non-Abelian case in \Sec{subsec:NonAbelian}.   Using the super-trick of background superfields, we will be able to derive important one-loop effects including the Konishi anomaly.  We end with a brief discussion of spontaneous gauge symmetry breaking.  

\subsection{Gauge Redundancy and Gauge Invariance}
\label{subsec:GaugeRedundancy}

In order to describe a massless spin-1 field in non-SUSY theories, we introduce a gauge redundancy under which
\be
\label{eq:gaugeredundancy}
A_\mu(x) \to A_\mu(x) + \partial_\mu \alpha(x)
\ee
describes the same physics.  It is worth remembering that \Eq{eq:gaugeredundancy} is not a true symmetry of the theory (though we often call it a ``gauge symmetry"), since it does not package physical states into multiplets with shared properties.  Indeed, one can gauge fix the theory (e.g. to Coulomb gauge) to remove the redundancy at the expense of making the theory look non-Lorentz-invariant.  The purpose of gauge redundancy is to allow for a manifestly local and Lorentz-invariant  description of a massless spin-1 field.

The same thing will happen in SUSY lagrangians with massless spin-1 fields, except in order to make SUSY manifest, we will have to introduce \emph{even more} redundancy.  To see the reason for this, consider the action of a $\U(1)$ gauge transformation acting on a chiral superfield:
\be
\bPhi \rightarrow e^{i q \alpha(x)} \bPhi.
\ee
As discussed in \Sec{subsec:whatissuper}, because $\alpha(x)$ is a spacetime-dependent scalar and not a superfield, $\bPhi$ is no longer a superfield after this gauge transformation.  In order to maintain manifest SUSY, we would like to promote the gauge parameter to a full superfield
\be
\label{eq:newgauge}
\bPhi \rightarrow e^{q\bOmega} \bPhi.
\ee
To make sure that $\bPhi$ remains a chiral superfield, we need $\boldsymbol{\Omega}$ to also be a chiral superfield ($\Dbar\boldsymbol{\Omega}=0$), whose lowest imaginary component is the ordinary gauge parameter.  A theory that is invariant under \Eq{eq:newgauge} will have a high degree of redundancy, but we can always gauge fix the theory to remove the extra redundant modes.

With this augmented gauge transformation, the kinetic term in \Eq{eq:chiralkinetic} is clearly not invariant under $\U(1)$ transformations:
\be
\bPhi^\dagger \bPhi \to \bPhi^\dagger e^{q (\boldsymbol{\Omega}^\dagger +\boldsymbol{\Omega})} \bPhi.
\ee
However, we can compensate for this by introducing a vector superfield $\bV = \bV^\dagger$ that transforms under a gauge transformation according to 
\be
\label{eq:Vredundancy}
\bV \rightarrow \bV - \frac{\bOmega + \bOmega^\dagger}{2}.
\ee
In this way, the gauge-invariant kinetic term is given by
\be
\label{eq:chiralgaugekineticterm}
\mathcal{L}_{\rm kinetic} = \int d^4 \theta \, \bPhi^\dagger e^{2 q \bV} \bPhi.
\ee
The vector superfield $\bV$ contains a spin-1 gauge field with the desired gauge transformation property in \Eq{eq:gaugeredundancy}.  It also contains a spin-1/2 gaugino, an auxiliary $D$ component, as well as a number of redundant components that can removed by choosing the appropriate gauge fixing for $\bOmega$.

\subsection{The Vector Multiplet}
\label{subsec:VectorMultiplet}

A generic vector superfield satisfying $\bV = \bV^\dagger$ can be written as
\be
\label{eq:genericmultiplet}
\bV =\quad
\begin{array}{c|ccc}
& \theta^{0}& \theta^{1}& \theta^{2}\\
\hline
\rule[-3mm]{0mm}{.8cm}
\thetabar^{0}&c&\chi_\alpha&N\\
\thetabar^{1}&\overline{\chi}^\alphadot&  A_\mu& \overline{\lambda}^\alphadot- \frac{i}{2} (\sigmabar^\mu \partial_\mu \chi)^\alphadot \\
\thetabar^{2}& \phantom{N} N^\dagger \phantom{N} &\lambda_\alpha - \frac{i}{2} (\sigma^\mu \partial_\mu \overline{\chi})_\alpha & \frac{1}{2} D - \frac{1}{4} \Box c \\
\end{array}
\ee

where $c$ is a real scalar field, $\chi$ is a spin-1/2 Weyl fermion, $N$ is a complex scalar field, $A_\mu$ is a spin-1 gauge boson, $\lambda$ is a spin-1/2 gaugino, and $D$ is a real scalar auxiliary field.\footnote{It is unfortunate that the same symbol $D$ is used for the SUSY-covariant derivative.}  Many of these components are redundant, since they can be removed by performing the transformation in \Eq{eq:Vredundancy} with
\be
 \label{eq:redundantmodes}
\bOmega = \{ c , \sqrt{2} \chi_\alpha , 2 N \}.
\ee

This is known as fixing to Wess-Zumino gauge, where the remaining components of $\bV$ are
\be
\label{eq:vectorinWZgauge}
\bV_{\text{WZ}}=\quad
\begin{array}{c|ccc}
& \theta^{0}& \theta^{1}& \theta^{2}\\
\hline
\rule[-3mm]{0mm}{.8cm}
\thetabar^{0}&0&0&0\\
\thetabar^{1}&0&A_\mu&\overline{\lambda}^\alphadot\\
\thetabar^{2}&0&\lambda_\alpha&\frac{1}{2}D\\
\end{array}
\ee
The only remaining redundancy in $\bOmega$ that \Eq{eq:redundantmodes} does not fix is $\bOmega = i \alpha (x)$, which acts like $A_\mu \to A_\mu + \partial_\mu \alpha(x)$ because\footnote{This explains the need for the funny factor of 2 in \Eq{eq:Vredundancy}.}
\be
\bOmega = i \alpha (y)  \quad \Rightarrow \quad \frac{\bOmega + \bOmega^\dagger}{2} = - \thetabar\sigmabar^\mu \theta\partial_\mu \alpha (x).
\ee
Thus, the physical modes in the vector multiplet are a massless spin-1 gauge field $A_\mu$ as well as the spin-1/2 gaugino $\lambda$ and the auxiliary field $D$.

Wess-Zumino gauge explicitly breaks SUSY in the sense that \Eq{eq:vectorinWZgauge} is not a superfield, because it has ``arbitrary'' zeroed out entries (see \Sec{subsec:whatissuper}).  That said, after a SUSY transformation, one can perform a compensating gauge transformation to restore the Wess-Zumino form.  For determining gauge-invariant and SUSY-invariant lagrangians, we treat $\bV$ as a true superfield, but when expressing $\bV$ in components, we will always use Wess-Zumino gauge.

If instead we wished to describe a massive spin-1 particle, then the modes in $\bOmega$ are no longer redundant and there are additional propagating modes.  We will discuss this further in \Sec{subsec:spontaneouslybrokengauge}.

\subsection{Gauge-Invariant SUSY Lagrangians}
\label{subsec:GaugeSUSYlagrangians}

Armed with the gauge transformation properties of chiral and vector multiplets, we can write down gauge-invariant SUSY lagrangians.  We have already argued for the form of the gauge-invariant kinetic term for chiral multiplets in \Eq{eq:chiralgaugekineticterm}.  Expanding out in (Wess-Zumino gauge) components,
\be
\label{eq:gaugeLkinetic}
\begin{split}
\mathcal{L}_{\text{kinetic}}&=\int d^4\theta \,\bPhi^\dagger e^{ 2q \bV}\bPhi\\
&=\nabla _\mu\phi^\dagger\nabla^\mu\phi+i\overline{\psi}\sigmabar^\mu\nabla_\mu\psi+F^\dagger F\\
&\quad -\sqrt{2}q(\phi^\dagger\psi\lambda+\phi\overline{\psi}\overline{\lambda})+ q\phi^\dagger\phi D , \\
\end{split}
\ee
where $\nabla_\mu=\partial_\mu-i q A_\mu$ is the familiar gauge-covariant derivative for a field of charge $q$.\footnote{We are working in a non-canonical normalization where the gauge coupling appears in \Eq{eq:gaugekinetic}.} In the last line of \Eq{eq:gaugeLkinetic}, the first term corresponds to matter couplings to gauginos, and the second term will affect the scalar potential after integrating out the auxiliary field $D$.  Interestingly, SUSY has related the gauge boson coupling to the gaugino coupling (and the size of the $D$-term potential).

Next, we wish to write down the SUSY version of the field strength tensor.  We need an object that is simultaneously gauge invariant and a proper superfield and has the field strength $F_{\mu \nu}$ as one of its components.  While $\bV$ itself is a vector superfield, it is not gauge invariant.\footnote{If there were no gauge symmetry, i.e. a massive gauge boson, a term like $\int d^4\theta \, m^2 \bV^2$ would be allowed.  See \Sec{subsec:spontaneouslybrokengauge}.}   To form a gauge-invariant superfield, we use the super-trick from \Sec{subsec:IntroD} of creating new superfields by acting with the SUSY-covariant derivative.  Under a gauge transformation and using the fact that $\bOmega^\dagger$ is anti-chiral ($D_\alpha \bOmega^\dagger = 0$)
\be
D_\alpha \bV \to D_\alpha \bV+\frac{D_\alpha\bOmega}{2},
\ee
so while this is not gauge invariant, we have removed much of the gauge dependence.  Adding $\Dbar$ and using \Eq{eq:twoDsmakeaP},
\be
\Dbar_{\betadot}D_\alpha \bV \to \Dbar_{\betadot} D_\alpha \bV+i(\sigma\cdot\partial)_{\alpha\betadot}\bOmega.
\ee
The resulting object is still not gauge invariant, but adding one more $\Dbar$ will do the job:
\be
-\frac{1}{4}\Dbar^2 D_\alpha \bV\rightarrow -\frac{1}{4}\Dbar^2 D_\alpha \bV+ i (\sigma\cdot\partial)_{\alpha\betadot}\Dbar^{\betadot}\Omega = -\frac{1}{4}\Dbar^2 D_\alpha \bV.
\ee
Thus, we define the gauge-invariant chiral superfield
\be
\bWalpha \equiv -\frac{1}{4}\Dbar^2 D_\alpha \bV.
\ee
This superfield carries a Lorentz index (as anticipated in \Sec{subsec:SUSYmultiplets}), and it is manifestly chiral ($\Dbar^\betadot \bWalpha = 0$) because $\Dbar^3 = 0$.  Despite using the symbol $\bW$, $\bWalpha$ is unrelated to the superpotential from \Eq{eq:WZsuperpotential}.

Because $\bWalpha$ is a chiral multiplet, we can express its components using the $y^\mu$ coordinate from \Eq{eq:y}.  Because it is gauge invariant, the components in any gauge are
\be
\label{eq:Walphacomponents}
\begin{tabular}{r | c c c c }
& $\theta^{0}$ & $\theta^{1}$ & $\theta^{2}$ \\
\hline
$\bWalpha(y^\mu, \theta)$  & $\lambda_\alpha$ & $\theta_\alpha D+\frac{i}{2}(\sigma^\mu\sigmabar^\nu\theta)_\alpha F_{\mu\nu}$ & $i\theta^2(\sigma^\mu\partial_\mu\lambda^\dagger)_\alpha$
\end{tabular}
\ee\\
where $F_{\mu\nu}=\partial_\mu A_\nu-\partial_\nu A_\mu$ is the ordinary field strength.

In order to get the gauge kinetic term, we need a term quadratic in $\bWalpha$ (so it is automatically gauge-invariant) and Lorentz-invariant. The only option is
\be
\int d^2\theta \, \bW^\alpha \bWalpha,
\ee
whose component expansion is 
\be
\label{eq:Wsquaredcomponents}
2i\overline{\lambda}\sigmabar^\mu\partial_\mu\lambda+ D^2 - \frac{1}{2}F_{\mu\nu}F^{\mu\nu} +\frac{i}{4}\epsilon^{\mu\nu\rho\tau}F_{\mu\nu}F_{\rho\tau}.
\ee
The first term is the gaugino kinetic term, the second term shows that $D$ is an auxiliary field, the third term is the gauge boson kinetic term, and the last term corresponds to the CP-violating $\Theta$ term.  To get from \Eq{eq:Walphacomponents} to \Eq{eq:Wsquaredcomponents}, we have used the sigma matrix trace relation:
\be
\text{Tr}(\sigma^\mu\sigmabar^\nu\sigma^\rho\sigmabar^\tau)=2(\eta^{\mu\nu}\eta^{\rho \tau}-\eta^{\mu\rho}\eta^{\nu \tau}+\eta^{\mu \tau}\eta^{\nu\rho}+i\epsilon^{\mu\nu\rho \tau}).
\ee

Forming a hermitian lagrangian and introducing an overall normalization, the SUSY gauge kinetic term is 
\be
\label{eq:gaugekinetic}
\mathcal{L}_{\text{kinetic,gauge}}=\int d^2\theta \,\left(\frac{1}{4g^2}-\frac{i\Theta_{\rm CP}}{32\pi^2} \right)\bW^\alpha \bWalpha+\hc
\ee
whose component expansion is
\be
 - \frac{1}{4 g^2}F_{\mu\nu}F^{\mu\nu}+\frac{1}{g^2} i\overline{\lambda}\sigmabar^\mu\partial_\mu\lambda+ \frac{1}{2 g^2} D^2 + \frac{\Theta_{\rm CP}}{64\pi^2}\epsilon^{\mu\nu\rho \tau}F_{\mu\nu}F_{\rho \tau}. \label{eq:WalphaWalphacomponents}
\ee
We are using the non-canonical normalization where the gauge coupling appears in the gauge boson kinetic term (and by SUSY, also in the gaugino kinetic term and auxiliary $D$ term), not in its couplings to matter.  This normalization will be of great help when discussing background superfields in \Sec{subsec:BackgroundField}.  To use canonical normalization instead, simply redefine 
\begin{align}
\bV & \rightarrow g \bV,
\end{align}
or equivalently do the same for every component of $\bV$.  The $\Theta_{\rm CP}$ term in \Eq{eq:WalphaWalphacomponents} corresponds to a total derivative and has no physical relevance for an Abelian gauge symmetry (but defines the vacuum in a non-Abelian theory).

Putting the pieces together, the lagrangian for a renormalizable Wess-Zumino model with a $\U(1)$ gauge symmetry is
\be
\begin{split}
\mathcal{L}_{\rm WZ}=&\int d^4\theta \,\sum_i \bPhi^i e^{ 2q_i \bV}\bPhi_i^\dagger\\
&+\int d^2\theta \, \bW(\bPhi_i)+\hc\\
&+\int d^2\theta \,\frac{1}{4g^2}\bW^\alpha \bWalpha+\hc
\end{split}
\ee
where the field $\bPhi_i$ has charge $q_i$ and the superpotential $\bW(\bPhi_i)$ must be $\U(1)$ invariant.  After integrating out the $F$ and $D$ auxiliary fields (and passing to a canonical basis for the gauge fields), the scalar potential of the theory is
\be
\label{eq:scalarpotential}
V = F^{i\dagger} F^i + \frac{1}{2}D^2 = W_i W^{\dagger i} + \frac{1}{2}g^2 \left| \sum_i q_i \phi^{\dagger}_i \phi^i  \right|^2.
\ee

\subsection{Additional Gauge-Invariant Operators}
\label{eq:additional}

In the spirit of \Sec{subsec:dim56}, we would like to write down additional higher-dimension (and gauge-invariant) operators that would appear in a low energy effective theory.  By power counting
\be
[\bV] = 0, \qquad [\bWalpha] = 3/2,
\ee
and recall that $[d^2 \theta] = 1$ and $[d^4 \theta] = 2$.

First, using \Eq{eq:shiftByChiral}, the following dimension-2 term is actually gauge invariant in the Abelian case
\be\label{eq:FI-term}
\mathcal{L}_{\rm FI} = \int d^4\theta \, \kappa \bV.
\ee
This is called a Fayet-Iliopoulos (FI) term, and we will encounter it again in \Sec{subsec:dterm} when we discuss SUSY breaking.

Another potential gauge-invariant term at dimension 4 is 
\begin{align}
\label{eq:linearWalpha}
 \int d^4\theta \, \D^\alpha \bWalpha = 0.
\end{align} 
There are two ways to see why this vanishes.  First, this is a total SUSY-covariant derivative, so it vanishes by \Eq{eq:superintbyparts}.  Second, the combination $\bL \equiv \D^\alpha \bWalpha$ is a linear superfield because $D^2 \bL = 0$ and $\bL^\dagger = \bL$ (see \Eq{eq:linearmultiplet}).  The action $\int d^4\theta \, \bL$ vanishes since $\int d^4\theta \,\bL = - \frac{1}{4} \int d^2\theta \, \Dbar^2 \bL = 0$.

At dimension 5, there are  two types of terms, but only one of them is non-zero:\footnote{Because $\D^\alpha \bWalpha$ is a linear superfield, $\int d^4 \theta \, \D^\alpha \bWalpha \bPhi^\dagger$ is just the complex conjugate of the second term.}
\be
\frac{1}{\Lambda} \int d^2\theta \, \bPhi \bW^\alpha \bWalpha, \qquad \frac{1}{\Lambda} \int d^4 \theta \, \D^\alpha \bWalpha \bPhi = 0. \label{eq:lineardim5term}
\ee
As with the superpotential and K\"ahler potential, it is convenient to introduce the gauge kinetic function $\bF(\bPhi^i)$ to capture the first type of term
\be
\int d^2\theta \, \bF(\bPhi^i) \bW^\alpha \bWalpha.
\ee
The gauge kinetic function is a holomorphic function of chiral superfields (and therefore a chiral superfield itself), whose lowest real component is $1/4g^2$.  We will address $\bF(\bPhi^i)$ extensively in \Secs{subsec:BackgroundField}{subsec:Konishi}, with subtleties related to holomorphy discussed in \Sec{subsec:holomorphytroubles}.

Here are some representative terms at dimension 6:
\begin{align}
\frac{1}{\Lambda^2} \int d^2 \theta \, \bPhi_1 \bPhi_2 \bW^\alpha \bWalpha, & \qquad \frac{1}{\Lambda^2} \int d^4 \theta \, \D^\alpha \bWalpha \bPhi_1^\dagger \bPhi_2, & \\
\frac{1}{\Lambda^2} \int d^4 \theta \, \bW^\alpha \sigma^\mu_{\alpha \alphadot} \partial_\mu \bW^{\dagger \alphadot}, & \qquad
\frac{1}{\Lambda^2} \int d^4 \theta \, \bPhi^\dagger \bW^\alpha \bWalpha.
\end{align}
The first term is already captured by the gauge kinetic function $\bF(\bPhi^i)$, while the others are genuinely new terms which would appear in any realistic effective lagrangian.

\subsection{The Non-Abelian Case}
\label{subsec:NonAbelian}

Thus far, we have only discussed Abelian gauge theories.  Non-Abelian groups involve more algebra, but the basic physics is the same.  For each gauge boson $A_\mu^a$ where $a$ labels the gauge generator $T^a$, we need a separate vector multiplet $\bV^a$ and gauge transformation parameter $\bOmega^a$:
\be
\bV^a\rightarrow \bV^a- \frac{\bOmega^a+\bOmega^{a\dagger}}{2}+\dots
\ee
Using the shorthand $\bV \equiv \bV^aT^a$ and $\bOmega=\bOmega^aT^a$, the full non-linear gauge transformation is
\be
e^{2\bV}\rightarrow e^{-\bOmega^\dagger}e^{2\bV} e^{-\bOmega}.
\ee
The gauge covariant object
\be
\bWalpha \equiv-\frac{1}{8}\Dbar^2 \left(e^{-2\bV}D_\alpha e^{2\bV} \right)
\ee
transforms in the adjoint representation of the gauge group as 
\be
\bWalpha\rightarrow e^{\bOmega}\bWalpha e^{-\bOmega},
\ee
while chiral matter fields transform as
\be
\bPhi \to e^{\bOmega^a T^a_{\Phi}}\bPhi,
\ee
where $T^a_{\Phi}$ are the generators corresponding to the appropriate representation of $\bPhi$.

The combination $\text{Tr}(\bW^\alpha\bW_\alpha)=\frac{1}{2}\bW^{\alpha a}\bW_{\alpha}^a$
is gauge invariant\footnote{We use the usual normalization for generators $\text{Tr}(T^aT^b)=\frac{1}{2}\delta^{ab}$.} and the gauge kinetic lagrangian is given by
\be
\mathcal{L}_{\text{gauge}}=\int d^2\theta \,\left(\frac{1}{4g^2}-\frac{i\Theta_{\rm CP}}{32\pi^2}\right)\bW^{\alpha a} \bW_{\alpha}^a+\hc
\ee
while the matter kinetic terms are
\be
\mathcal{L}_{\text{kinetic}}=\int d^4\theta \,\bPhi^\dagger e^{2\bV}\bPhi.
\ee
Note that the $\Theta_{\rm CP}$ term is physical for a generic non-Abelian gauge group.  Also, notice that the Fayet-Iliopoulos term from \Eq{eq:FI-term} is not allowed for non-Abelian gauge groups.

For doing phenomenological studies (such as for the SUSY SM) the typical starting point is:
\be
\begin{split}
\mathcal{L}=&\int d^4\theta \, \bK(\bPhi^{\dagger\i} e^{2\bV},\bPhi^j)\\
&+\int d^2\theta \, \bW(\bPhi^i)+\hc\\
&+\int d^2\theta \, \bF_{ab}(\bPhi^i)\bW^{\alpha a}\bW_{\! \alpha}^b +\hc\\
&+D_\alpha,\partial_\mu\textrm{ terms}.
\end{split}
\ee
where $\bK$ is the K\"ahler potential, $\bW$ the superpotential, and $\bF_{ab}$ is the gauge kinetic function (now with generator indices).

\subsection{Super-trick \#3:  Background Superfields}
\label{subsec:BackgroundField}

Our third super-trick is to use the background field method to calculate otherwise tedious one-loop effects directly in superspace.  In non-SUSY theories, background fields are a powerful technique to calculate effective actions for scalar fields \cite{Abbott:1981ke}.  The basic idea is to treat any scalar $\phi(x)$ as if it were a constant $\phi_0$ and calculate the effective action in the background $\phi_0$.  Up to derivative terms like $\partial_\mu \phi$, the resulting action for the $\phi_0$ background can be lifted to an action for the full $\phi(x)$.

In SUSY theories, this technique becomes even more powerful since the action for a constant $\phi_0$ background can be lifted to an action for a full superfield $\bPhi$.  Of course, in general the lagrangian can depend on
\be
\mathcal{L}(\bPhi,\partial_\mu \bPhi, D_\alpha \bPhi, \ldots),
\ee
and the background field method misses dependence involving derivatives on superfields.  In practice, this is not too much of a limitation, since the K\"ahler potential, superpotential, and gauge kinetic functions are functions only of superfields and not their derivatives.\footnote{An important exception shows up in the case of SUSY breaking, where $\vev{D^2 \bPhi} \not = 0$, so the background field method can miss certain important SUSY-breaking effects.}

As an example of using the background field method, consider a gauge theory with a massive ``quark'' $\bQ\bQ^c$ coupled to a chiral superfield $\bX$:
\be
\label{eq:Wqqc}
\bW= m \bQ \bQ^c + \lambda \bX \bQ \bQ^c.
\ee
Assuming that $\vev{\bX} = 0$, the quark is heavy and we can integrate it out of the spectrum. The tree-level equations of motion for $\bQ$ and $\bQ^c$ simply tell us that $\vev{\bQ} = \vev{\bQ^c} = 0$, so we need to go to loop-level to find a non-trivial effective Lagrangian below the quark mass threshold $m$.  In particular, we would like to know if any operator of the type
\be
\label{eq:HigherOp}
\frac{1}{\Lambda}\int d^2\theta \, \bX\bW^\alpha \bWalpha
\ee
is generated after integrating out $\bQ\bQ^c$.  Using the background field method, we can replace $\bX$ with the background scalar field $x_0$ and ask whether $\int d^2\theta \, x_0 \bW^\alpha \bWalpha$ is generated.  This in turn is equivalent to asking whether the gauge coupling for the $\U(1)$ gauge boson has any dependence on $x_0$.

Consider the running of the gauge coupling at one loop.  Suppose $g(\Lambda)$ is the value of the gauge coupling at some high energy scale $\Lambda$ above the quark mass scale $m$.  Then at some energy $\mu<m$, the low energy gauge coupling will be
\be
\label{eq:gIR}
\frac{1}{g^2(\mu,m)}=\frac{1}{g^2(\Lambda)}+\frac{b_0}{8\pi^2}\log\frac{\Lambda}{m}+\frac{b_1}{8\pi^2}\log\frac{m}{\mu}.
\ee
Here, $b_0$ is the one-loop beta-function at energies $m<E<\Lambda$, which includes $\bQ,\bQ^c$ as degrees of freedom, while $b_1$ is the beta-function at energies $\mu<E<m$ where $\bQ$, $\bQ^c$ have been integrated out.
In particular, the $b$ coefficients are defined by
\be
\frac{d\alpha}{d\log{E}}=\frac{b}{2\pi}\alpha^2 
\ee
at the appropriate scale $E$, where $\alpha=\frac{g^2}{4\pi}$.

Turning on a background value of $x_0$ is equivalent to shifting $m \to m + \lambda x_0$.  Using the background field method, we can lift $m$ to have dependence on the full superfield $\bX$ via
\be
m \to m + \lambda \bX.
\ee
This in turn allows us to analytically continue the gauge coupling into superspace \cite{Giudice:1997ni,Giudice:1998bp,ArkaniHamed:1998kj} by promoting the gauge coupling to a chiral superfield (i.e~the gauge kinetic function):
\be
\bF(\bX) = \frac{1}{4g^2(\mu,m+\lambda \bX)}=\frac{1}{4g^2(\mu,m)}-\frac{\lambda (b_0-b_1)}{32\pi^2}\frac{\bX}{m}+\mathcal{O}\left(\frac{\bX^2}{m^2}\right).
\ee
Indeed, the operator of \Eq{eq:HigherOp} is generated with a coefficient proportional to the differences of beta functions.
\be
\label{eq:Leff}
\mathcal{L}_{\text{eff}}\supset\ \int d^2\theta \,\left(\frac{1}{4 g^2(\mu, m)}-\frac{\lambda (b_0-b_1)}{32\pi^2}\frac{\bX}{m}\right)\bW^\alpha \bWalpha+\hc
\ee

A few comments are in order about this result:
\begin{enumerate}
\item If $\langle F_X\rangle\neq 0$, then a gaugino mass is generated in \Eq{eq:Leff}.  Indeed this is an example of soft SUSY-breaking terms generated through gauge mediation. In \Refs{Giudice:1997ni,Giudice:1998bp,ArkaniHamed:1998kj}, the background superfield method was used extensively to recover different types of soft SUSY-breaking terms generated by integrating out some high-energy dynamics, including two-loop (and higher) effects.

\item The background superfield method only catches leading terms in the effective Lagrangian. In particular terms involving $D_\alpha \bX$ and $\partial_\mu \bX$ can not be recovered with this method, as $D_\alpha m=0$ and $\partial_\mu m=0$.

\item Since $m$ is just a real parameter, one might wonder why we made the replacement $m \to m + \lambda \bX$ as opposed to $m\rightarrow m+\lambda \bX^\dagger$ or $m \to \sqrt{(m + \lambda \bX)(m + \lambda \bX^\dagger)}$.  Ultimately, holomorphy of the gauge kinetic function forbids any alternative replacements, though we will discuss subtleties of this argument in \Sec{subsec:holomorphytroubles}. 

\item The background superfield method has nothing in particular to do with SUSY and it can be used in ordinary non-SUSY theories as well. For instance, if we want to calculate the coupling of the Higgs boson to
two gluons in the SM, we can apply the same procedure as above. In this case, $\mu\simeq m_h\simeq 125$ GeV and in running from the high scale $\Lambda$ down to the Higgs mass, the top mass $m_t=\lambda_t v_{\rm EW}$ is the only threshold we cross.  The low energy effective lagrangian is
\be
\mathcal{L}_{\rm eff} = - \frac{1}{2 g_S^2(\mu,m_t)} \mathrm{Tr} \left(G^{\mu \nu} G_{\mu \nu} \right),
\ee
where $G_{\mu \nu}$ is the gluon field strength and
\be
\frac{1}{g_S^2(\mu,m_t)}=\frac{1}{g_S^2(\Lambda)}+\frac{b_0}{8\pi^2}\log\frac{\Lambda}{m_t}+\frac{b_1}{8\pi^2}\log\frac{m_t}{\mu}.
\ee
The top mass depends on the background (physical) Higgs boson $h_0$ as
\be
m_t\longrightarrow m_t+\frac{\lambda_t}{\sqrt{2}}h_0,
\ee
so after integrating out the top quark we generate an operator
\be
\mathcal{L}_{\text{eff}}\supset\frac{\sqrt{2}g_S^2(m_h)}{48\pi^2v_{\rm EW}} h_0 \mathrm{Tr} \left(G^{\mu \nu} G_{\mu \nu} \right),
\ee 
where $G^{\mu\nu}$ is now canonically normalized. This operator gives the leading contribution to the Higgs-gluon-gluon vertex,
with contributions coming from loops of lighter particles of mass $m$ suppressed by powers of $m/m_h$.  It exhibits the famous {\it non-decoupling} effect that the top contribution is independent of $\lambda_t$ for $m_t \gg m_{h_0}$.
\end{enumerate}

\subsection{The Konishi Anomaly}
\label{subsec:Konishi}

We can use the background field method to understand other loop-level effects, such as the Konishi anomaly \cite{Clark:1979te,Konishi:1983hf}.  Recall that in a non-SUSY $\U(1)$ gauge theory with fermions $\psi$ ($\psi^c$) with charges $+1$ ($-1$), the chiral rotation
\be
\psi\rightarrow e^{+i\alpha}\psi, \qquad \psi^c\rightarrow e^{+i\alpha}\psi^c,
\ee
is anomalous.  This means that if we perform the chiral rotation with a constant value of $\alpha$, we must include an anomaly term to recover the same physics.
\be
\label{eq:chiralanomaly}
\mathcal{L}(\psi, \psi^c)\rightarrow\mathcal{L}(e^{+i\alpha}\psi, e^{+i\alpha}\psi^c)+\frac{\alpha}{64\pi^2}\epsilon^{\mu\nu\rho\tau}F_{\mu\nu}F_{\rho\tau}.
\ee

We can use the background field method to understand what happens if $\alpha\rightarrow \boldsymbol{\alpha}$ is promoted to a superfield.  Consider the chiral rotation
\be
\bPhi\rightarrow e^{\boldsymbol{\alpha}}\bPhi, \qquad \bPhi^c\rightarrow e^{\boldsymbol{\alpha}}\bPhi^c,
\ee
where $\boldsymbol{\alpha}$, $\bPhi$, and $\bPhi^c$ are all chiral multiplets.  This is a valid field redefinition since it leaves the one-particle asymptotic states unchanged.  However, this chiral rotation is anomalous, so to reflect the anomaly (while maintaining manifest SUSY) we must have
\be
\label{eq:Konishianomaly}
\mathcal{L}(\bPhi, \bPhi^c)\rightarrow\mathcal{L}(e^{\boldsymbol{\alpha}}\bPhi, e^{\boldsymbol{\alpha}}\bPhi^c)+\frac{1}{16\pi^2}\int d^2\theta \, \boldsymbol{\alpha} \bW^\alpha \bWalpha+\hc
\ee
The last term, which is a generalization of the familiar chiral anomaly, is known as the Konishi anomaly.  The imaginary component of $\boldsymbol{\alpha}$ corresponds to \Eq{eq:chiralanomaly}, while the other components give new effects required by SUSY.

To see the Konishi anomaly in action, consider the lagrangian
\be
\begin{split}
\mathcal{L}&=\int d^4\theta \,(\bQ^\dagger e^{2\bV}\bQ+\bQ^{c\dagger}e^{-2\bV}\bQ^c)\left(1+\frac{\bX}{\Lambda}+\frac{\bX^\dagger}{\Lambda}+\cdots\right)\\
&\quad +\left(\int d^2\theta \, m\bQ\bQ^c+\int d^2\theta \,\frac{1}{4g^2}\bW^\alpha \bWalpha +\hc\right).\\
\end{split}
\ee
When we integrate out $\bQ$ and $\bQ^c$ at the mass scale $m$, we might wonder if a one loop $\bX\bW^\alpha \bWalpha$ coupling is generated.
Since $\bX$ and $\bX^\dagger$ only appear in the combination $\bX+\bX^\dagger$, holomorphy of the gauge kinetic
function would forbid such a term. We can easily check this using the Konishi anomaly.  By performing a field redefinition
\be
\bQ\rightarrow\bQ e^{-\bX/\Lambda}, \quad \bQ^c\rightarrow\bQ^c e^{-\bX/\Lambda},
\ee 
we can remove the linear term $\bX+\bX^\dagger$ in the K\"ahler potential, but this changes the superpotential
\be
m\bQ\bQ^c \to me^{-2\bX/\Lambda}\bQ\bQ^c.
\ee 
In addition, because of the Konishi anomaly, we also get a term 
\be
\frac{1}{4g^2(\Lambda)}\rightarrow\frac{1}{4g^2(\Lambda)}-\frac{1}{16\pi^2}\frac{\bX}{\Lambda}.
\ee
Integrating out $\bQ,\bQ^c$ using the background field method, at energies $\mu<m$ we have
\be
\begin{split}
& \frac{1}{4g^2(\mu, me^{-2\bX/\Lambda})}-\frac{1}{16\pi^2}\frac{\bX}{\Lambda}\\
&\qquad  =\frac{1}{4g^2(\mu,m)}-\frac{b_0-b_1}{32\pi^2}\frac{2\bX}{\Lambda}+\mathcal{O}(\bX^2)-\frac{1}{16\pi^2}\frac{\bX}{\Lambda}\\
&\qquad  =\frac{1}{4g^2(\mu,m)}+\mathcal{O}(\bX^2),
\end{split}
\ee
where $b_0-b_1=-1$ for integrating out $\bQ$ and $\bQ^c$.\footnote{The beta function contributions from a chiral multiplet include those from both the scalar and fermion components.}  By properly including the Konishi anomaly, we get no $\bX \bW^\alpha \bWalpha$ coupling as expected from holomorphy.\footnote{If $m=0$ there is no cancellation between the Konishi anomaly and the gauge coupling threshold effect, and the resulting term is related to an effect called ``anomaly mediation" \cite{Randall:1998uk,Giudice:1998xp}.}

\subsection{Holomorphy and its Subtleties}
\label{subsec:holomorphytroubles}

Holomorphy is a powerful tool to constrain the form of terms in the $\int d^2 \theta$ part of a SUSY action.  Because the  $\int d^2 \theta$ term only leads to a SUSY-invariant action if the integrand is a chiral superfield, this dramatically reduces the possibilities for what can appear in $\int d^2 \theta$ terms.  

The most celebrated applications of this principle are in superpotential non-renormalization theorems \cite{Grisaru:1979wc,Seiberg:1993vc}. The superpotential $\bW$ must be a function of chiral superfields only and therefore must be holomorphic.  The same restrictions also apply to couplings if one promotes them to spurions, such that a coupling $\lambda$ but not its complex conjugate $\overline{\lambda}$ can appear in the superpotential.  If one also charges the $\lambda$ spurion under $U(1)_R$ and ordinary global symmetries, there cannot be arbitrary functions of $\lambda$ in the superpotential.  In fact, by ensuring that the theory has sensible limits, one can show that no new appearances of the couplings in the superpotential are induced by quantum corrections.  That is, couplings in the superpotential do not run under perturbative renormalization group flow (though there may be non-perturbative corrections due to instantons).

Similar arguments apply to gauge theories, whose kinetic Lagrangian can also be expressed as a $d^2 \theta$ integral.  We can promote the gauge coupling (and $\Theta_{\rm CP}$ angle) to a spurion $\bS$
\begin{align}
\mathcal{L} & \supset \int d^2 \theta \, \bS \bW^\alpha \bWalpha .
\end{align}
In general, one might think that $\bS$ could change into an arbitrary chiral multiplet under renormalization group flow.  However, the lowest imaginary component of $\bS$ is $\Theta_{\rm CP}$, which cannot have any effect on physics in perturbation theory.\footnote{Non-perturbative corrections to the gauge coupling due to instantons are allowed.}  Insisting on a well-defined  zero-coupling limit ($g \rightarrow 0$, $\bS \rightarrow \infty$), the only allowed change under renormalization group flow is \cite{ArkaniHamed:1998kj}
\be
\bS \to \bS + \text{constant}, \label{eq:Soneloop}
\ee
which is how the gauge coupling runs at one loop (see \Eq{eq:gIR}).  Higher-loop contributions are prohibited by holomorphy plus $\Theta_{\rm CP}$, so the beta function is saturated at one loop.\footnote{Note that we cannot give $\bV$ any global $U(1)_R$ charge as we did for fields and couplings in the superpotential.  In any interacting gauge theory, $\bV$ always appears in the form $e^{\bV}$ somewhere, so $\bV$ cannot have $R$-charge.  This similarly implies $\bS$ cannot have $R$-charge, so we cannot use such arguments to rule out the constant in \Eq{eq:Soneloop}.}

The subtlety of using such holomorphy arguments, though, is that some effects that would be forbidden in $\int d^2 \theta$ terms can appear in $\int d^4 \theta$ terms.  For example, in the lagrangian
\be
\int d^4 \theta \, \bPhi^\dagger \bPhi + \left(\int d^2 \theta \,  (m \bPhi^2 + \lambda \bPhi^3) + \hc \right),
\ee
we know by holomorphy that the (holomorphic) $m$ and $\lambda$ do not run at any loop order.  However, since the K\"ahler potential is not holomorphic, there is no restriction on wavefunction renormalization
\be
\label{eq:Zfactor}
\bPhi^\dagger \bPhi \to Z(\mu) \bPhi^\dagger \bPhi,
\ee
and via a field rescaling $\bPhi \to \bPhi/\sqrt{Z(\mu)}$, the physical values of $m$ and $\lambda$ do run
\be
m \to \frac{m}{Z(\mu)}, \quad \lambda \to \frac{\lambda}{Z(\mu)^{3/2}}.
\ee
Of course, holomorphy is still extremely powerful for telling us that the combination $m^3/\lambda^2$ is invariant.

A similar subtlety relates to the argument that the (holomorphic) gauge coupling only runs at one-loop.  Consider the effect of the field rescaling $\bPhi \to \bPhi/\sqrt{Z(\mu)}$.  Via the Konishi anomaly in \Eq{eq:Konishianomaly}, scaling $\bPhi$ by a real number changes the effective gauge coupling, and since $Z(\mu)$ runs at all loop order, the (physical) gauge coupling must as well.  The solution to this is contained in the NSVZ beta function \cite{Novikov:1983uc,Novikov:1985rd,Novikov:1985ic}, which differentiates the canonical gauge coupling from the holomorphic gauge coupling \cite{Shifman:1986zi,Shifman:1991dz,Dine:1994su,ArkaniHamed:1997mj}.

With respect to the arguments presented in \Secs{subsec:BackgroundField}{subsec:Konishi} using background field methods, we used holomorphy to argue that the gauge kinetic function should only be a function of $\bX$ and not $\bX^\dagger$, which let us unambiguously continue the background value $x_0 \to \bX$.  This logic (which is still correct) assumed that we had to write the gauge kinetic term in the form $\int d^2 \theta \, \bF \bW^\alpha \bWalpha$.  Consider, though, an alternative way to write the gauge kinetic term
\be
\label{eq:altgaugekinetic}
\int d^2\theta \, \frac{1}{4g^2} \bW^\alpha \bWalpha+\hc = \int d^4\theta \,\frac{1}{4g^2}\left(\bW^\alpha D_\alpha \bV+\bW^\dagger_{\! \alphadot}\Dbar^{\alphadot}\bV\right), 
\ee
which can be verified by using the equivalence of $\int d^4 \theta$ and $\int d^2 \theta\left( -\frac{1}{4}\Dbar^2\right)$.  This second expression is fully gauge-invariant (for the same reason that the second term in \Eq{eq:lineardim5term} vanished), so one might worry that we could analytically continue $1/g^2$ in an alternative way.  Luckily, this argument does not hold and the right-hand side of \Eq{eq:altgaugekinetic} is not gauge-invariant for a generic superfield-valued $1/g^2$, so the logic in \Secs{subsec:BackgroundField}{subsec:Konishi} is still correct, albeit for subtle reasons.  

Finally, note that we can write down a gauge-invariant, non-local expression in superspace
\be
\int d^4\theta \,\frac{1}{4g^2}\bW^\alpha\frac{\Dbar^2}{8\Box}\bWalpha,
\ee
which can be shown to be equivalent to \Eq{eq:altgaugekinetic} using the chiral projector from \Eq{eq:chiralprojector}.  Here, $1/g^2$ \emph{can} be lifted to a real vector superfield, in which case the correct analytic continuation is not $m \to m + \lambda \bX$ but $m \to \sqrt{(m + \lambda \bX)(m + \lambda \bX^\dagger)}$.  Because of the logarithmic structure of the one-loop beta function, however, this way of writing the gauge kinetic term yields the same answer as \Eq{eq:Leff}.  The reason for the equivalence is ultimately due to holomorphy, albeit now in a hidden form.  See \Ref{ArkaniHamed:1998kj} for further discussion.

\subsection{Spontaneously Broken Gauge Theories}
\label{subsec:spontaneouslybrokengauge}

Before tackling SUSY-breaking theories in the next lecture, we want to briefly mention spontaneously broken gauge theories.  Consider the SUSY lagrangian of a gauge theory with a chiral multiplet $\bQ$ ($\bQ^c$) with $\U(1)$ charge $+1$ ($-1$), and a neutral chiral multiplet $\bN$.  The superpotential for this theory is
\be
\bW = \lambda \bN \left(\bQ \bQ^c - \mu^2 \right).
\ee
By the $\bN$ equation of motion (assuming SUSY is to remain unbroken)
\be
\frac{\partial \bW}{\partial \bN} = \lambda(\bQ  \bQ^c - \mu^2),
\ee
so for this to equal 0, both $\bQ$ and $\bQ^c$ must get vacuum expectation values which break the $\U(1)$ symmetry.  (In fact, by the $D$-term equations of motion, $\vev{\bQ} = \vev{\bQ^c}$.)  This will yield a spontaneously broken massive gauge theory.

To better understand the physics, consider the field redefinitions
\be
\bQ \equiv (\mu + \bR)e^\bPhi, \qquad \bQ^c \equiv (\mu + \bR)e^{-\bPhi},
\ee
where $\bR$ and $\bPhi$ are chiral multiplets.  The superpotential in these fields is
\be
\bW = 2 \lambda \mu \bN \bR + \lambda \bN \bR^2,
\ee
showing that $\bN$ and $\bR$ get a mass of $2 \lambda \mu$.  Note that the field $\bPhi$ does not appear in the superpotential, since under the gauge transformation of \Eq{eq:newgauge}, it transforms as a shift
\be
\bPhi \rightarrow \bPhi + \bOmega.
\ee
Focusing only on the (as of yet massless) $\bPhi$ field, the kinetic terms for $\bQ$ and $\bQ^c$ become
\be
\int d^4 \theta \, \bQ^\dagger e^{2\bV} \bQ + \bQ^{c\dagger} e^{-2\bV} \bQ^c \to \int d^4 \theta \, \mu^2 \left(e^{2\bV + \bPhi^\dagger + \bPhi} +  e^{-(2\bV + \bPhi^\dagger + \bPhi)}\right).
\ee
We can use our gauge redundancy with $\bOmega = \bPhi$ to set all of the components of $\bPhi$ to zero, resulting in
\be
\int d^4 \theta \, 4 \mu^2 \bV^2 + \ldots
\ee
As we can no longer use the gauge redundancy to go to Wess-Zumino gauge, this is a mass term for all of the components of the vector multiplet.  It contains a massive gauge boson, a Dirac fermion (formed from the two Weyl fermions in $\bV$, $\chi$ and $\lambda$), and a real scalar $c$.  Going to canonical normalization, they all have a mass of $2 g \mu$.  Effectively, the now-massive spin-1 gauge boson has ``eaten'' the lowest imaginary component of $\bPhi$, leaving one real scalar that gets the same mass because of SUSY.

%% file: jthaler_TASI_Lecture4.tex
\section{SUSY Breaking and Goldstinos}
\label{sec:lecture4}

In this lecture, we discuss the physics of spontaneous SUSY breaking, including the standard SUSY breaking paradigm of a hidden sector coupled to a visible sector.  After giving a few explicit examples of SUSY breaking theories, we highlight the important role of the goldstino and demonstrate a super-trick to calculate the goldstino couplings to matter.  We conclude with a brief discussion of SUSY breaking in SUGRA.

\subsection{Spontaneous SUSY Breaking}
\label{subsec:operatorlanguage}

Thus far, we have emphasized the development of SUSY from a superspace lagrangian point of view.  To understand the basics of SUSY breaking, it is helpful to think in a state/operator language.  If the ground state $\left|0\right>$ of a theory preserves SUSY, this means that a SUSY transformation leaves the vacuum invariant
\be
e^{-i\epsilon \Q-i\bar{\epsilon}\Qbar} \left|0\right> = \left|0\right>,
\ee
or equivalently
\be
\label{eq:SUSYQequalszero}
\Q^\alpha \left|0\right> = 0, \quad \Qbar^\alphadot \left|0\right> = 0  \quad (\text{SUSY vacuum}).
\ee
If the ground state spontaneously breaks SUSY, this means that the vacuum shifts under a SUSY transformation,
\be
\Q^\alpha \left|0\right> \not= 0   \quad (\text{SUSY-breaking vacuum}).
\ee
We would like to find simple criteria to determine whether or not SUSY is spontaneously broken.

It is instructive to consider the vacuum energy of the theory.  Using the SUSY algebra in \Eq{eq:SUSYalgebra}, the Hamiltonian $H = P_0$ of a SUSY theory is
\be
\label{eq:Hdefinition}
H = \frac{1}{4} (\Qbar_1 Q_1 + Q_1 \Qbar_1 + \Qbar_2 Q_2 + Q_2 \Qbar_2 ).
\ee
If SUSY is unbroken, then by \Eq{eq:SUSYQequalszero}
\be
\vev{H} = \left<0\right| H \left|0\right> = 0 \quad (\text{SUSY vacuum}),
\ee
so the vacuum energy is zero.  The converse is also true, such that a zero vacuum energy implies $\Q^\alpha \left|0\right> = 0$ and SUSY is unbroken.  In contrast, if the vacuum energy is non-zero, then SUSY is spontaneously broken in the vacuum.  In fact, because each term in \Eq{eq:Hdefinition} is an operator squared,
\be
\vev{H} > 0 \quad (\text{SUSY-breaking vacuum}),
\ee
so spontaneous SUSY breaking corresponds to a strictly positive vacuum energy.

Our world is clearly not supersymmetric (otherwise we would see sparticles around!), so if SUSY is realized in nature, it must be spontaneously broken.  By the above logic, one might conclude that $\vev{H}>0$ in our universe.  Observational measures of the cosmological constant reveal that $\vev{H} \approx 0$, though, posing a conundrum.  As we will explain in \Sec{subsec:AdSSUSY}, our universe must have an underlying anti-de Sitter (AdS) SUSY, and flat space corresponds to a large breaking of that AdS$_4$ SUSY.  To a good approximation, though, we can still use the flat space SUSY algebra for understanding the physics of SUSY and SUSY breaking.

Beyond the vacuum energy, there is one other important way to test if SUSY is broken. Because the mass-squared operator $P^2$ commutes with $Q_\alpha$ and $\Qbar_{\betadot}$, components of an irreducible SUSY multiplet must have the same mass when SUSY is unbroken.  More formally, mass is a Casimir invariant of the SUSY-extended Poincar\'{e} algebra.  Therefore, mass-squared splittings between states in the SUSY multiplet is evidence for (flat space) SUSY breaking.

Beyond simple mass splittings, there are other possible signatures of spontaneous SUSY breaking that can appear as terms in the low-energy effective theory.  These are denoted ``soft SUSY-breaking terms"  because when considered by themselves, they break SUSY in a way that does not  introduce quadratic divergences.  We will discuss these soft terms in more detail in \Sec{subsec:Gcouplings}. 
\subsection{The Vacuum Energy}

Now that we know that the vacuum energy offers a robust test for SUSY breaking, we can return to our lagrangian point of view.  The vacuum energy in a global SUSY theory is governed by the scalar potential.  As we saw in \Eq{eq:scalarpotential}, for a renormalizable gauge theory
\be
V = F^{i \dagger}F^i + \frac{1}{2} D^a D^a, 
\ee
where (for canonically normalized gauge fields)
\begin{align}
F^i&=-W^{\dagger i}, &D^a=-g(\phi^*T^a\phi).
\end{align}
If SUSY is broken, at least one $F$-term or $D$-term must have a non-zero expectation value.  Conversely, if all $\vev{F^i}$ and $\vev{D^a}$ are zero, then SUSY is unbroken.

Including the effects of a K\"ahler potential and gauge kinetic function, the vacuum energy generalizes to
\be
V = g_{i \j}F^iF^{\dagger\j} + \frac{1}{2}(\text{Re}f_{ab}) D^a D^b,
\ee
where now
\be
F^i=-g^{i\j}W^\dagger_\j, \qquad D^a=-(\text{Re}f^{-1})^{ab}(K_iT_b\phi^i).
\ee
Here $f_{ab}$ is the gauge kinetic function for non-canonically normalized gauge fields, $K_i=\frac{\partial K}{\partial\phi^i}$, and the K\"ahler metric $g^{i \j}$ was introduced in \Sec{subsec:Kahler}.  Even in this more general case, we see that to find SUSY-breaking theories, we simply need to find scenarios where $\vev{F^i} \not = 0$ and/or $\vev{D^a} \not = 0$.  Adding SUSY-covariant derivatives gives additional contributions to the vacuum energy, but does not change the requirement that SUSY breaking requires a non-zero vev for at least one auxiliary field.

\subsection{The Standard SUSY-Breaking Paradigm}
\label{subsec:paradigm}

Before talking about explicit models that break SUSY, we do want to make a connection to some physics of phenomenological relevance.  Clearly, if SUSY is realized in nature, it must be spontaneously broken, since we need a mass splitting between SM particles and their superpartners.  Crucially, though, experimental bounds imply that most of the superpartners should be heavier than their SM counterparts (with a key exception being the top squark).

For {\it renormalizable tree-level} theories that spontaneously break SUSY, there is a supertrace sum rule that says that \cite{Ferrara:1979wa}
\begin{align}
\label{eq:Str}
\textrm{STr}(m^2) & \equiv \sum_s (-1)^{2s} (2s +1 ) \textrm{Tr}(m_s^2)=-2g_a\text{Tr}(T^a)\vev{D^a} = 0,
\end{align}
where $s$ represents the spin of the particle.\footnote{This last equality is obvious for a non-Abelian gauge theory with $\text{Tr}(T^a)$ = 0.  For a $\U(1)$ gauge group, the sum of the hypercharges must vanish to avoid the gravitational anomaly.}  Consider the MSSM with flavor conservation (i.e. no mixing between scalars of different generations) and with no additional broken $\U(1)$ symmetries involved in SUSY breaking. For the first generation of squarks, for example, since
\be
\text{Tr}(\sigma^3)=0\quad\text{and} \quad Y_{\tilde{u}_L}+Y_{\tilde{u}^*_R}+Y_{\tilde{d}_L} + Y_{\tilde{d}_R} =0,
\ee
\Eq{eq:Str} decouples, leading to the relation
\be
m^2_{\tilde{u}_R}+m^2_{\tilde{u}_L}+m^2_{\tilde{d}_R} + m^2_{\tilde{d}_L}=2(m^2_u + m^2_d).
\ee
If $\SU(3)_C$ is to remain unbroken, this would imply light (MeV) scale superpartners, in conflict with observation.  Similar arguments exist in the presence of large flavor mixings \cite{Dimopoulos:1981zb}, even apart from the dangerous flavor-changing neutral currents they would introduce.

\begin{figure}[t]
\begin{center}
\psfig{file=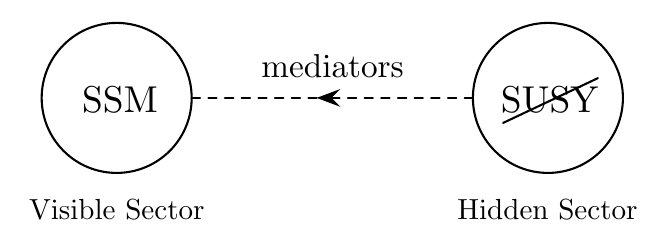,width=3in}
\end{center}
\caption{Standard paradigm of a SUSY-breaking hidden sector coupled to the SUSY SM via mediators.}
\label{fig:SUSYbreaking}
\end{figure}

For these reasons, the standard SUSY-breaking paradigm is for SUSY to be broken in a ``hidden sector'', and the effects of SUSY breaking communicated to the SUSY SM (the ``visible sector'') via loop processes or higher-dimension operators.  We draw this schematically as in \Fig{fig:SUSYbreaking}.
The effect of SUSY breaking on the visible sector is obviously important.  But the phenomenological implications of the SUSY-breaking sector itself are typically meager (apart from the goldstino to be discussed in \Sec{eq:goldstino}).  For this reason, we will not focus much on specific models of SUSY breaking.  In fact, the super-trick in \Sec{subsec:nonlineargoldstino} is aimed at abstracting the most important features of the hidden sector.

\subsection{The Polonyi Model}
\label{subsec:polonyi}

The absolute simplest model of SUSY breaking is the Polonyi model of one chiral multiplet $\bX = \{\phi, \psi, F\}$:
\begin{align}
\mathcal{L} & = \int d^4 \theta \, \bX^\dagger \bX + \int d^2 \theta \, f \bX + \rm h.c. \\
& = \phi^* \Box \phi + i \overline{\psi} \sigmabar^\mu \partial_\mu \psi + F^\dagger F + f F + f^* F^\dagger.
\end{align}
After using the equation of motion $F = -f ^*$, the scalar potential is simply
\be
\label{eq:polonyivacuum}
V = |f|^2,
\ee  
so for any non-zero choice of $f$, SUSY is broken.  

This should seem surprising, since apart from the vacuum energy, $\bX$ looks just like a free massless SUSY multiplet.  In particular, the mass of $\phi$ and $\psi$ are the same.  However, this model does indeed break SUSY.  To convince you, consider adding a higher-dimensional term to the K\"ahler potential
\begin{align}
\bK & = \bX^\dagger \bX - \frac{(\bX^\dagger \bX)^2}{4 \Lambda^2}, \\
\bW & = f \bX.
\end{align}
The resulting scalar potential is
\be
V = F^* F \left(1 - \frac{\phi^\dagger \phi}{\Lambda^2} \right) + f F + f^* F^* = \frac{|f|^2}{1 - \frac{\phi^* \phi}{\Lambda^2}}.
\ee
This potential is minimized at $\vev{\phi} = \vev{\phi^\dagger} = 0$, yielding the vacuum energy $\vev{V} = |f|^2$ as in \Eq{eq:polonyivacuum}.  Expanding about the potential minimum, the scalar and fermion fields have masses
\be
m^2_\phi  = \frac{|f|^2}{\Lambda^2}, \qquad m_\psi = 0.
\ee
For any finite value of $\Lambda$, we now see the mass splitting between bosons and fermions expected from SUSY breaking.  Interestingly, the fermion is still massless.  In fact, as we will see in \Sec{eq:goldstino}, any SUSY-breaking theory has a goldstino, a massless Goldstone fermion arising from spontaneous SUSY breaking.  

\subsection{Obstructions to Generic $F$-term Breaking}
\label{subsec:Fobstructions}

The Polonyi model will be our template for SUSY-breaking models, since it contains most of the relevant physics.  While we will not focus on explicit models of SUSY breaking, we do wish to convey that achieving generic $F$-term SUSY breaking is non-trivial.  

Consider coupling a Polonyi field $\bX$ to quarks $\bQ$, $\bQ^c$:
\be
\bW = f \bX + \lambda \bX \bQ \bQ^c.
\ee
You might naively think that $\vev{F_X} = - f^*$ and one would generate SUSY mass splittings for $\bQ,\bQ^c$.  But instead, the vacuum shifts and SUSY stays unbroken!  In particular, the set of equations
\be
\begin{split}
\bW_X &=f + \lambda \bQ \bQ^c=0,\\
\bW_Q&=\lambda \bX \bQ^c=0,\\
\bW_{Q^c}&=\lambda \bX \bQ=0,
\end{split}
\ee
admits solutions with $\vev{X}=0$ and $\vev{Q Q^c}=-f/\lambda$. At those points, all $F$-terms vanish so the scalar potential $V=0$ and SUSY is unbroken.  With respect to the standard SUSY-breaking paradigm in \Sec{subsec:paradigm}, this tendency for SUSY to get restored in the presence of interactions makes it challenging to construct explicit models of SUSY breaking and mediation.

In general, if you have $N$ chiral multiplets $\bX_i$ and $N$ equations of the type $\partial \bW / \partial \bX_i = 0$ to solve, then there is usually a solution and SUSY is unbroken in the true vacuum.   To avoid this outcome, one can consider broken SUSY in a metastable vacuum \cite{Intriligator:2007py}.  Alternatively, one has to arrange the dynamics in such a way that the above equations cannot be simultaneously satisfied.  In \Ref{Nelson:1993nf}, the presence of an $R$-symmetry is shown to be a necessary condition for spontaneous SUSY breaking, while the spontaneous breaking of such an $R$-symmetry is sufficient to ensure spontaneous SUSY breaking. 
 
\subsection{$D$-term Breaking}
\label{subsec:dterm}

In a generic SUSY-breaking theory, non-zero $F$-terms will be accompanied by nonzero $D$-terms.  For non-Abelian gauge theories, we can make the stronger statement that non-zero $F$-terms are required in order for the $D$-terms to be non-zero, such that there is no pure $D$-term SUSY breaking for non-Abelian gauge groups \cite{Wess:1992cp}.

For an Abelian gauge group, one can get $D$-term breaking without $F$-term breaking by introducing the Fayet-Iliopoulos term \cite{Fayet:1974jb,Fayet:1974pd} from \Eq{eq:FI-term} 
\begin{align}
\mathcal{L} & \supset \int d^2 \theta \, \bW^\alpha \bWalpha - \int d^4 \theta \, \kappa \bV \supset \frac{1}{2} D^2 - \kappa D .
\end{align}
After solving the $D$ equation of motion, we find
\be
D = \kappa, \qquad V = \frac{1}{2}\kappa^2.
\ee
The vacuum energy is non-vanishing, so SUSY is broken.  We can see this in the spectrum by coupling this to a pair of oppositely-charged chiral superfields with a superpotential mass term
\begin{align}
\bW = m \bQ \bQ^c.
\end{align}
There is no way to make all three auxiliary fields ($D$, $F_Q$, and $F_{Q^c}$) vanish simultaneously, so SUSY is spontaneously broken.  One can confirm that for $m > \sqrt{2 g \kappa}$, the minimum of the potential is still at the origin and the gauge symmetry remains unbroken.  The gaugino and gauge boson remain massless and there is one Dirac fermion (or two Weyl fermions) of mass $m$, but the scalars now have masses of\footnote{Since this is a renormalizable theory, the supertrace sum rule of \Eq{eq:Str} indeed holds.}
\begin{align}
m^2_{Q,Q^c} & = m^2 \pm 2 g \kappa.
\end{align}

However, Fayet-Iliopoulos terms are hard to incorporate into realistic SUGRA theories.  In SUGRA, Fayet-Iliopoulos terms correspond to having a gauged $\U(1)_R$ symmetry \cite{Barbieri:1982ac,Ferrara:1983dh}.  As we will see in \Sec{subsec:AdSSUSY}, demanding a $\U(1)_R$ symmetry makes it difficult to achieve the negligible cosmological constant observed in nature.  

\subsection{The Goldstino}
\label{eq:goldstino}

For any model of SUSY breaking, there is one irreducible prediction:  the goldstino.  Just as a spontaneously broken global symmetry always gives rise to a Goldstone boson, spontaneous SUSY breaking always gives rise  to a Goldstone fermion.  There are a couple of different ways to see the emergence of the goldstino.

In the operator language of \Sec{subsec:operatorlanguage}, consider a vacuum that breaks SUSY, such that after performing a SUSY transformation, the vacuum state is changed
\be
e^{-i\epsilon \Q-i\bar{\epsilon}\Qbar} \left|0\right> = \left|\epsilon\right>.
\ee
However, because the Hamiltonian ($P_0$) commutes with SUSY ($\Q$, $\Qbar$), the (fermionic) state $\left|\epsilon\right>$ has exactly the same energy as the vacuum state.  If we now imagine performing a space-time-dependent SUSY transformation
\be
e^{-i \eta(x) \Q-i\bar{\eta}(x) \Qbar} \left|0\right> = \left|\eta(x)\right>,
\ee
then the state $\left|\eta(x)\right>$ will have a higher energy than the vacuum.  But by making $\eta(x)$ closer and closer to a constant $\epsilon$ (i.e.~by making $\eta(x)$ a longer and longer wavelength perturbation), we get closer and closer to the vacuum energy.  This implies that there is (at least) one gapless fermionic excitation in the theory, namely the goldstino.

A more mechanical way to see the need for a goldstino is to consider the scalar potential (with a trivial K\"ahler metric and no gauge interactions, for simplicity):
\begin{align}
\label{eq:simplepotential}
V & = W^{\dagger i} W_i.
\end{align}
If we have SUSY breaking in a stable vacuum, then $\vev{V} \neq 0$ but $\frac{\partial V}{\partial \phi^i} = 0$ for all $\phi^i$:
\begin{align}
\frac{\partial V}{\partial \phi^j} & = W^{\dagger i} W_{i j} = 0.
\end{align}
This can be satisfied in two ways.  Either $\vev{W^{\dagger i}} = 0$ for all $i$ (but then there is no SUSY breaking), or $W_{i j}$ has a zero eigenvalue in the $W^{\dagger i}$ direction.  But from \Eq{eq:finalWZ}, we see that $W_{i j}$ is the fermion mass matrix!  So there must be a massless mode
\be
\eta = \frac{1}{F_{\text{tot}}} W^{\dagger i} \chi_i,
\ee
namely the goldstino. Here, $F_{\text{tot}}\equiv\vev{V}^{1/2}$. This logic is easily extended beyond \Eq{eq:simplepotential} to arbitrary (perturbative) SUSY-breaking theories, and the expression with a non-trivial K\"ahler metric and gauge kinetic terms is
\be
\label{eq:goldstinodirection}
\eta = \frac{1}{F_{\text{tot}}}\left(F^i\chi_i+\frac{1}{\sqrt{2}}D_a\lambda^a\right),
\ee
where $F^i=-g^{i\j}W^\dagger_\j$, $D_a=-(K_iT_a\phi^i)$, and $\lambda^a$ are the gauginos.  Now $F_{\text{tot}}$ includes the contribution from both $F$-terms and $D$-terms.  We will encounter an even slicker way to see the goldstino in the next subsection.  

Thus, the presence of a massless goldstino is guaranteed for global SUSY breaking in flat space.  In \Sec{subsec:AdSSUSY}, we will discuss that in SUGRA, the goldstino (from broken AdS$_4$ SUSY) is eaten to form the longitudinal component of the gravitino (the superpartner of the graviton).

\subsection{Supertrick \#4: Non-linear Goldstino Multiplet}
\label{subsec:nonlineargoldstino}

Our last super-trick is aimed at understanding the properties and interactions of the goldstino. As inspiration for this super-trick, consider the analogy of a Higgs doublet breaking a global $\SU(2)$ symmetry.  The Higgs boson gets a vev
\begin{align}
\vev{h} & = \left(\begin{array}{c} 0 \\ v \end{array} \right).
\end{align}
Because of this spontaneous symmetry breaking, there must be Goldstone bosons, and we can identify them by performing the broken $\SU(2)$ symmetry on the vacuum:
\begin{align}
h_{\rm NL} & = U \left(\begin{array}{c} 0 \\ v \end{array} \right), \qquad  U \equiv e^{i \pi^a T^a/v}.
\end{align}
$h_{\rm NL}$ furnishes a non-linear representation of the $\SU(2)$ global symmetry, and we can recover the effective chiral Lagrangian by writing down the most generic interactions of $h_{\rm NL}$ (or $U$).  The advantage of using this non-linear realization is that we can ignore the massive physical Higgs modes.  Indeed, we can use $h_{\rm NL}$ even if $\SU(2)$-breaking is triggered not by a perturbative Higgs doublet but by nonperturbative strong dynamics.

The same can be done for spontaneous SUSY breaking \cite{Rocek:1978nb,Lindstrom:1979kq,Komargodski:2009rz}.
Consider one multiplet that gets a non-zero $F$ component to break SUSY,
\begin{align}
\vev{\bX} & = \theta^2 F.
\end{align}
In general, $\bX$ will have a scalar component, but if SUSY is broken, we expect this state to be heavy whereas the goldstino is massless.  To identify the goldstino direction, we can perform the broken SUSY transformation on the vacuum via
\begin{align}
\theta^\alpha & \rightarrow \theta^\alpha + \eta^\alpha, \\
y^\mu & \rightarrow y^\mu + 2 i \overline{\eta} \sigmabar^\mu \theta, 
\end{align}
where $\eta = \eta(x)$ is the (unnormalized) goldstino. If we treat $F$ as a non-dynamical constant in space-time\cite{Cheung:2010mc}, the second line is irrelevant.  Normalizing the goldstino via $\eta \to \eta/\sqrt{2} F$, we have the following non-linear representation of SUSY:
\be
\label{eq:Xnl}
\begin{split}
\bX_{\rm NL}  & = \left(\theta + \frac{\eta}{\sqrt{2} F} \right)^2 F\\
&=\frac{\eta^2}{2F}+\sqrt{2}\theta\eta+\theta^2F.
\end{split}
\ee
In \Sec{subsec:Gcouplings}, we will couple $\bX_{\rm NL}$ to visible sector fields, treating $F$ as a constant background field to derive soft mass terms and goldstino couplings.  Even if there are multiple non-zero $F$-terms and $D$-terms in the hidden sector, the goldstino mode can be described by a single $\bX_{\rm NL}$, since any non-zero $F$-term (or $D$-term) can be captured by multiplying $\bX_{\rm NL}$ (or $\bX_{\rm NL}^\dagger \bX_{\rm NL}$) by an overall constant.

An alternative interpretation of \Eq{eq:Xnl} is as a constrained superfield satisfying $\bX_{\rm NL}^2 = 0$ (the analog of $U^\dagger U = 1$ in the chiral Lagrangian) \cite{Komargodski:2009rz}.  Here, $F$ is not a constant background and must be solved for dynamically.  Neglecting terms with additional derivatives, the unique lagrangian we can write for $\bX_{\rm NL}$ is
\begin{align}
\mathcal{L}_{\rm NL} & = \int d^4 \theta \,  \bX^\dagger_{\rm NL} \bX_{\rm NL} + \int d^2 \theta \, f \bX_{\rm NL} + \hc \label{eq:goldstinolagrangiansuperspace} \\
& = \frac{\overline{\eta}^2}{2 F^\dagger} \Box \frac{\eta^2}{2 F} + i \overline{\eta} \sigmabar^\mu \partial_\mu \eta + F^\dagger F + f F + F^\dagger f^*.
\end{align}
Solving for the $F$ equation of motion is more complicated because of the first term, but doing so would recover terms with eight goldstinos and four derivatives that are present in the Akulov-Volkov lagrangian \cite{Volkov:1973ix}. It is consistent to neglect those terms however, since terms that we have already neglected involving $\D_\alpha \bX_{\rm NL}$ and  $\partial_\mu \bX_{\rm NL}$ would play a role at that order.  Using the equation of motion $F =  - f^*$, the goldstino lagrangian is
\be
\label{eq:goldstinolagrangian}
\mathcal{L}_{\rm NL} = \frac{\overline{\eta}^2}{2 f^\dagger} \Box \frac{\eta^2}{2 f} + i \overline{\eta} \sigmabar^\mu \partial_\mu \eta - |f|^2 + \ldots
\ee
This represents a massless goldstino with vacuum energy $\vev{V} = |f|^2 $ and goldstino decay constant $f$.  One can think of this as the low energy effective lagrangian one obtains after integrating out all of the (heavy) dynamics of the SUSY-breaking sector.

Using the non-linear goldstino multiplet to describe broken SUSY at low energies has a number of advantages.  The main advantage is that one does not have to worry about the (usually irrelevant) details of how one actually breaks SUSY in the hidden sector.  Whether SUSY is broken by one field or many, there is only one true goldstino, and its physical modes are completely contained in $\bX_{\rm NL}$.  We can also ignore the details of how the sgoldstino (scalar partner of the goldstino) is stabilized.  

There is an important distinction between treating $\bX_{\rm NL}$ as a constrained superfield $\bX_{\rm NL}^2 = 0$ versus as an expansion around a constant $\theta^2 F$.  With constrained superfields, we are guaranteed that the vacuum energy will be correlated with the goldstino decay constant (as necessitated by the SUSY algebra), whereas with fixed $F$, this is not the case.  On the other hand, if we treat $F$ as fixed, then we can write down direct couplings between the hidden and visible sectors without worrying about the vacuum changing.  In particular, we saw in \Sec{subsec:Fobstructions} that $F$-terms can shift and restore SUSY, leading to pathologies in the constrained superfield formalism.  For this reason, we prefer to treat $F$ as non-dynamical in $\bX_{\rm NL}$.

\subsection{Goldstino Couplings to Matter}
\label{subsec:Gcouplings}

We now use the super-trick of non-linear goldstinos to understand the leading interactions between the hidden sector and the visible sector in the standard SUSY-breaking paradigm.  As we mentioned, there is very little hope to directly see states in the hidden sector, apart from their impact on soft terms for the visible sector.  The one exception is the goldstino (eaten by the gravitino), which is generically light.  Using our super-trick, we will show that the leading couplings of the goldstino to visible fields are universal and determined by SUSY-breaking soft terms in the visible sector \cite{Clark:1996aw,Fayet:1977vd,Fayet:1979yb}.

Let $\vev{\bX_{\rm NL}} = \theta^2 F_{\rm hid}$ represent SUSY breaking in the hidden sector.  Because $\bX_{\rm NL}^2 = 0$ (and ignoring $\D_\alpha \bX_{\rm NL}$, $\partial_\mu \bX_{\rm NL}$ terms), the possible interactions between the hidden sector and visible sector are quite restricted.  Note that $[\bX_{\rm NL}] =1$ as for an ordinary chiral multiplet.  Consider first the coupling of $\bX_{\rm NL}$ to the gauge kinetic term (as anticipated in \Sec{subsec:BackgroundField})
\be
\label{eq:XNLtogauginos}
\mathcal{L} \supset -\int d^2 \theta \, \frac{\bX_{\rm NL}}{2 \Lambda} W^\alpha W_\alpha + \textrm{h.c.}
\ee
Expanding in components yields various types of terms.  The most important terms are 
\begin{align}
\mathcal{L} & \supset -\frac{1}{2} m_{\lambda} \lambda \lambda + \frac{im_\lambda}{\sqrt{2} F_{\rm hid}} \lambda \sigma^{\mu \nu} \eta F_{\mu \nu}+  \hc
\end{align}
where $\sigma^{\mu\nu}$ is defined in \Eq{eq:sigmamunu}. The first term is a mass term for the gaugino
\be
m_\lambda \equiv \frac{F_{\rm hid}}{\Lambda},
\ee
which clearly breaks SUSY since it splits the gaugino mass from the gauge boson mass.  The second term is a goldstino-gaugino-gauge boson coupling proportional to the gaugino mass $m_\lambda$.  In fact, it is generically true that the goldstino couples proportional to soft masses (with deviations controlled by $\D_\alpha \bX_{\rm NL}$, $\partial_\mu \bX_{\rm NL}$ terms).  

There are additional terms in \Eq{eq:XNLtogauginos} which are important for consistency of the SUSY-breaking theory.  If $D$ in the visible sector gets a non-zero vacuum expectation value, then there are two additional mass terms
\be
\label{eq:goldstinomixing}
\mathcal{L}\supset \frac{D}{\sqrt{2}\Lambda}\eta\lambda+\frac{D^2}{8\Lambda F_{\rm hid}}\eta^2+\hc
\ee
The first term is a mixing term between the goldstino and the gaugino.  This arises because the true goldstino points in the direction of \Eq{eq:goldstinodirection}, which includes contributions from both the visible sector and hidden sector.  In general, such a mixing term will lift the mass of the goldstino, but because of the second term, the fermion mass matrix indeed has a zero eigenvalue.  This highlights the usefulness of the non-linear goldstino multiplet, since without the $\eta^2$ in $\bX_{\rm NL}$, the massless goldstino would only show up after adjusting the vacuum structure of the theory.

We can generalize \Eq{eq:XNLtogauginos} to account for all possible interactions between visible sector fields and the hidden sector surrogate $\bX_{\rm NL}$.  Restricting to terms that are visible sector renormalizable, the possible terms are\footnote{Terms like $(\bX_{\rm NL} + \bX_{\rm NL}^\dagger)\bPhi^\dagger \bPhi$ can be removed by a field redefinition $\bPhi \to \bPhi - \bPhi \bX_{\rm NL}$.  When doing that field redefinition, make sure to remember the Konishi anomaly from \Sec{subsec:Konishi}!  This field redefinition yields an asymmetry in \Eq{eq:GoldstinoSoft} such that there are explicit goldstino couplings to the $D$-term but not to the $F$-term.}
\begin{align}
\mathcal{L}&\supset -\int d^4\theta \, \frac{\widetilde{m}_i^2}{F^2} \bX_{\rm NL}^\dagger\bX_{\rm NL}\bPhi_i^\dagger e^{2 \bV}\bPhi_i-\bigg(\int d^2\theta \, \frac{m_{\lambda}}{2F_{\rm hid}}\bX_{\rm NL}\bW^{\alpha a}\bW_\alpha^a \nonumber \\
&\quad + \frac{{C}_{i}}{F_{\rm hid}}\bX_{\rm NL}\bPhi_i + \frac{{B}_{ij}}{2F_{\rm hid}}\bX_{\rm NL}\bPhi_i\bPhi_j + \frac{{A}_{ijk}}{6F_{\rm hid}}\bX_{\rm NL}\bPhi_i\bPhi_j\bPhi_k+\hc\bigg). \label{eq:goldstinocouplingsuperspace}
\end{align}
This yields the following soft SUSY-breaking terms in the visible sector lagrangian 
\be
\mathcal{L}_{\rm soft} = -\widetilde{m}_i^2|\phi_i|^2-\left(\frac{m_\lambda}{2}\lambda^a\lambda^a + C_i \phi_i + \frac{{B}_{ij}}{2}\phi_i\phi_j+\frac{{A}_{ijk}}{6}\phi_i\phi_j\phi_k+\hc\right),
\ee
Each of these terms can be interpreted as follows:
\begin{itemize}
\item $\widetilde{m}^2_i$ gives a mass to the boson (but not the fermion) in a chiral multiplet.
\item $m_\lambda$ is a mass term for gauginos.
\item $C_i$ is a source term for scalars.
\item $B_{i j}$ gives a mass splitting between the scalar and the pseudoscalar in a chiral multiplet.
\item $A_{i j k}$ is a (holomorphic) three-point interaction for scalars.
\end{itemize}

As these soft terms arose by coupling the goldstino multiplet $\bX_{\rm NL}$ to visible sector fields, each of these soft terms also has an associated coupling of a single goldstino
\be
\label{eq:GoldstinoSoft}
\begin{split}
\mathcal{L}_{\rm \eta} &= \frac{1}{F_{\rm hid}}\eta\bigg(\widetilde{m}_i^2\psi_i\phi_i^\dagger + C_i \psi_i +{B}_{ij}\psi_i\phi_j+\frac{{A}_{ijk}}{2}\psi_i\phi_j\phi_k\\
&\quad+\frac{im_\lambda}{\sqrt{2}}\sigma^{\mu\nu}\lambda^a F_{\mu\nu}^a+\frac{m_\lambda}{\sqrt{2}}\lambda^a D^a\bigg)+\hc
\end{split}
\ee
As anticipated, these goldstino couplings are proportional to soft masses, suppressed by an overall factor of $1/F_{\rm hid}$.  Along with the SUSY interactions, the soft terms and goldstino couplings lead to the dominant phenomenology of the SUSY SM.  In particular, \Eq{eq:GoldstinoSoft} allows sparticles to decay to the corresponding particle and a goldstino (eaten by the gravitino).

If one chooses to impose a global $R$-symmetry on the lagrangian (as in \Sec{subsec:globalsymmetries}), using $\bX_{\rm NL}$ makes it simple to determine whether the various soft terms contained in \Eq{eq:goldstinocouplingsuperspace} respect that $R$-symmetry.  Clearly, the soft scalar mass terms $\widetilde{m}^2_i$ respect any $R$-symmetry, but the other terms are less certain. As $\bX_{\rm NL}$ appears as a linear term in the superpotential (see \Eq{eq:goldstinolagrangiansuperspace}), it must have $R$-charge 2.  This implies that the gaugino mass terms $m_\lambda$ violate any $R$-symmetry; recall that $\bV$ has $R$-charge $0$, so $\bWalpha$ has $R$-charge $+1$.  The $A$, $B$, and $C$ terms respect the $R$-symmetry only if the chiral superfields they multiply have net vanishing $R$-charge.  In many models, $\bX_{\rm NL}$ multiplies the same terms that arise in an $R$-respecting superpotential ($R$-charge 2), in which case the resulting soft terms do not respect the $R$-symmetry.

There are more exotic ways to couple $\bX_{\rm NL}$ to matter fields using SUSY-covariant derivatives.  Consider a term in the lagrangian of a non-Abelian gauge field
\be
\int d^2 \theta \, \text{Tr}(\bPhi \bW^\alpha) \bWalpha\!\!',
\ee
where $\bPhi$ is a chiral multiplet in the adjoint representation and
\be
\bWalpha\!\!' = -\frac{1}{4} \Dbar^2 D_\alpha (\bX^\dagger_{\rm NL} \bX_{\rm NL}).
\ee
Because $\vev{\bWalpha\!\!'} = \theta_\alpha |F|^2$, this term induces a (SUSY-breaking) Dirac mass term between the Weyl fermion in $\bPhi$ and the gaugino in $\bW^\alpha$.  Such terms appear in theories of supersoft SUSY breaking \cite{Dine:1992yw,Fox:2002bu}.  Note that these Dirac mass terms for gauginos (unlike the Majorana mass terms discussed above) can respect a global $R$-symmetry if $\bPhi$ has vanishing $R$-charge.

\subsection{The Supercurrent}
\label{eq:supercurrent}

Using the non-linear goldstino multiplet, we found that goldstino couplings were directly related to soft terms.  There is a more formal way of seeing this same effect using conservation of the supercurrent.

The supercurrent is the Noether current associated with SUSY transformations \cite{Wess:1973kz,deWit:1975nq}
\be
\label{eq:SuperCurrent}
\begin{split}
j^\mu_\alpha & =(\sigma^\nu \sigmabar^\mu \psi^i)_\alpha \nabla_\nu \phi^{\dagger}_i + i (\sigma^\mu \overline{\psi}_i)_\alpha W^{\dagger i} \\
& \quad \, +\frac{1}{2 \sqrt{2}} (\sigma^\nu \sigmabar^\rho \sigma^\mu \overline{\lambda}^a) F^a_{\nu \rho} + \frac{i}{\sqrt{2}} g \phi^\dagger T^a \phi (\sigma^\mu \overline{\lambda}^a)_\alpha.
\end{split}
\ee
Note that the supercurrent has an extra $\alpha$-index to match the SUSY generator $\Q_\alpha$.  Conservation of the supercurrent implies $\partial_\mu j^\mu_\alpha = 0$.

We can isolate the goldstino contribution to the supercurrent via \cite{Fayet:1977vd,Fayet:1979yb}
\be
j^{\mu}_\alpha=j^{\mu,\text{matter}}_\alpha-iF_{\rm tot} \left(\sigma^\mu\bar\eta\right)_\alpha,
\ee
where $F_{\rm tot} = \sqrt{|F_{\rm vis}|^2 + |F_{\rm hid}|^2}$ also includes any SUSY breaking in the visible sector.  Conservation of the full supercurrent implies
\be
\label{eq:TotalSupercurrent}
\partial_\mu j^{\mu}_\alpha=0=\partial_\mu j^{\mu,\text{matter}}_\alpha-iF_{\rm tot} \left(\sigma^\mu\partial_\mu\bar\eta\right)_\alpha.
\ee
As expected, because SUSY in the visible sector is broken, $\partial_\mu j^{\mu,\text{matter}}_\alpha\neq 0$. If we interpret \Eq{eq:TotalSupercurrent} as an equation of motion for the goldstino, this implies that in addition to \Eq{eq:goldstinolagrangian}, the goldstino lagrangian must contain
\begin{align}
\label{eq:Gsupercurrent}
\mathcal{L}_\eta & \supset -\frac{1}{F_{\rm tot}}\eta\partial_\mu j^{\mu,\text{matter}}+\hc
\end{align}
This is called a Goldberger-Treiman relation from the analogous relation for couplings of Goldstone bosons of spontaneously broken global symmetries to matter currents \cite{Goldberger:1958zz}.

For a massless on-shell goldstino
\be
(\partial_\mu\eta)\sigma^\mu=0,
\ee
so after integration by parts, the second and fourth terms in \Eq{eq:SuperCurrent} are irrelevant for \Eq{eq:Gsupercurrent}, which reduces to
\be
\begin{split}
\mathcal{L}_\eta&\supset \frac{1}{F_{\rm tot}}\eta\bigg((\Box\psi^i)\phi_i^\dagger-\psi^i\Box\phi_i^\dagger-\frac{1}{\sqrt{2}}\sigma^{\nu\rho}(\sigma^\mu\partial_\mu\overline{\lambda}^a)F^a_{\nu\rho}\\
&\quad +\frac{1}{\sqrt{2}}\sigma^\rho\overline{\lambda}^a\partial^\nu F_{\nu\rho}^a\bigg)+\hc
\end{split}
\ee
Using equations of motion for the visible sector fields, we find that the three-point couplings of the goldstino are proportional to \emph{physical} mass differences:
\be
\begin{split}
\mathcal{L} & \supset \frac{m_{\phi^i}^2 - m_{\psi^i}^2}{F_{\rm tot}} \eta \psi^i \phi^\dagger_i + \frac{B_{i j}}{F_{\rm tot}} \eta \chi^i \psi^j  \frac{i m_\lambda}{\sqrt{2} F_{\rm tot}} \eta \sigma^{\mu \nu} \lambda^a F^a_{\mu \nu},
\end{split}
\ee
where, for simplicity, we have assumed unbroken gauge groups.  We see that this result exactly reproduces \Eq{eq:GoldstinoSoft} in a non-trivial way.  One advantage of the supercurrent method is that it automatically accounts for effects of goldstino mixing terms like \Eq{eq:goldstinomixing}.

\subsection{The Gravitino and AdS SUSY}
\label{subsec:AdSSUSY}

For the last topic of these lectures, we would like to go beyond global SUSY to talk a bit about supergravity (SUGRA).  We have seen that in global SUSY, SUSY breaking leads to a goldstino.  In supergravity, the goldstino is eaten by the gravitino to become its longitudinal components, with a mass given by
\be
\label{eq:gravitinomassaftertuning}
m_{3/2} = \frac{F_{\rm tot}}{\sqrt{3} M_{\rm Pl}},
\ee 
where $M_{\rm Pl}$ is the reduced Planck constant.  This is sufficiently confusing that we want to explain (in words) what really happens, though a proof would require a real lecture on SUGRA.

In global SUSY, SUSY breaking implies $\vev{V} > 0$.  Because our universe has $\vev{V} \approx 0$ though, we cannot understand SUSY breaking using the flat space SUSY algebra alone.  Even before talking about any details of SUGRA, we can see a possible way out if SUSY could be realized in AdS$_4$ space!  AdS space is a solution to Einstein's equations with negative vacuum energy
\be
\Lambda_{\rm AdS} = - 3 \frac{M_{\rm Pl}^2}{\lambda^2_{\rm AdS}},
\ee
where $\lambda_{\rm AdS}$ is the AdS curvature.  Whenever SUSY is broken, the vacuum energy increases, so if it were possible to finely balance the AdS curvature against SUSY breaking, we could have zero vacuum energy with (AdS) SUSY breaking. 

Indeed, there does exist a global (a.k.a.~rigid) AdS$_4$ SUSY algebra where $\lambda_{\rm AdS}$ is fixed, but $M_{\rm Pl} \to \infty$.  At finite $M_{\rm Pl}$ (namely SUGRA), space-time is dynamical so the cosmological constant depends on the vacuum structure of the theory.  This happens via the scalar potential (with a trivial K\"ahler metric and no gauge interactions, for simplicity)
\begin{align}
V & \simeq |F|^2 - 3 \frac{|W|^2}{M_{\rm Pl}^2}, \label{eq:SUGRApotential}
\end{align}
where $W$ is the superpotential itself.  When SUGRA is unbroken, $\vev{F} = 0$ but the superpotential $W$ contributes to the cosmological constant, yielding the AdS curvature $\lambda_{\rm AdS} = M_{\rm Pl}^2/|W|$.  Even with no SUSY breaking, the gravitino (graviton partner) has a mass parameter\footnote{Despite having a mass parameter, it only has two physical polarizations, a consequence of the AdS$_4$ little group.}
\be
\label{eq:gravitinomass}
m_{3/2}  = \frac{|W|}{M_{\rm Pl}^2}  = \frac{1}{\lambda_{\rm AdS}}.
\ee
So $m_{3/2}$ is \emph{not} an order parameter for SUSY breaking, rather it measures the curvature of SUSY AdS space.\footnote{Note that $m_{3/2}$ is an order parameter for $R$-symmetry breaking, since the gravitino has non-vanishing $R$-charge.}

Thus, we can achieve SUSY breaking with zero cosmological constant, if we delicately balance $\vev{F} \neq 0$ against the AdS curvature, which from \Eq{eq:SUGRApotential} implies
\be
F_{\rm tot} = \sqrt{3} \frac{W}{M_{\rm Pl}} \quad (\text{zero cosmological constant}).
\ee
Because SUSY is broken, there is a goldstino, but in SUGRA it is eaten to form the longitudinal components of the gravitino (with the same mass $m_{3/2}$ as \Eq{eq:gravitinomass}).  At zero cosmological constant, we recover the claimed formula in \Eq{eq:gravitinomassaftertuning}.  Crucially, the order parameter for SUSY breaking is $F$, and $m_{3/2}$ is related to $F$ only by a fine-tuning.

\begin{figure}[t]
\centering
\psfig{file=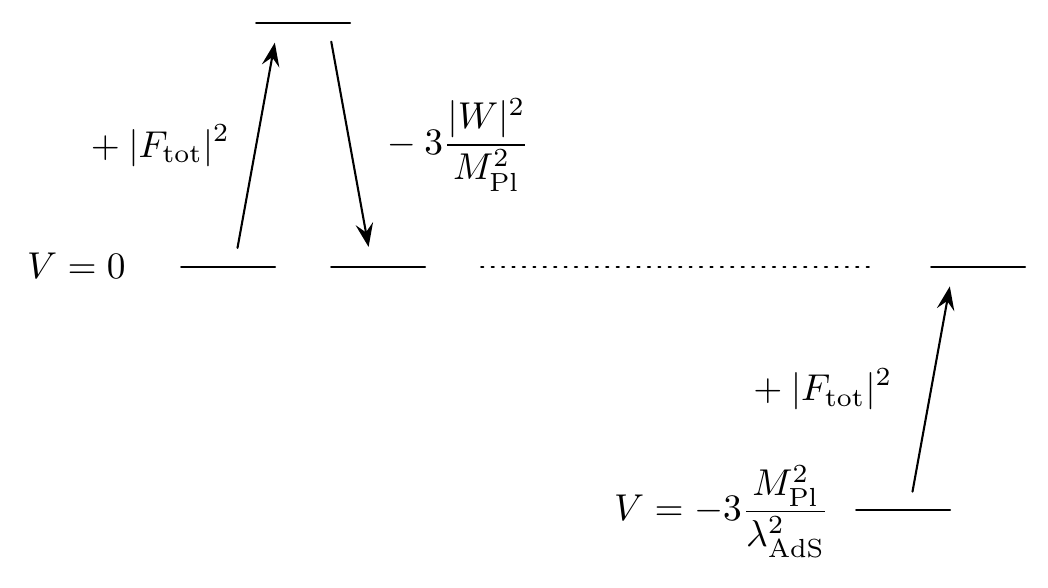,width=4.5in}
\caption{Two ways to think about achieving SUSY breaking with $V\simeq0$.  The second picture makes clear the underlying AdS$_4$ algebra.}
\label{fig:AdsLifting}
\end{figure}

We can therefore picture SUSY breaking in two different ways as shown in \Fig{fig:AdsLifting}.  We can either think about flat space SUSY being spontaneously broken to yield $V = |F_{\rm tot}|^2$, then finely adjusting $\vev{W}$ to return to $V\simeq0$.  Alternatively, we can think about starting with an AdS$_4$ SUSY algebra that would yield $V = -3 M_{\rm Pl}^2/\lambda_{\rm AdS}^2$, but the vacuum spontaneously breaks SUSY to yield $V\simeq0$.

We prefer to think in terms of uplifting AdS$_4$ space, because it clarifies certain SUGRA effects that would seem mysterious in flat space.  Unlike in flat space, $\{Q_\alpha, Q^\beta \} \not=0$ in AdS space.  In a pseudo-flat space language, the AdS$_4$ SUSY algebra is given by \cite{Adams:2011vw,deWit:1999ui,Keck:1974se,Zumino:1977av,Ivanov:1980vb}
\begin{align}
\{Q_\alpha, \Qbar_{\betadot} \} & = -2 \sigma^\mu_{\alpha \betadot} P_\mu, \\
\{Q_\alpha, Q^\beta \} & = - 2 i \lambda_{\rm AdS}^{-1} (\sigma^{\mu \nu})_\alpha{}^\beta M_{\mu \nu},
\end{align}
where $M_{\mu \nu}$ still satisfy the algebra of the (flat space) Lorentz generators in \Eqs{eq:MMcoMM}{eq:PcoMM}.\footnote{One should regard these indices as local Lorentz indices, and the vielbein must be used to convert to the more typical Einstein indices of AdS space.}  Because we are actually in AdS space, the ``translation'' generators $P_\mu$ are modified 
\begin{align}
[P_\mu, P_\nu] & = - i \lambda_{\rm AdS}^{-2} M_{\mu \nu},\\
[Q_\alpha, P_\mu] & = \frac{1}{2} \lambda_{\rm AdS}^{-1} \sigma^\mu_{\alpha \alphadot}\Qbar^\alphadot. 
\end{align}
All other commutators are the same as in flat space.  Because AdS SUSY has a different algebra than flat space SUSY, there are correspondingly different properties, a few of which we will mention here:
\begin{itemize}
\item In global AdS SUSY, bosons and fermions in the same multiplet can be split by an amount proportional to $\lambda_{\rm AdS}^{-1}$!  This happens already at tree level, giving rise to a $\mu/B_\mu$ problem in theories where $m_{3/2}$ is large \cite{Randall:1998uk}.
\item Global AdS SUSY has a boundary, and in order to maintain SUSY in AdS, certain loop effects on the boundary must be compensated by counterterms in the bulk \cite{Gripaios:2008rg}. This has an impact in SUGRA, because once SUSY is broken and AdS is uplifted to flat space, the bulk counterterms remain, giving soft masses without SUSY breaking.  This contributes to an effect known as anomaly mediation \cite{Randall:1998uk,Giudice:1998xp,D'Eramo:2012qd}.
\item When global AdS SUSY is broken, the goldstino is not massless as it is in flat space.  Rather, conservation of the AdS supercurrent implies that the goldstino has a mass of $2 \lambda_{\rm AdS}^{-1}$ (note the factor of 2!) \cite{Cheung:2010mc,McArthur:2013wv,Zumino:1977av}.  If SUSY is broken by $N$ independent sectors, then there is a corresponding multiplicity of $N$ ``goldstini'' \cite{Cheung:2010mc}.  One linear combination of these is eaten by the gravitino of $m_{3/2}$, and the other $N-1$ pseudo-goldstinos get a mass of $2 m_{3/2}$ at tree level.
\end{itemize}
Clearly, the structure of SUGRA and AdS$_4$ SUSY is quite rich, but beyond the level of these lectures.

%% file: jthaler_TASI_Master.bbl
\providecommand{\href}[2]{#2}\begingroup\raggedright\begin{thebibliography}{10}

\bibitem{Salam:1974yz}
A.~Salam and J.~Strathdee, {\it {Supergauge Transformations}},  {\em
  Nucl.Phys.} {\bf B76} (1974) 477--482.

\bibitem{Ferrara:1974ac}
S.~Ferrara, J.~Wess, and B.~Zumino, {\it {Supergauge Multiplets and
  Superfields}},  {\em Phys.Lett.} {\bf B51} (1974) 239.

\bibitem{Wess:1992cp}
J.~Wess and J.~Bagger, {\em {Supersymmetry and supergravity}}.
\newblock 1992.

\bibitem{Martin:1997ns}
S.~P. Martin, {\it {A Supersymmetry primer}},
  \href{http://xxx.lanl.gov/abs/hep-ph/9709356}{{\tt hep-ph/9709356}}.

\bibitem{Terning:2006bq}
J.~Terning, {\em {Modern supersymmetry: Dynamics and duality}}.
\newblock 2006.

\bibitem{Luty:2005sn}
M.~A. Luty, {\it {2004 TASI lectures on supersymmetry breaking}},
  \href{http://xxx.lanl.gov/abs/hep-th/0509029}{{\tt hep-th/0509029}}.

\bibitem{Dumitrescu:2011zz}
T.~T. Dumitrescu and Z.~Komargodski, {\it {Aspects of supersymmetry and its
  breaking}},  {\em Nucl.Phys.Proc.Suppl.} {\bf 216} (2011) 44--68.

\bibitem{Weinberg:2000cr}
S.~Weinberg, {\em {The quantum theory of fields. Vol. 3: Supersymmetry}}.
\newblock 2000.

\bibitem{Freedman:2012zz}
D.~Z. Freedman and A.~Van~Proeyen, {\em {Supergravity}}.
\newblock 2012.

\bibitem{Binetruy:2006ad}
P.~Binetruy, {\em {Supersymmetry: Theory, experiment and cosmology}}.
\newblock 2006.

\bibitem{Signer:2009dx}
A.~Signer, {\it {ABC of SUSY}},  {\em J.Phys.} {\bf G36} (2009) 073002,
  [\href{http://xxx.lanl.gov/abs/0905.4630}{{\tt arXiv:0905.4630}}].

\bibitem{Drees:1996ca}
M.~Drees, {\it {An Introduction to supersymmetry}},
  \href{http://xxx.lanl.gov/abs/hep-ph/9611409}{{\tt hep-ph/9611409}}.

\bibitem{Baer:2006rs}
H.~Baer and X.~Tata, {\em {Weak scale supersymmetry: From superfields to
  scattering events}}.
\newblock 2006.

\bibitem{Argyres}
P.~Argyres, {\em {Introduction to Supersymmetry}}.
\newblock 1996.

\bibitem{Dine:2007zp}
M.~Dine, {\em {Supersymmetry and string theory: Beyond the standard model}}.
\newblock 2007.

\bibitem{ArkaniHamed:2005px}
N.~Arkani-Hamed, G.~L. Kane, J.~Thaler, and L.-T. Wang, {\it {Supersymmetry and
  the LHC inverse problem}},  {\em JHEP} {\bf 0608} (2006) 070,
  [\href{http://xxx.lanl.gov/abs/hep-ph/0512190}{{\tt hep-ph/0512190}}].

\bibitem{Cheung:2010mc}
C.~Cheung, Y.~Nomura, and J.~Thaler, {\it {Goldstini}},  {\em JHEP} {\bf 1003}
  (2010) 073, [\href{http://xxx.lanl.gov/abs/1002.1967}{{\tt
  arXiv:1002.1967}}].

\bibitem{Cheung:2011jq}
C.~Cheung, F.~D'Eramo, and J.~Thaler, {\it {The Spectrum of Goldstini and
  Modulini}},  {\em JHEP} {\bf 1108} (2011) 115,
  [\href{http://xxx.lanl.gov/abs/1104.2600}{{\tt arXiv:1104.2600}}].

\bibitem{Craig:2012yd}
N.~Craig, M.~McCullough, and J.~Thaler, {\it {The New Flavor of Higgsed Gauge
  Mediation}},  {\em JHEP} {\bf 1203} (2012) 049,
  [\href{http://xxx.lanl.gov/abs/1201.2179}{{\tt arXiv:1201.2179}}].

\bibitem{Craig:2012di}
N.~Craig, M.~McCullough, and J.~Thaler, {\it {Flavor Mediation Delivers Natural
  SUSY}},  {\em JHEP} {\bf 1206} (2012) 046,
  [\href{http://xxx.lanl.gov/abs/1203.1622}{{\tt arXiv:1203.1622}}].

\bibitem{D'Eramo:2012qd}
F.~D'Eramo, J.~Thaler, and Z.~Thomas, {\it {The Two Faces of Anomaly
  Mediation}},  {\em JHEP} {\bf 1206} (2012) 151,
  [\href{http://xxx.lanl.gov/abs/1202.1280}{{\tt arXiv:1202.1280}}].

\bibitem{Cheung:2011jp}
C.~Cheung, F.~D'Eramo, and J.~Thaler, {\it {Supergravity Computations without
  Gravity Complications}},  {\em Phys.Rev.} {\bf D84} (2011) 085012,
  [\href{http://xxx.lanl.gov/abs/1104.2598}{{\tt arXiv:1104.2598}}].

\bibitem{Bertolini:2011tw}
D.~Bertolini, K.~Rehermann, and J.~Thaler, {\it {Visible Supersymmetry Breaking
  and an Invisible Higgs}},  {\em JHEP} {\bf 1204} (2012) 130,
  [\href{http://xxx.lanl.gov/abs/1111.0628}{{\tt arXiv:1111.0628}}].

\bibitem{Dreiner:2008tw}
H.~K. Dreiner, H.~E. Haber, and S.~P. Martin, {\it {Two-component spinor
  techniques and Feynman rules for quantum field theory and supersymmetry}},
  {\em Phys.Rept.} {\bf 494} (2010) 1--196,
  [\href{http://xxx.lanl.gov/abs/0812.1594}{{\tt arXiv:0812.1594}}].

\bibitem{Haag:1974qh}
R.~Haag, J.~T. Lopuszanski, and M.~Sohnius, {\it {All Possible Generators of
  Supersymmetries of the s Matrix}},  {\em Nucl.Phys.} {\bf B88} (1975) 257.

\bibitem{Zumino:1979et}
B.~Zumino, {\it {Supersymmetry and Kahler Manifolds}},  {\em Phys.Lett.} {\bf
  B87} (1979) 203.

\bibitem{Graham:2009gr}
P.~W. Graham and S.~Rajendran, {\it {A Domino Theory of Flavor}},  {\em
  Phys.Rev.} {\bf D81} (2010) 033002,
  [\href{http://xxx.lanl.gov/abs/0906.4657}{{\tt arXiv:0906.4657}}].

\bibitem{Dobrescu:2010mk}
B.~A. Dobrescu and P.~J. Fox, {\it {Uplifted supersymmetric Higgs region}},
  {\em Eur.Phys.J.} {\bf C70} (2010) 263--270,
  [\href{http://xxx.lanl.gov/abs/1001.3147}{{\tt arXiv:1001.3147}}].

\bibitem{Ibe:2010ig}
M.~Ibe, A.~Rajaraman, and Z.~Surujon, {\it {Does Supersymmetry Require Two
  Higgs Doublets?}},  \href{http://xxx.lanl.gov/abs/1012.5099}{{\tt
  arXiv:1012.5099}}.

\bibitem{Davies:2011mp}
R.~Davies, J.~March-Russell, and M.~McCullough, {\it {A Supersymmetric One
  Higgs Doublet Model}},  {\em JHEP} {\bf 1104} (2011) 108,
  [\href{http://xxx.lanl.gov/abs/1103.1647}{{\tt arXiv:1103.1647}}].

\bibitem{Abbott:1981ke}
L.~Abbott, {\it {Introduction to the Background Field Method}},  {\em Acta
  Phys.Polon.} {\bf B13} (1982) 33.

\bibitem{Giudice:1997ni}
G.~Giudice and R.~Rattazzi, {\it {Extracting supersymmetry breaking effects
  from wave function renormalization}},  {\em Nucl.Phys.} {\bf B511} (1998)
  25--44, [\href{http://xxx.lanl.gov/abs/hep-ph/9706540}{{\tt
  hep-ph/9706540}}].

\bibitem{Giudice:1998bp}
G.~Giudice and R.~Rattazzi, {\it {Theories with gauge mediated supersymmetry
  breaking}},  {\em Phys.Rept.} {\bf 322} (1999) 419--499,
  [\href{http://xxx.lanl.gov/abs/hep-ph/9801271}{{\tt hep-ph/9801271}}].

\bibitem{ArkaniHamed:1998kj}
N.~Arkani-Hamed, G.~F. Giudice, M.~A. Luty, and R.~Rattazzi, {\it
  {Supersymmetry breaking loops from analytic continuation into superspace}},
  {\em Phys.Rev.} {\bf D58} (1998) 115005,
  [\href{http://xxx.lanl.gov/abs/hep-ph/9803290}{{\tt hep-ph/9803290}}].

\bibitem{Clark:1979te}
T.~Clark, O.~Piguet, and K.~Sibold, {\it {THE ABSENCE OF RADIATIVE CORRECTIONS
  TO THE AXIAL CURRENT ANOMALY IN SUPERSYMMETRIC QED}},  {\em Nucl.Phys.} {\bf
  B159} (1979) 1.

\bibitem{Konishi:1983hf}
K.~Konishi, {\it {Anomalous Supersymmetry Transformation of Some Composite
  Operators in SQCD}},  {\em Phys.Lett.} {\bf B135} (1984) 439.

\bibitem{Randall:1998uk}
L.~Randall and R.~Sundrum, {\it {Out of this world supersymmetry breaking}},
  {\em Nucl.Phys.} {\bf B557} (1999) 79--118,
  [\href{http://xxx.lanl.gov/abs/hep-th/9810155}{{\tt hep-th/9810155}}].

\bibitem{Giudice:1998xp}
G.~F. Giudice, M.~A. Luty, H.~Murayama, and R.~Rattazzi, {\it {Gaugino mass
  without singlets}},  {\em JHEP} {\bf 9812} (1998) 027,
  [\href{http://xxx.lanl.gov/abs/hep-ph/9810442}{{\tt hep-ph/9810442}}].

\bibitem{Grisaru:1979wc}
M.~T. Grisaru, W.~Siegel, and M.~Rocek, {\it {Improved Methods for
  Supergraphs}},  {\em Nucl.Phys.} {\bf B159} (1979) 429.

\bibitem{Seiberg:1993vc}
N.~Seiberg, {\it {Naturalness versus supersymmetric nonrenormalization
  theorems}},  {\em Phys.Lett.} {\bf B318} (1993) 469--475,
  [\href{http://xxx.lanl.gov/abs/hep-ph/9309335}{{\tt hep-ph/9309335}}].

\bibitem{Novikov:1983uc}
V.~Novikov, M.~A. Shifman, A.~Vainshtein, and V.~I. Zakharov, {\it {Exact
  Gell-Mann-Low Function of Supersymmetric Yang-Mills Theories from Instanton
  Calculus}},  {\em Nucl.Phys.} {\bf B229} (1983) 381.

\bibitem{Novikov:1985rd}
V.~Novikov, M.~A. Shifman, A.~Vainshtein, and V.~I. Zakharov, {\it {Beta
  Function in Supersymmetric Gauge Theories: Instantons Versus Traditional
  Approach}},  {\em Phys.Lett.} {\bf B166} (1986) 329--333.

\bibitem{Novikov:1985ic}
V.~Novikov, M.~A. Shifman, A.~Vainshtein, and V.~I. Zakharov, {\it
  {Supersymmetric Instanton Calculus (Gauge Theories with Matter)}},  {\em
  Nucl.Phys.} {\bf B260} (1985) 157--181.

\bibitem{Shifman:1986zi}
M.~A. Shifman and A.~Vainshtein, {\it {Solution of the Anomaly Puzzle in SUSY
  Gauge Theories and the Wilson Operator Expansion}},  {\em Nucl.Phys.} {\bf
  B277} (1986) 456.

\bibitem{Shifman:1991dz}
M.~A. Shifman and A.~Vainshtein, {\it {On holomorphic dependence and infrared
  effects in supersymmetric gauge theories}},  {\em Nucl.Phys.} {\bf B359}
  (1991) 571--580.

\bibitem{Dine:1994su}
M.~Dine and Y.~Shirman, {\it {Some explorations in holomorphy}},  {\em
  Phys.Rev.} {\bf D50} (1994) 5389--5397,
  [\href{http://xxx.lanl.gov/abs/hep-th/9405155}{{\tt hep-th/9405155}}].

\bibitem{ArkaniHamed:1997mj}
N.~Arkani-Hamed and H.~Murayama, {\it {Holomorphy, rescaling anomalies and
  exact beta functions in supersymmetric gauge theories}},  {\em JHEP} {\bf
  0006} (2000) 030, [\href{http://xxx.lanl.gov/abs/hep-th/9707133}{{\tt
  hep-th/9707133}}].

\bibitem{Ferrara:1979wa}
S.~Ferrara, L.~Girardello, and F.~Palumbo, {\it {A General Mass Formula in
  Broken Supersymmetry}},  {\em Phys.Rev.} {\bf D20} (1979) 403.

\bibitem{Dimopoulos:1981zb}
S.~Dimopoulos and H.~Georgi, {\it {Softly Broken Supersymmetry and SU(5)}},
  {\em Nucl.Phys.} {\bf B193} (1981) 150.

\bibitem{Intriligator:2007py}
K.~A. Intriligator, N.~Seiberg, and D.~Shih, {\it {Supersymmetry breaking,
  R-symmetry breaking and metastable vacua}},  {\em JHEP} {\bf 0707} (2007)
  017, [\href{http://xxx.lanl.gov/abs/hep-th/0703281}{{\tt hep-th/0703281}}].

\bibitem{Nelson:1993nf}
A.~E. Nelson and N.~Seiberg, {\it {R symmetry breaking versus supersymmetry
  breaking}},  {\em Nucl.Phys.} {\bf B416} (1994) 46--62,
  [\href{http://xxx.lanl.gov/abs/hep-ph/9309299}{{\tt hep-ph/9309299}}].

\bibitem{Fayet:1974jb}
P.~Fayet and J.~Iliopoulos, {\it {Spontaneously Broken Supergauge Symmetries
  and Goldstone Spinors}},  {\em Phys.Lett.} {\bf B51} (1974) 461--464.

\bibitem{Fayet:1974pd}
P.~Fayet, {\it {Supergauge Invariant Extension of the Higgs Mechanism and a
  Model for the electron and Its Neutrino}},  {\em Nucl.Phys.} {\bf B90} (1975)
  104--124.

\bibitem{Barbieri:1982ac}
R.~Barbieri, S.~Ferrara, D.~V. Nanopoulos, and K.~Stelle, {\it {SUPERGRAVITY, R
  INVARIANCE AND SPONTANEOUS SUPERSYMMETRY BREAKING}},  {\em Phys.Lett.} {\bf
  B113} (1982) 219.

\bibitem{Ferrara:1983dh}
S.~Ferrara, L.~Girardello, T.~Kugo, and A.~Van~Proeyen, {\it {RELATION BETWEEN
  DIFFERENT AUXILIARY FIELD FORMULATIONS OF N=1 SUPERGRAVITY COUPLED TO
  MATTER}},  {\em Nucl.Phys.} {\bf B223} (1983) 191.

\bibitem{Rocek:1978nb}
M.~Rocek, {\it {Linearizing the Volkov-Akulov Model}},  {\em Phys.Rev.Lett.}
  {\bf 41} (1978) 451--453.

\bibitem{Lindstrom:1979kq}
U.~Lindstrom and M.~Rocek, {\it {CONSTRAINED LOCAL SUPERFIELDS}},  {\em
  Phys.Rev.} {\bf D19} (1979) 2300--2303.

\bibitem{Komargodski:2009rz}
Z.~Komargodski and N.~Seiberg, {\it {From Linear SUSY to Constrained
  Superfields}},  {\em JHEP} {\bf 0909} (2009) 066,
  [\href{http://xxx.lanl.gov/abs/0907.2441}{{\tt arXiv:0907.2441}}].

\bibitem{Volkov:1973ix}
D.~Volkov and V.~Akulov, {\it {Is the Neutrino a Goldstone Particle?}},  {\em
  Phys.Lett.} {\bf B46} (1973) 109--110.

\bibitem{Clark:1996aw}
T.~Clark and S.~Love, {\it {Goldstino couplings to matter}},  {\em Phys.Rev.}
  {\bf D54} (1996) 5723--5727,
  [\href{http://xxx.lanl.gov/abs/hep-ph/9608243}{{\tt hep-ph/9608243}}].

\bibitem{Fayet:1977vd}
P.~Fayet, {\it {Mixing Between Gravitational and Weak Interactions Through the
  Massive Gravitino}},  {\em Phys.Lett.} {\bf B70} (1977) 461.

\bibitem{Fayet:1979yb}
P.~Fayet, {\it {Scattering Cross-Sections of the Photino and the Goldstino
  (Gravitino) on Matter}},  {\em Phys.Lett.} {\bf B86} (1979) 272.

\bibitem{Dine:1992yw}
M.~Dine and D.~MacIntire, {\it {Supersymmetry, naturalness, and dynamical
  supersymmetry breaking}},  {\em Phys.Rev.} {\bf D46} (1992) 2594--2601,
  [\href{http://xxx.lanl.gov/abs/hep-ph/9205227}{{\tt hep-ph/9205227}}].

\bibitem{Fox:2002bu}
P.~J. Fox, A.~E. Nelson, and N.~Weiner, {\it {Dirac gaugino masses and
  supersoft supersymmetry breaking}},  {\em JHEP} {\bf 0208} (2002) 035,
  [\href{http://xxx.lanl.gov/abs/hep-ph/0206096}{{\tt hep-ph/0206096}}].

\bibitem{Wess:1973kz}
J.~Wess and B.~Zumino, {\it {A Lagrangian Model Invariant Under Supergauge
  Transformations}},  {\em Phys.Lett.} {\bf B49} (1974) 52.

\bibitem{deWit:1975nq}
B.~de~Wit and D.~Z. Freedman, {\it {On Combined Supersymmetric and Gauge
  Invariant Field Theories}},  {\em Phys.Rev.} {\bf D12} (1975) 2286.

\bibitem{Goldberger:1958zz}
M.~Goldberger and S.~Treiman, {\it {Conserved Currents in the Theory of Fermi
  Interactions}},  {\em Phys.Rev.} {\bf 110} (1958) 1478--1479.

\bibitem{Adams:2011vw}
A.~Adams, H.~Jockers, V.~Kumar, and J.~M. Lapan, {\it {N=1 Sigma Models in
  $AdS_4$}},  {\em JHEP} {\bf 1112} (2011) 042,
  [\href{http://xxx.lanl.gov/abs/1104.3155}{{\tt arXiv:1104.3155}}].

\bibitem{deWit:1999ui}
B.~de~Wit and I.~Herger, {\it {Anti-de Sitter supersymmetry}},  {\em Lect.Notes
  Phys.} {\bf 541} (2000) 79--100,
  [\href{http://xxx.lanl.gov/abs/hep-th/9908005}{{\tt hep-th/9908005}}].

\bibitem{Keck:1974se}
B.~Keck, {\it {An Alternative Class of Supersymmetries}},  {\em J.Phys.} {\bf
  A8} (1975) 1819--1827.

\bibitem{Zumino:1977av}
B.~Zumino, {\it {Nonlinear Realization of Supersymmetry in de Sitter Space}},
  {\em Nucl.Phys.} {\bf B127} (1977) 189.

\bibitem{Ivanov:1980vb}
E.~Ivanov and A.~S. Sorin, {\it {SUPERFIELD FORMULATION OF OSP(1,4)
  SUPERSYMMETRY}},  {\em J.Phys.} {\bf A13} (1980) 1159--1188.

\bibitem{Gripaios:2008rg}
B.~Gripaios, H.~D. Kim, R.~Rattazzi, M.~Redi, and C.~Scrucca, {\it {Gaugino
  mass in AdS space}},  {\em JHEP} {\bf 0902} (2009) 043,
  [\href{http://xxx.lanl.gov/abs/0811.4504}{{\tt arXiv:0811.4504}}].

\bibitem{McArthur:2013wv}
I.~McArthur, {\it {Goldstino superfields in $AdS_4$}},  {\em JHEP} {\bf 1304}
  (2013) 124, [\href{http://xxx.lanl.gov/abs/1301.4842}{{\tt
  arXiv:1301.4842}}].

\bibitem{Kugo:1982mr}
T.~Kugo and S.~Uehara, {\it {IMPROVED SUPERCONFORMAL GAUGE CONDITIONS IN THE
  N=1 SUPERGRAVITY YANG-MILLS MATTER SYSTEM}},  {\em Nucl.Phys.} {\bf B222}
  (1983) 125.

\bibitem{Kugo:1982cu}
T.~Kugo and S.~Uehara, {\it {CONFORMAL AND POINCARE TENSOR CALCULI IN N=1
  SUPERGRAVITY}},  {\em Nucl.Phys.} {\bf B226} (1983) 49.

\bibitem{Gates:1983nr}
S.~Gates, M.~T. Grisaru, M.~Rocek, and W.~Siegel, {\it {Superspace Or One
  Thousand and One Lessons in Supersymmetry}},  {\em Front.Phys.} {\bf 58}
  (1983) 1--548, [\href{http://xxx.lanl.gov/abs/hep-th/0108200}{{\tt
  hep-th/0108200}}].

\bibitem{Sundrum:2011ic}
R.~Sundrum, {\it {From Fixed Points to the Fifth Dimension}},  {\em Phys.Rev.}
  {\bf D86} (2012) 085025, [\href{http://xxx.lanl.gov/abs/1106.4501}{{\tt
  arXiv:1106.4501}}].

\end{thebibliography}\endgroup
